%% file: main-ieee.tex
\documentclass[10pt,journal,compsoc]{IEEEtran}

\ifCLASSOPTIONcompsoc
\usepackage[nocompress]{cite}
\else
\usepackage{cite}
\fi
\usepackage{hyperref}
\usepackage{pgf-pie}
\usepackage{graphicx}
\usepackage{standalone}
\usepackage{amsmath}
\usepackage{xcolor} 
\usepackage{subfig}
\usepackage{pgfplots}
\usepackage{pgfplotstable}
\usepackage{adjustbox}
\usepackage{booktabs}
\usepackage{csquotes}
\usepackage{multirow}
\usepackage[figuresright]{rotating}
\usepackage[most]{tcolorbox}
\usepackage{colortbl}
\usepackage{xcolor}
\usepackage{balance}
\usepackage{url}
\usepackage{ifxetex,ifluatex}
\usepackage{etoolbox}
\usepackage{tikz}

\usepackage{framed}

\DeclareMathAlphabet\mathbfcal{OMS}{cmsy}{b}{n}
\pgfplotsset{compat=newest}

\usepackage{amssymb}
\usepackage{pifont}

\newcommand{\cmark}{\textcolor{teal}{\ding{51}}}%
\newcommand{\xmark}{\textcolor{red}{\ding{55}}}%

\definecolor{steel}{rgb}{0, 0.2, 0.9} 

\newtcolorbox{quotebox}{colback=black!10,boxrule=0.4pt,colframe=black,fonttitle=\bfseries,top=2pt,bottom=2pt}

\newcommand{\keybox}[1]{
\begin{tcolorbox}[leftrule=1mm,toprule=0mm,bottomrule=0mm,left=1pt,right=2pt,top=2pt,bottom=2pt]
\em #1
\end{tcolorbox}
}

\newcounter{findingcount}
\input{tablefigures}
\ifCLASSINFOpdf
\else
\fi
\makeatletter
\tcbset{
    myhbox/.style 2 args={%
        enhanced, 
        breakable,
        colback=white,
        colframe=blue!30!black,
        attach boxed title to top left={yshift*=-\tcboxedtitleheight}, 
        title={#2},
        boxed title size=title,
        boxed title style={%
            sharp corners, 
            rounded corners=northwest, 
            colback=tcbcolframe, 
            boxrule=0pt,
        },
        underlay boxed title={%
            \path[fill=tcbcolframe] (title.south west)--(title.south east) 
                to[out=0, in=180] ([xshift=5mm]title.east)--
                (title.center-|frame.east)
                [rounded corners=\kvtcb@arc] |- 
                (frame.north) -| cycle; 
        },
        #1
    }
}   
\makeatother

\newtcolorbox{myhbox}[2][]{%
    myhbox={#1}{#2}
}

\DeclareMathOperator*{\argmin}{argmin}

\mathchardef\mhyphen="2D
\newcommand{\vect}[1]{\boldsymbol{#1}}
\DeclareMathAlphabet\mathbfcal{OMS}{cmsy}{b}{n}

\newcolumntype{P}[1]{>{\centering\arraybackslash}m{#1}}
\newcolumntype{Y}{>{\centering\arraybackslash}X}

\input{Figures/RQ5/data}

   \usepackage[utf8]{inputenc}
   \usepackage[T1]{fontenc}
   \newcommand*\quotefont{\fontfamily{LinuxLibertineT-LF}} 

\newcommand*\quotesize{30} 
\newcommand*{\openquote}
   {\tikz[remember picture,overlay,xshift=-3ex,yshift=-0.5ex]
   \node (OQ) {\quotefont\fontsize{\quotesize}{\quotesize}\selectfont``};\kern0pt}

\newcommand*{\closequote}[1]
  {\tikz[remember picture,overlay,xshift=3ex,yshift={#1}]
   \node (CQ) {\quotefont\fontsize{\quotesize}{\quotesize}\selectfont''};}

\colorlet{shadecolor}{white}

\newcommand*\shadedauthorformat{\emph} 

\newcommand*\authoralign[1]{%
  \if#1l
    \def\authorfill{}\def\quotefill{\hfill}
  \else
    \if#1r
      \def\authorfill{\hfill}\def\quotefill{}
    \else
      \if#1c
        \gdef\authorfill{\hfill}\def\quotefill{\hfill}
      \else\typeout{Invalid option}
      \fi
    \fi
  \fi}
\newenvironment{shadequote}[2][l]%
{\authoralign{#1}
\ifblank{#2}
   {\def\shadequoteauthor{}\def\yshift{-2ex}\def\quotefill{\hfill}}
   {\def\shadequoteauthor{\par\authorfill\shadedauthorformat{#2}}\def\yshift{1ex}}
\begin{snugshade}\begin{quote}\openquote}
{\shadequoteauthor\quotefill\closequote{\yshift}\end{quote}\end{snugshade}}

\begin{document}
	%
	
	

 	\title{Accuracy Can Lie: On the Impact of Surrogate Model in Configuration Tuning}
	
	
	
	
	\author{
		Pengzhou~Chen,~\IEEEmembership{}
		Jingzhi~Gong,~\IEEEmembership{}
		Tao Chen~\IEEEmembership{} %
		
	\IEEEcompsocitemizethanks{
		    \IEEEcompsocthanksitem Corresponding author: Tao Chen (email: t.chen@bham.ac.uk).
			\IEEEcompsocthanksitem Pengzhou Chen is with the School of Computer Science and Engineering, University of Electronic Science and Technology of China, China, 610056.
            \IEEEcompsocthanksitem Jingzhi Gong is with the School of Computer Science, University of Leeds, UK, LS2 9JT.
			\IEEEcompsocthanksitem Tao Chen is with the School of Computer Science, University of Birmingham, UK, B15 2TT.
		}
		\thanks{}
        }

	
	\IEEEtitleabstractindextext{%
		\begin{abstract}
     
     To ease the expensive measurements during configuration tuning, it is natural to build a surrogate model as the replacement of the system, and thereby the configuration performance can be cheaply evaluated. Yet, a stereotype therein is that the higher the model accuracy, the better the tuning result would be, or vice versa. This ``accuracy is all'' belief drives our research community to build more and more accurate models and criticize a tuner for the inaccuracy of the model used. However, this practice raises some previously unaddressed questions, e.g., are the model and its accuracy really that important for the tuning result? Do those somewhat small accuracy improvements reported (e.g., a few \% error reduction) in existing work really matter much to the tuners? What role does model accuracy play in the impact of tuning quality? To answer those related questions, in this paper, we conduct one of the largest-scale empirical studies to date---running over the period of 13 months $24\times7$---that covers 10 models, 17 tuners, and 29 systems from the existing works while under four different commonly used metrics, leading to 13,612 cases of investigation. Surprisingly, our key findings reveal that the accuracy can lie: there are a considerable number of cases where higher accuracy actually leads to no improvement in the tuning outcomes (up to 58\% cases under certain setting), or even worse, it can degrade the tuning quality (up to 24\% cases under certain setting). We also discover that the chosen models in most proposed tuners are sub-optimal and that the required \% of accuracy change to significantly improve tuning quality varies according to the range of model accuracy. Deriving from the fitness landscape analysis, we provide in-depth discussions of the rationale behind, offering several lessons learned as well as insights for future opportunities. Most importantly, this work poses a clear message to the community: we should take one step back from the natural ``accuracy is all'' belief for model-based configuration tuning.

		\end{abstract}

		\begin{IEEEkeywords}
		Search-based software engineering, software configuration tuning, performance optimization, configurable systems, model-based optimization, heuristic algorithms, hyperparameter tuning, fitness landscape analysis.
	\end{IEEEkeywords}}

	\maketitle

	\IEEEdisplaynontitleabstractindextext

	%
	\IEEEpeerreviewmaketitle

\input{introduction}

\input{background}

\input{methodology}

\input{results}

\input{discussion}

\input{lessons}


\input{threats}

\input{related}

\input{conclusion}

\section*{Acknowledgments}
This work was supported by an NSFC Grant (62372084) and a UKRI Grant (10054084).


\bibliographystyle{IEEEtranS}
\bibliography{IEEEabrv,reference}

\begin{IEEEbiography}[{\includegraphics[width=1in,height=1.25in,clip,keepaspectratio]{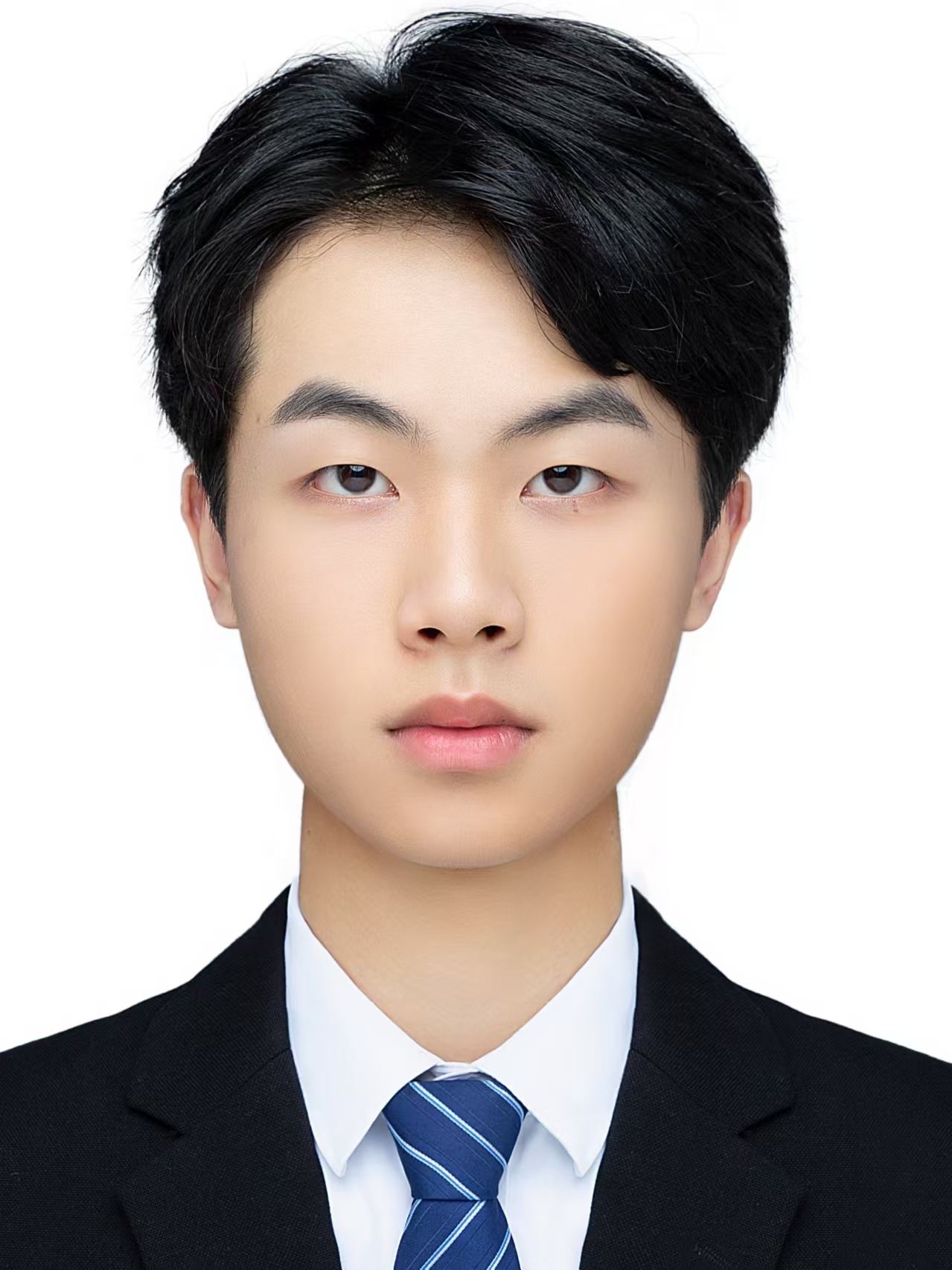}}]{Pengzhou Chen} received the B.Eng. degree in Computer Science from the University of Electronic Science and Technology of China in 2024. As a member of IDEAS Lab, he is currently working towards the M.Sc. degree, with the School of Computer Science and Engineering, University of Electronic Science and Technology of China. His research interests include search-based software engineering, performance modeling, and machine learning. 
\end{IEEEbiography}

\begin{IEEEbiography}[{\includegraphics[width=1in,height=1.25in,clip,keepaspectratio]{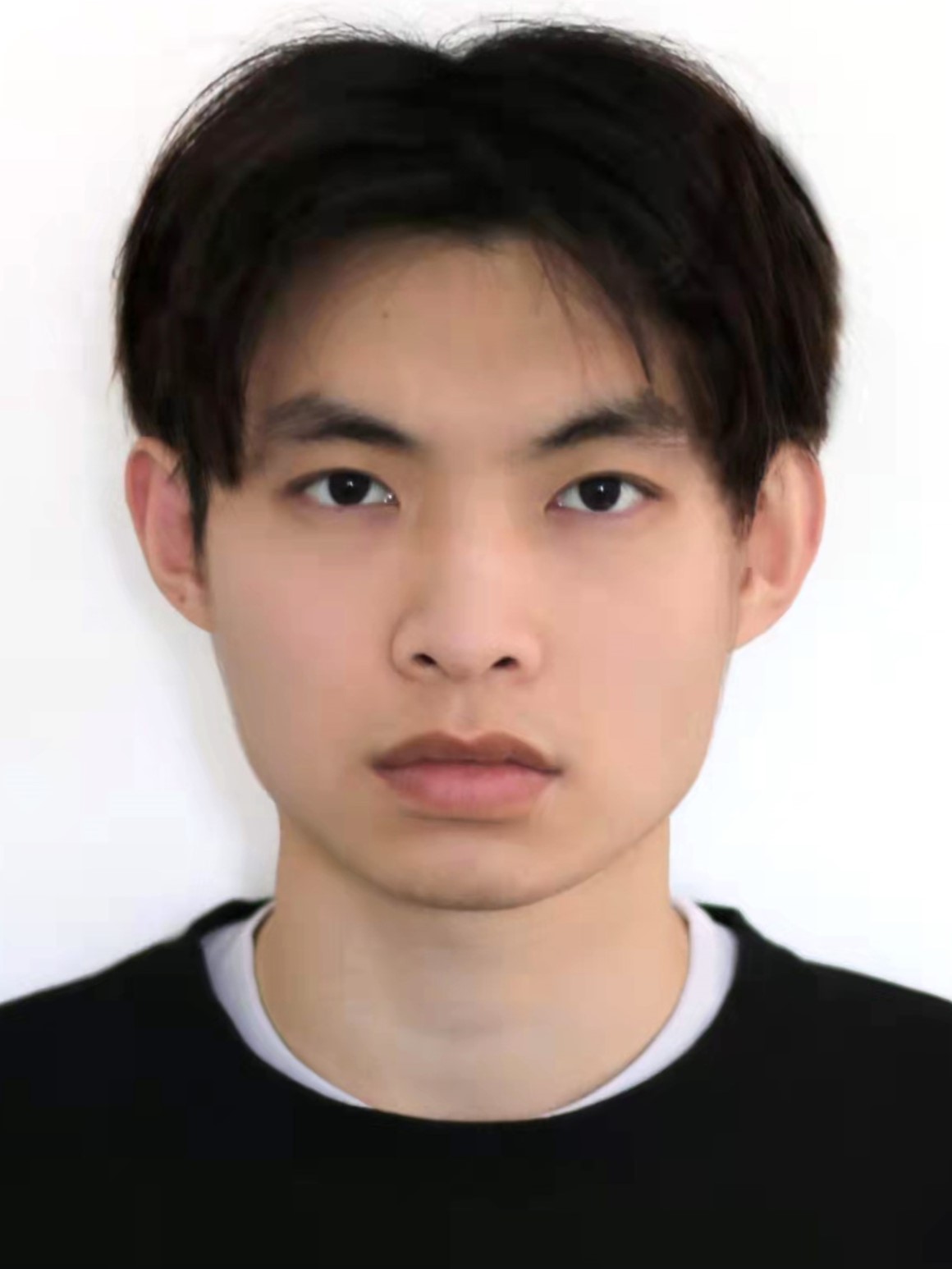}}]{Jingzhi Gong} received the Ph.D. degree from the Loughborough University, U.K., in 2024. As a member of the IDEAS Lab, his doctoral research has been published in major Software Engineering conferences/journals including FSE, TSE, TOSEM, and MSR. His research interests mainly include configuration performance modeling, machine/deep learning, and software engineering. He is currently a postdoctoral research fellow at the University of Leeds, U.K., and TurinTech AI, U.K., developing cutting-edge code optimization approaches using large language models and compiler-based techniques. 
\end{IEEEbiography}

\begin{IEEEbiography}[{\includegraphics[width=1in,height=1.25in,clip,keepaspectratio]{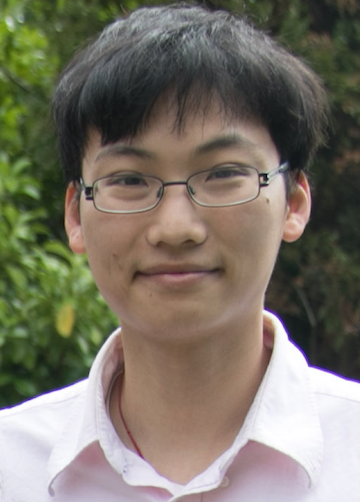}}]{Tao Chen} directs the Intelligent Dependability Engineering for Adaptive Software Laboratory (IDEAS Lab), conducting cutting-edge research on the intersection between AI and Software Engineering. Currently, his research interests include performance engineering, self-adaptive systems, search-based software engineering, data-driven software engineering, and their interplay with machine learning and computational intelligence. His work has been regularly published in all major Software Engineering conferences/journals (ICSE, FSE, ASE, TOSEM, and TSE) and has been supported by projects worth over $\pounds$1 million from external funding bodies. He currently serves as an Associate Editor for ACM Transactions on Autonomous and Adaptive Systems and a Program Committee member for many prestigious conferences.
\end{IEEEbiography}

\end{document}

%% file: tablefigures.tex
\usepackage{algorithm2e}
\usepackage{xcolor}

\SetAlFnt{\footnotesize}

\SetAlCapSty{xAlCapSty}

\SetCommentSty{xCommentSty}

\SetNlSty{mynlfont}{}{} 

\LinesNumbered

\SetSideCommentRight

\DontPrintSemicolon

\RestyleAlgo{algoruled}

\usepackage[autolanguage]{numprint}

\newcommand*\np[2][z]{
\ifx z#1%
$\numprint{#2}$%
\else%
$\numprint[#1]{#2}$%
\fi\xspace%
}

\usepackage{fp}
\newcommand{\ShowAbsoluteNumber}[1]{%
\ifnum #1<10%
{\hspace*{0pt}#1}%
\else%
\ifnum #1<100%
{\hspace*{0pt}#1}%
\else%
\ifnum #1<1000%
{\hspace*{0pt}#1}%
\else%
{\numprint{#1}}%
\fi%
\fi%
\fi%
}

\newcommand{\ShowPercentage}[2]{%
\FPeval\percentage{round(#1/#2*100,0)}%
\FPeval\percentageOneDecimal{round(#1/#2*100,1)}%
\ifnum \percentage=0%
{\np[\%]{\FPprint{percentageOneDecimal}}}%
\else%
\ifnum \percentage<10%
{\np[\%]{\FPprint{percentageOneDecimal}}}%
\else%
{\np[\%]{\FPprint{percentageOneDecimal}}}%
\fi%
\fi%
\xspace
}

\newcommand{\ShowPercentageTwo}[2]{%
\FPeval\percentagetwo{round(#1/#2*100,0)}%
\FPeval\percentageTwoDecimal{round(#1/#2*100,2)}%
\ifnum \percentagetwo=0%
{\np[\%]{\FPprint{percentageTwoDecimal}}}%
\else%
\ifnum \percentagetwo<10%
{\np[\%]{\FPprint{percentageTwoDecimal}}}%
\else%
{\np[\%]{\FPprint{percentageTwoDecimal}}}%
\fi%
\fi%
\xspace
}

\newlength\BARSIZE  
\newcommand{\inlinechart}[2]{%
\FPeval{\BLACKBARSIZE}{#1/#2}\textcolor{black!80}{\rule{\BLACKBARSIZE\BARSIZE}{1.6ex}}%
\FPeval{\BLACKBARSIZE}{1 - (#1/#2)}\textcolor{black!10}{\rule{\BLACKBARSIZE\BARSIZE}{1.6ex}}%
}

\newcommand{\pinlinechart}[3]{%
\FPeval{\BLACKBARSIZE}{#1/#2}\textcolor{steel!#3}{\rule{\BLACKBARSIZE\BARSIZE}{1.6ex}}%
\FPeval{\BLACKBARSIZE}{1 - (#1/#2)}\textcolor{black!10}{\rule{\BLACKBARSIZE\BARSIZE}{1.6ex}}%
}

\newcommand{\ninlinechart}[3]{%
\FPeval{\BLACKBARSIZE}{1 - (#1/#2)}\textcolor{black!10}{\rule{\BLACKBARSIZE\BARSIZE}{1.6ex}}%
\FPeval{\BLACKBARSIZE}{#1/#2}\textcolor{red!#3}{\rule{\BLACKBARSIZE\BARSIZE}{1.6ex}}%
}

\newcommand*\percent[5][v]{%
\ifx q#1%
    \np{#2}/\np{#3}(\ShowPercentage{#2}{#3})\else%
\ifx w#1%
    \np{#2}(\ShowPercentage{#2}{#3})\else%
\ifx m#1%
    \np{#2}%
    \inlinechart{#2}{#3}\else%
\ifx t#1%
    \FPprint{#2}%
    \hspace*{0.5ex}%
    \inlinechart{#2}{#3}\else%
\ifx p#1%
    \FPprint{#4}%
    \hspace*{0.5ex}%
    \pinlinechart{#2}{#3}{#5}\else%
\ifx n#1%
    \FPprint{#4}%
    \hspace*{0.5ex}%
    \ninlinechart{#2}{#3}{#5}\else%
\ifx c#1%
    \inlinechart{#2}{#3}{#5}%
\else%
    \np{#2}%
    \ifx r#1%
        /\np{#3}%
    \fi%
    \hspace*{0.5ex}(\ShowPercentage{#2}{#3}) %
    \inlinechart{#2}{#3}%
    \xspace
\fi\fi\fi\fi\fi%
}

%% file: Figures/RQ5/data.tex
  \pgfplotstableread{
x         y    y-max  y-min
A 0.7000880604631059 1.0801880574521463 1.0801880574521463
B 6.142070913249965 4.802509230288999 4.802509230288999
C 9.275438851045308 7.15610640061132 7.15610640061132
D 12.99717189129753 7.797263165110128 7.797263165110128
E 23.518011359744726 11.726846274270576 11.726846274270576
F 21.418979943802956 15.140817225439962 15.140817225439962
G 22.574429791802256 13.030649755965026 13.030649755965026
H 30.5213561256203 18.189024701359216 18.189024701359216
I 38.31806490738754 23.2170575269255 23.2170575269255
J 35.07876246570294 24.96630225069647 24.96630225069647
K 67.98714576416815 57.61285979837609 57.61285979837609
}{\mytable}
  \pgfplotstableread{
x         y    y-max  y-min
A 0.5626829474043065 0.8953848486000578 0.8953848486000578
B 4.199363071531841 4.7711156882199255 4.7711156882199255
C 4.685307957958996 5.635375121877837 5.635375121877837
D 8.183684117094666 4.686275336680698 4.686275336680698
E 9.19728487060809 6.793228658693461 6.793228658693461
F 20.30709631791353 11.639060920926068 11.639060920926068
G 23.490604668950734 21.476298098481582 21.476298098481582
H 25.272117999377652 19.96240431232658 19.96240431232658
I 15.343508566105642 12.599021623179993 12.599021623179993
J 27.345847305564746 17.40571248246213 17.40571248246213
K 55.569063723553015 50.15056521404109 50.15056521404109
}{\mytablee}

  \pgfplotstableread{
x         y    y-max  y-min
A 0.8109780752638777 1.3843790882509972 1.3843790882509972
B 5.458778695974042 4.626983197553234 4.626983197553234
C 7.743869069043876 6.775732245308482 6.775732245308482
D 13.3581002319833 9.93687156078057 9.93687156078057
E 18.904799921498658 13.143552462016892 13.143552462016892
F 23.74506294938417 16.129802259588562 16.129802259588562
G 25.754499286634285 16.512320387875896 16.512320387875896
H 30.53977059771995 17.720843537442573 17.720843537442573
I 36.813873234553306 21.055425619127575 21.055425619127575
J 41.25087658622556 17.571510897554457 17.571510897554457
K 60.117017233029756 55.60162992874211 55.60162992874211
}{\mytablea}

  \pgfplotstableread{
x         y    y-max  y-min
A 0.7615056763155121 1.2117567909496905 1.2117567909496905
B 3.357936383678686 2.8176274462654822 2.8176274462654822
C 6.974739211118797 7.367532176269756 7.367532176269756
D 10.70870594776044 7.817426977487917 7.817426977487917
E 15.814315540464344 11.46514973438296 11.46514973438296
F 20.128821914960763 15.27352142297676 15.27352142297676
G 25.018477416181618 21.02612052121554 21.02612052121554
H 39.54813656464516 23.58899580071828 23.58899580071828
I 39.40703580168083 24.334393346241733 24.334393346241733
J 27.870110294269708 18.287662427127955 18.287662427127955
K 56.80279102864985 50.80433530613973 50.80433530613973
}{\mytableaa}

  \pgfplotstableread{
x         y    y-max  y-min
A 0.812054479943299 1.3248803340155324 1.3248803340155324
B 9.112860952960375 4.362728318903163 4.362728318903163
C 14.888806587943712 6.952933462369999 6.952933462369999
D 23.173467101324626 9.694392648015544 9.694392648015544
E 33.184528400798094 13.730871745461082 13.730871745461082
F 28.914259280062595 18.27917973756212 18.27917973756212
G 41.54716114303943 10.191061180553957 10.191061180553957
H 11.39912271121439 9.030842361786107 9.030842361786107
I 66.49815448491339 13.323834065845459 13.323834065845459
J 80.59720572749404 18.250723011555113 18.250723011555113
K 72.58143347510882 39.132547595757664 39.132547595757664
}{\mytableb}

  \pgfplotstableread{
x         y    y-max  y-min
A 0.6784815701590656 1.0243756213403368 1.0243756213403368
B 8.150357398836482 3.991569116425162 3.991569116425162
C 8.70860469026286 6.569968999200625 6.569968999200625
D 17.98777842633074 7.430945114127391 7.430945114127391
E 13.67340742453421 13.759919174674115 13.759919174674115
F 5.955037039002232 0.0001 0.0001
G 0.0001 0.0001 0.0001
H 0.0001 0.0001 0.0001
I 31.881203659058954 17.895319602840992 17.895319602840992
J 93.73742626820898 5.039803362277204 5.039803362277204
K 46.956692333727794 29.054058679968815 29.054058679968815

}{\mytablebb}

  \pgfplotstableread{
x         y    y-max  y-min
A 1.1087408925903586 1.441817259504485 1.441817259504485
B 10.086669081798162 5.025539621648598 5.025539621648598
C 15.562676252157123 8.618516758115932 8.618516758115932
D 23.117510719123345 10.090752806980237 10.090752806980237
E 31.416162310651885 13.896283394074954 13.896283394074954
F 31.873118984883547 16.634670638328277 16.634670638328277
G 46.44617275183983 14.202562811753635 14.202562811753635
H 50.612397579520874 27.51876745000294 27.51876745000294
I 68.22456201120298 14.82071438501148 14.82071438501148
J 75.37202627519096 24.978801156781543 24.978801156781543
K 85.92845394667201 45.47990191223782 45.47990191223782
}{\mytablec}

  \pgfplotstableread{
x         y    y-max  y-min
A 1.0954455069534752 1.6613418490195684 1.6613418490195684
B 9.686648161369826 4.450824309250104 4.450824309250104
C 15.495885885137564 9.158059667812982 9.158059667812982
D 14.826273463368246 9.087392027663842 9.087392027663842
E 17.166758146396134 14.43737702370175 14.43737702370175
F 20.46337108014706 3.831228120052696 3.831228120052696
G 49.937682362448335 0.0001 0.0001
H 0.0001 0.0001 0.0001
I 64.20880153218013 20.755454025491797 20.755454025491797
J 87.85172801993649 12.716587114670649 12.716587114670649
K 48.78534412842534 32.12611825679194 32.12611825679194
}{\mytablecc}

  \pgfplotstableread{
x         y    y-max  y-min
A 0.6349723381866571 0.6832979145618729 0.6832979145618729
B 1.8912546995576045 2.5710331045927615 2.5710331045927615
C 4.8366819314109915 5.3889335584612175 5.3889335584612175
D 7.840873309332455 7.517992879096022 7.517992879096022
}{\rmytable}
  \pgfplotstableread{
x         y    y-max  y-min
A 1.1097960054100735 0.8943369182617936 0.8943369182617936
B 1.3831119379255665 1.5010891663249837 1.5010891663249837
C 3.9958789691677805 4.335980936440258 4.335980936440258
D 6.614375312663061 6.935482583742389 6.935482583742389
}{\rmytablee}

  \pgfplotstableread{
x         y    y-max  y-min
A 1.0312996813585766 0.8395356907761508 0.8395356907761508
B 1.8113333436275276 2.150200227052395 2.150200227052395
C 4.688559773722409 5.0068269217738735 5.0068269217738735
D 8.325481488198136 8.348556421094242 8.348556421094242
}{\rmytablea}

  \pgfplotstableread{
x         y    y-max  y-min
A 1.1188561350484985 0.8805912738279683 0.8805912738279683
B 1.4972076372124448 1.9672822562922319 1.9672822562922319
C 4.241504146842288 4.501380091388367 4.501380091388367
D 6.047917161333886 6.261234791647532 6.261234791647532
}{\rmytableaa}

  \pgfplotstableread{
x         y    y-max  y-min
A 1.3044187866134767 1.4108586733412283 1.4108586733412283
B 4.795305224973517 3.9117007106430477 3.9117007106430477
C 10.021579105605598 7.6191302704808175 7.6191302704808175
D 11.678700334009532 11.22140553032058 11.22140553032058
}{\rmytableb}

  \pgfplotstableread{
x         y    y-max  y-min
A 0.9018783686237544 1.333485521411199 1.333485521411199
B 2.94949926110997 2.936392296176108 2.936392296176108
C 7.9985845078899835 8.93285098095569 8.93285098095569
D 11.999655022600544 9.41901229259781 9.41901229259781
}{\rmytablebb}

  \pgfplotstableread{
x         y    y-max  y-min
A 1.5809401046329081 1.6109194799826883 1.6109194799826883
B 6.273619334146393 4.595134232174214 4.595134232174214
C 10.642833480255323 7.054848654200045 7.054848654200045
D 14.833238956446301 10.475431420039818 10.475431420039818
}{\rmytablec}

  \pgfplotstableread{
x         y    y-max  y-min
A 0.6055489276122876 0.6520745747968063 0.6520745747968063
B 3.6645457823336094 2.914269663051915 2.914269663051915
C 8.015823676018222 7.040676840131879 7.040676840131879
D 10.725780198138574 9.676402981539175 9.676402981539175
}{\rmytablecc}

%% file: introduction.tex
\IEEEraisesectionheading{\section{Introduction}\label{sec:introduction}}

\IEEEPARstart{M}{odern} software systems are often designed with great flexibility, containing a vast number of configuration options to satisfy diverse needs~\cite{DBLP:journals/tse/SayaghKAP20,DBLP:conf/sigsoft/XuJFZPT15}. Yet, this flexibility comes with a cost: it has been shown that the system configuration, if not set appropriately, can cause devastating issues to the performance (e.g., throughput, runtime, or latency), leaving the full potential of a system untapped. For example, a study found that 59\% of the severe performance bugs worldwide are caused by poor configuration~\cite{DBLP:conf/esem/HanY16}; Jamshidi and Casale~\cite{DBLP:conf/mascots/JamshidiC16} reveal that the best configuration of streaming software \textsc{Storm} lead to a throughput which is $480 \times$ better than that of the default one. Therefore, configuration tuning that aims to search for the optimal configuration according to a performance attribute at deployment time is of high importance during the software quality assurance phase.

However, simply traversing the entire configuration space to find the optimum is unlikely to succeed, due to two reasons: (1) the number of configuration options, and hence the resulting configuration space, has been increasing dramatically~\cite{DBLP:conf/sigsoft/XuJFZPT15}. For instance, even after the performance-sensitive configuration options have been elicited, the compiler \textsc{SaC} still involves 59 options, leading to a space of possible configurations up to $3.14 \times 10^{28}$. (2) Measuring the configurations on the system demands considerable time and resources and hence it is profoundly expensive~\cite{flash,DBLP:conf/sigsoft/0001L21,DBLP:conf/sigsoft/0001L24,DBLP:conf/mascots/JamshidiC16}. For example, Chen and Li~\cite{DBLP:conf/sigsoft/0001L21,DBLP:conf/sigsoft/0001L24} report that it can take up to 166 minutes to measure a single configuration on the database system \textsc{MariaDB}.

Over the past decades, researchers have designed sophisticated heuristic tuners to address the above challenges, rooting from different research communities such as databases~\cite{bestconfig}, algorithms~\cite{lopez2016irace}, machine learning~\cite{Sway}, and big data~\cite{conex}. Examples include \texttt{BestConfig}~\cite{bestconfig} that leverages local search with recursive bounds and \texttt{ConEx}~\cite{conex} that uses Markov Chain Monte Carlo sampling. Those approaches, termed \textit{model-free tuners}, share one common ground: they solely rely on direct measurement of the systems in guiding the tuning, despite the expectation that, with the help of intelligently designed heuristic, the number of measurements would be drastically reduced for finding the (near-)optimal configuration.

Given the expensive measurement in configuration tuning, a perhaps more natural resolution is to build a surrogate model---the mathematical function that reflects the correlation between configuration and performance---as a replacement of the system itself, hence mitigating expensive profiling directly. Indeed, one recent survey~\cite{DBLP:journals/tse/SayaghKAP20} on software configuration concludes that \textit{``a large body of work tries to model the configuration of an
application, then use this model to suggest an optimal configuration''.} Approaches that follow this thread, namely \textit{model-based tuners}, can be aligned with one of the following two categories:

\begin{itemize}
    \item \textbf{Batch model-based tuners, e.g.,~\cite{shi2024efficient,DBLP:journals/tosem/ChenLBY18,DBLP:journals/jss/ChengYW21,DBLP:journals/tsc/ChenB17,DBLP:conf/kbse/BaoLXF18,DBLP:journals/taco/LiL22,DBLP:conf/ijcai/AnsoteguiMSST15,DBLP:conf/asplos/YuBQ18}:} This tuner type refers to the model-free tuners where the direct system measurement is seamlessly replaced by a surrogate model, which is trained under a batch of previously measured configuration samples in advance. The model prediction steers the tuning process while no other measurements are involved throughout the tuning.
    \item \textbf{Sequential model-based tuners, e.g.,~\cite{DBLP:conf/cloud/WangYYWLSHZ21,BOCA,flash,Ottertune,restune,SMAC}:} This is the type of tuner that sophisticatedly synergizes a model with the tuner, in which the model is updated progressively with newly measured configurations and it also predicts configuration performance to influence the search direction in tuning.
\end{itemize}

Since the models guide the tuning directions within model-based tuners, it is natural to believe that their accuracy (measured by some forms of prediction errors~\cite{flash,DeepPerf}) to the real system is crucial. Indeed, while model-based tuners are rarely compared with model-free tuners in a fair manner, it is not uncommon to see similar claims below:

\begin{shadequote}[l]{--- Wang et al.~\cite{DBLP:conf/cloud/WangYYWLSHZ21}}
\textit{Shortly after 28 samplings, the fine-tuned regression can accurately fit the target. This explains why the tuner can find the global optimum in a few shots.}
\end{shadequote}

\noindent This is a clear example where the accuracy of the model is used to explain its usefulness to the configuration tuning. At the same time, the claim of insufficient model accuracy is also commonly the motivation for proposing a new model to be equipped with a model-based tuner:

\begin{shadequote}[l]{--- Yu et al.~\cite{DBLP:conf/asplos/YuBQ18}}
\textit{We see that the average errors of models built by Response Surface, Artificial Neural Network, Support Vector Machine, and Random Forest are 23\%, 27\%, 14\%, and 18\%, respectively. We believe that performance models with such high errors cannot accurately identify the optimal configurations.}
\end{shadequote}




\noindent Yet, in another example, Zhu et al.~\cite{bestconfig} criticize the model-based tuners for the inaccuracy of the surrogate model therein, hence arguing that the model-free counterparts should be favored:

\begin{shadequote}[l]{--- Zhu et al.~\cite{bestconfig}}
\textit{Although the prediction is improving as the number of samples increases, Gaussian Process's predictions about best points are hardly accurate....methods like Co-Training Model Tree and Gaussian Process cannot output competitive configuration settings.}
\end{shadequote}

%


\begin{figure}[t!]
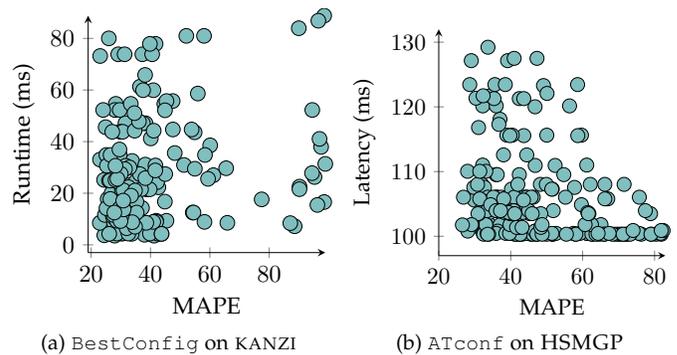

\centering
\subfloat[\texttt{BestConfig} on \textsc{kanzi}]{\includestandalone[width=0.49\columnwidth]{Figures/exp1}}
~\hfill
\subfloat[\texttt{ATconf} on \textsc{HSMGP}]{\includestandalone[width=0.49\columnwidth]{Figures/exp2}}

\caption{Exampled accuracy (MAPE) and achieved tuning performance under (a) batch and (b) sequential model-based tuners with different models. Each point denotes the results of a model under one run (10 models $\times$ 30 runs).} 
\label{fig:exp}
\end{figure}

The above, together with the other vast majority of work on configuration performance modeling that solely seeks to improve model accuracy~\cite{DeepPerf,DaL,HINNPerf,DECART,SPL,gong2024dividable}, implies that the community tends to believe \textbf{\textit{the accuracy is the key that strongly impacts whether mode-based tuners should be favored over their model-free counterparts (i.e., the model-free tuners), as it is the dominant factor in the usefulness of a model for configuration tuning}}.


Nevertheless, despite the widely followed general belief, we have never had a thorough understanding of how the models impact the tuning process. In fact, a few preliminary works~\cite{DBLP:conf/sigsoft/NairMSA17,flash} and our years of experience have hinted that the above belief can be inaccurate or even misleading. For example, Figure~\ref{fig:exp} shows the tuning performance led by models with diverse accuracy, measured by Mean Absolute Percentage Error (MAPE)\footnote{MAPE is a popular residual metric of accuracy that computes the percentage of absolute errors relative to the actual values~\cite{DeepPerf,DaL,SPL,DECART}.}, under the aforementioned two types of model-based tuners over 30 runs. Clearly, in Figure~\ref{fig:exp}a, it is difficult to say for sure that a better accuracy (smaller MAPE) would lead to better performance (smaller runtime). The correlation between model accuracy and the tuned configuration performance does not exhibit clear patterns, as they change in a rather non-linear and non-monotonic manner. In contrast, for the case in Figure~\ref{fig:exp}b, there is a clear trace that a better accuracy actually leads to worse performance (higher latency of the system), which is even more counter-intuitive to the general belief.

Therefore, it raises a question to the current research that follows the ``accuracy is all'' belief when leveraging a model to guide the tuning: what roles do the model and its accuracy play in the configuration tuning?

\subsection{Contributions}

To better understand the above doubt, in this paper, we conduct a systematic, large-scale empirical study that covers 10 models, 17 tuners, and 29 systems from the existing works while under four different metrics, leading to 13,612 cases of investigation. Our findings are surprising and even counter-intuitive, from which the key and most unexpected observation is probably \textbf{\textit{the accuracy can lie}}, suggesting that the current practice that relies on the ``accuracy is all'' belief is likely to be misleading. Therefore, the results conjecture that we should take one step back from using accuracy as the key driver of model-based configuration tuning. Specifically, we make several contributions to the community as follows.

\subsubsection{New Findings}

We reveal findings that have not been previously explored in a systematic way:

   \begin{enumerate}
       \item Compared with the model-free counterparts under a sufficient and fair budget, models are helpful for sequential model-based tuners with up to 72\% cases and a maximum of $5\times$ tuning improvements, but they tend to have marginal impacts or can even be harmful to the tuning quality of batch model-based tuners.

       \item The originally chosen model in sequential model-based tuners can be considerably improved in 46\% of the cases by simply switching to the other ``newly created'' tuner-model pairs; while for 50\% cases they have marginal difference.

       \item The most/least accurate model can only serve as an indication of the best/worst tuning quality in between 14\% and 45\% of the cases.

       \item The correlations between model accuracy and the goodness of tuning are far from being positively strong as implied in the belief: up to 97\% of the cases, they are mostly negligible, or sometimes, even negative, i.e., worse accuracy can lead to better tuning results.

       \item The necessary accuracy change to create significant tuning improvement varies depending on the range of the model's accuracy, e.g., for the model with a MAPE falls in $[30,40]$, it needs at least $13.3\%$ improvement for having considerably better tuning performance. Notably, the better the accuracy, the smaller the accuracy change is needed to considerably enhance the tuning.
   \end{enumerate}

\subsubsection{New Interpretations}

  By means of representative examples, we demonstrate a way to explain the reasons behind the most unexpected results from our empirical study, using the concepts/metrics from fitness landscape analysis~\cite{DBLP:series/sci/PitzerA12}. In particular, such an analysis provides new interpretations of the following questions observed from the results of the empirical study:

   \begin{enumerate}
       \item Why is the model useful (useless) to the tuning
quality?
\item Why better model accuracy does not always lead
to superior tuning quality?
\item Why does a better accuracy need a smaller change to
significantly influence the tuning?
   \end{enumerate}

\subsubsection{New Insights and Opportunities}

   Deriving from the observations and analysis, we summarize a few lessons learned that provide insights for future research opportunities in the field:

   \begin{enumerate}
       \item While models help to significantly reduce overhead, they are only useful for tuning quality under progressive updates when the budget is sufficient and fair, hence exploring efficient online model learning is a promising research direction.

        \item Manually selecting a model beforehand makes it difficult to find the optimal choice for tuning, hence the combination of model-tuner pair selection and tuning the configuration can form a new bi-level optimization problem that demands automation.

        \item The community should shift away from the accuracy-driven research for model-based configuration tuning. A new question would be: what other (efficient) proxies, alongside accuracy, can better measure the model usefulness for tuning quality. Further, models should incorporate code patterns that cause the landscape sparsity and ruggedness.

        \item It makes little sense to claim the benefit of a certain \% of accuracy improvement alone; at least, one should refer to the minimum accuracy changes discovered in this work. This calls for a more systematic procedure to examine and interpret the meaningfulness of the change in model accuracy.
   \end{enumerate}

\subsection{Open Science and Organization}

To promote open science practices, all source code, data, and supplementary materials of this work can be publicly accessed at our repository: \textcolor{blue}{\texttt{\url{https://github.com/ideas-labo/model-impact}}}.

The rest of this paper is organized as follows. Section~\ref{sec:background} introduces the necessary preliminaries and background of this work. Section~\ref{sec:methodology} describes our empirical research methodology. Section~\ref{sec:results} presents the results with a detailed analysis. Section~\ref{sec:discussion} discusses the possible rationale behind the findings, followed by a summary of the lessons learned and future opportunities in Section~\ref{sec:lessons}. Sections~\ref{sec:threats} and~\ref{sec:related} present the threats to validity and related work, respectively. Finally, Section~\ref{sec:conclusion} concludes the paper with pointers for future work.

%% file: Figures/exp1.tex
\begin{tikzpicture}
    \begin{axis}[
      width=5cm,
    height=5cm,
    ymin=-3,
    xmin=19,
        axis x line  = bottom,
    axis y line  = left,
        ylabel=Runtime (ms),
        xlabel=MAPE,
       legend style={at={(0.5,0.95),font=\fontsize{7}{8}\selectfont},
      anchor=north west,legend columns=1}]
      
    \addplot[only marks,mark=*,fill=teal!50,draw=black,mark size=3pt] plot coordinates {
    
(24.812198746538566, 9.342)
(24.4406346685552, 3.702)
(22.93029711297209, 8.508)
(24.366772793650863, 25.164)
(24.636791190469133, 11.59)
(24.895746037447854, 45.586)
(29.263212448505598, 33.974)
(28.812730953371517, 21.04)
(23.0011244207385, 33.032)
(27.65246652336049, 16.75)
(25.141027853784685, 34.83)
(25.57927206208428, 17.79)
(30.393423147935273, 4.404)
(28.309335689788274, 54.628)
(29.118480551626025, 73.83)
(34.212708408218894, 22.85)
(30.38013860448901, 4.368)
(26.017464772386095, 24.924)
(27.676165300072082, 8.706)
(33.97491955770421, 20.552)
(32.335657182481604, 8.786)
(27.348423211354024, 23.242)
(26.909040658386964, 20.066)
(27.36646501291073, 10.846)
(28.631891064581826, 35.11)
(25.320635412870963, 30.642)
(26.955491332702753, 12.522)
(27.908189619651253, 3.558)
(30.371831575185205, 21.132)
(45.37957169189639, 9.342)
(28.011206811037496, 52.258)
(51.28549867070367, 30.948)
(40.980984890575996, 59.912)
(47.47619384001364, 55.716)
(41.1807255070914, 3.558)
(40.10212905023548, 41.24)
(54.87193858834305, 43.614)
(29.506567168473197, 26.284)
(35.69301407761829, 10.61)
(44.65991856267758, 54.628)
(37.25485858011623, 73.83)
(52.065222879786056, 80.968)
(61.18714284681231, 27.0)
(59.385421963906424, 25.574)
(54.94067657524621, 29.482)
(41.268762185140446, 8.928)
(35.55910241460217, 23.242)
(41.7129467246486, 77.866)
(36.45828299882981, 61.278)
(65.48904399074735, 29.69)
(42.65471547501535, 22.306)
(31.553882172310082, 47.116)
(54.236487469653724, 12.522)
(53.85528322745831, 44.706)
(38.33300433467842, 47.116)
(97.4774298779933, 37.96)
(95.24500256793769, 26.284)
(94.50373266523455, 52.258)
(98.50451581595108, 88.796)
(94.41290504925458, 27.788)
(96.9391412184302, 41.052)
(90.20327995755885, 22.552)
(96.54339202859475, 86.832)
(98.51237380060283, 16.592)
(96.22833342343257, 15.462)
(98.88155681949338, 31.368)
(88.5638797069249, 7.2)
(89.95941698472812, 83.924)
(87.097357455979, 8.414)
(41.920188736786855, 9.342)
(29.6503513247359, 52.258)
(37.98964771873198, 3.908)
(37.50178203594697, 59.912)
(33.81478938545161, 32.494)
(45.38389942712412, 55.716)
(37.13237757378173, 14.302)
(42.21649108516205, 24.312)
(44.847142077229066, 16.75)
(29.64104653162637, 34.83)
(34.56839813710971, 10.61)
(44.88590275975257, 52.136)
(40.56360554811276, 73.83)
(58.03724183376613, 80.968)
(60.10393269908517, 38.558)
(36.736647161176876, 25.574)
(48.17032438496977, 35.576)
(41.90306345143956, 8.928)
(32.94551304670197, 23.242)
(39.80248723038383, 77.866)
(31.291406180784275, 7.2)
(38.185574614714554, 65.834)
(31.638228741540775, 30.642)
(54.475088150834836, 12.522)
(47.547920519007576, 44.706)
(37.434729917360414, 47.116)
(42.48153368169767, 4.322)
(65.89257648962075, 8.54)
(28.6760412501143, 4.368)
(38.655974494496945, 21.524)
(38.25314640053584, 31.088)
(35.07642891200249, 24.484)
(29.86710527968082, 43.804)
(23.046385723562672, 73.168)
(41.66872505224236, 30.762)
(30.60941199125008, 43.804)
(36.41701217938035, 27.414)
(39.74310181763552, 21.524)
(28.801811377535145, 24.484)
(36.81509464712813, 21.524)
(39.81538362970742, 4.368)
(34.65954306588965, 44.222)
(27.122187358418486, 43.804)
(40.9267422802006, 8.902)
(26.09120452607249, 4.27)
(33.67241410806113, 4.27)
(43.72913722589641, 4.27)
(44.972115145709765, 27.414)
(32.56694424087988, 4.368)
(43.73019978103711, 4.27)
(58.182126435155126, 8.902)
(32.92139647226776, 29.992)
(38.284362969258936, 25.84)
(32.57452302722534, 22.358)
(35.25501331513051, 16.75)
(58.35594903623424, 34.83)
(31.39922004219768, 29.306)
(41.0745453168195, 44.222)
(29.98498616643944, 27.516)
(35.48362904289497, 9.006)
(77.44830548733749, 17.62)
(38.59077772630179, 8.54)
(55.94932114183706, 58.622)
(90.18467870090593, 21.574)
(36.102323879240025, 47.842)
(35.33648903461744, 12.522)
(30.908196251943465, 21.132)
(28.63280966532725, 9.342)
(24.052756339409193, 52.258)
(26.388573783759604, 24.924)
(28.200849904891236, 25.164)
(27.560992854548346, 11.59)
(25.969962189547225, 79.998)
(31.966434115754733, 6.45)
(34.77769442070502, 21.04)
(25.16541567466938, 14.192)
(26.54733761289042, 34.83)
(26.467803175225075, 17.79)
(34.79921760458263, 7.016)
(33.36394585573893, 54.628)
(31.285786279001176, 73.83)
(33.45623613811042, 19.06)
(28.84485734238293, 25.574)
(34.59474029652522, 51.032)
(33.85378723165503, 8.786)
(29.54098050395429, 36.95)
(29.57803361795883, 14.094)
(30.15901603320341, 10.846)
(29.901337915830727, 9.232)
(30.985209233069, 13.754)
(28.200170752700615, 30.642)
(27.82083711571275, 12.522)
(41.617491064013585, 28.832)
(30.814862137191696, 17.138)
 };

       \end{axis}
    \end{tikzpicture}

%% file: Figures/exp2.tex
\begin{tikzpicture}
    \begin{axis}[
      width=5cm,
    height=5cm,
    ymin=97,
    xmin=20,
    ymax=132,
    xmax=83,
        axis x line  = bottom,
    axis y line  = left,
        ylabel=Latency (ms),
        xlabel=MAPE,
       legend style={at={(0.5,0.95),font=\fontsize{7}{8}\selectfont},
      anchor=north west,legend columns=1}]
      
    \addplot[only marks,mark=*,fill=teal!50,draw=black,mark size=3pt] plot coordinates {
    
(29.043880033319574, 127.142)
(31.287104905888146, 101.37)
(46.77238875565533, 120.012)
(31.059069406386957, 116.806)
(30.29463409842414, 136.531)
(32.22596227960025, 120.151)
(31.964602872992177, 108.019)
(34.07880248987546, 101.8)
(28.61555970132037, 123.445)
(39.99222509464475, 103.49)
(33.15029108809594, 101.789)
(58.59807377387378, 123.445)
(28.865501039343478, 143.065)
(34.110870898782366, 103.678)
(37.01815013194095, 118.088)
(33.06239714853794, 103.959)
(34.2544417588413, 121.714)
(46.81642566737095, 110.946)
(35.60438561378903, 123.445)
(28.01386232026644, 110.04)
(42.06288946552651, 103.49)
(30.344435013003835, 121.345)
(30.57630061076978, 137.005)
(29.881252455735268, 105.588)
(42.38814502042738, 121.278)
(59.67987164923878, 115.625)
(61.14506811667032, 111.004)
(36.152672010922295, 106.038)
(35.979102590267615, 120.151)
(61.66205764819965, 101.8)
(48.25122501646423, 108.019)
(36.11709629163363, 106.038)
(65.09244242046941, 101.8)
(39.137695034153886, 127.142)
(49.11362518451214, 123.291)
(71.52580278905214, 108.019)
(51.554885322623576, 107.284)
(45.27575103787626, 103.49)
(37.03232230564157, 101.754)
(51.828524825082845, 103.959)
(43.63450561496438, 121.278)
(62.11871037228251, 108.019)
(43.641324351607494, 115.557)
(43.40975941849526, 103.678)
(50.101530886721356, 122.076)
(38.93405872660842, 115.625)
(75.76465472977085, 106.038)
(46.93235903589024, 100.795)
(32.70793278466982, 107.551)
(56.36223049577907, 120.151)
(38.50283331808309, 123.445)
(46.26427936772098, 106.038)
(49.906497362709544, 108.019)
(66.03539745421753, 100.315)
(76.2894601827664, 100.315)
(80.44463228558686, 100.315)
(57.42561613209486, 100.444)
(32.363924285161346, 100.444)
(85.47612496710579, 100.315)
(48.11894388815926, 100.444)
(58.03648066267449, 100.315)
(88.28337549513205, 100.315)
(64.7019974787357, 100.315)
(72.39859091493487, 100.315)
(71.87351921376383, 100.315)
(61.20879576352131, 100.444)
(39.95829318652338, 100.444)
(41.16081095013042, 100.315)
(43.146149393276936, 100.315)
(69.95709209854326, 100.315)
(145.59760341834343, 100.315)
(45.64145359834381, 100.315)
(88.71428118908902, 100.315)
(75.61869678191847, 100.444)
(65.97107248811895, 100.315)
(36.37906676787963, 100.795)
(145.8866963166394, 100.444)
(92.20426739212746, 100.315)
(35.95983353833231, 100.444)
(103.54966917731213, 100.444)
(44.56616332947259, 100.444)
(51.595993648130275, 100.315)
(46.12830319390992, 100.315)
(115.68847843131338, 100.315)
(181.45816402010237, 100.315)
(137.26208890339478, 100.795)
(150.2782083968725, 100.315)
(170.0309624160105, 100.444)
(142.6646521217362, 100.444)
(182.186111213653, 100.315)
(149.89688664090104, 112.574)
(186.1130121030329, 100.444)
(179.37386332812088, 100.315)
(146.89748169461754, 100.795)
(151.76707391729676, 100.444)
(191.77341239212603, 100.444)
(168.48161421271325, 100.795)
(147.2440135296208, 100.315)
(121.48858836460835, 100.795)
(149.625845714721, 100.444)
(136.026393470545, 100.315)
(126.98308258393634, 100.315)
(155.76363939105104, 100.315)
(62.687530846477166, 100.315)
(118.48598110976639, 100.315)
(98.19405412886111, 100.315)
(100.19909235320401, 100.315)
(61.218313197851295, 100.444)
(81.85000005258905, 100.315)
(47.26442495161225, 100.315)
(79.07024767786606, 100.315)
(166.44395411215623, 100.315)
(95.82833906625281, 100.315)
(94.73820698729834, 100.444)
(81.69657132920365, 100.444)
(117.3317787792138, 100.315)
(83.36475054869244, 100.315)
(59.22751447758377, 100.444)
(69.19013248838642, 100.315)
(133.92957174532393, 100.315)
(95.08012120605875, 100.315)
(70.786833623702, 100.315)
(187.60660556427857, 100.315)
(119.68986004252145, 100.315)
(58.12122311630633, 100.315)
(99.39920666461369, 100.315)
(51.653736522337944, 100.315)
(73.19948732988544, 100.315)
(68.30023250990358, 100.315)
(42.37954785574849, 100.315)
(78.98707882656008, 100.315)
(63.61098588908958, 100.444)
(71.24756370076145, 100.315)
(32.48458691526287, 100.315)
(96.17965788759864, 100.315)
(46.71193028682411, 100.795)
(47.291043106822926, 100.315)
(106.20703593817184, 100.444)
(49.98936722422916, 100.315)
(62.58042332356168, 100.315)
(67.3664043058979, 100.444)
(52.61844850370303, 100.444)
(63.711180003698196, 100.315)
(33.85996052302852, 100.315)
(41.41662063169439, 100.315)
(86.53942293035752, 100.315)
(159.2070161799624, 100.315)
(38.65294820671153, 100.444)
(85.97406376781855, 100.444)
(95.24267754145103, 100.315)
(42.62703821737733, 100.315)
(39.690431693217874, 100.444)
(150.27162458760006, 100.315)
(87.6165255418915, 100.444)
(33.462993558459104, 100.444)
(119.95303757118305, 100.315)
(51.46686063500461, 100.444)
(55.47983339573859, 100.315)
(48.7464821445372, 100.315)
(34.17478219688383, 146.771)
(40.420608362469046, 115.625)
(47.30983822525421, 127.501)
(40.44767998204979, 103.959)
(35.36753134808025, 121.278)
(58.65569810514396, 115.557)
(37.99661867861505, 112.574)
(33.616471894151765, 106.038)
(66.98303505235965, 105.778)
(33.678970504041445, 129.219)
(48.391823916310834, 101.37)
(59.663414199886624, 104.879)
(39.78787873678814, 115.557)
(48.36937104001694, 103.49)
(32.608325343801845, 105.401)
(57.74751041621386, 107.551)
(41.89826846006895, 143.739)
(55.239297029511846, 106.038)
(37.31110329958548, 106.038)
(40.21193781703346, 107.284)
(50.037422266475026, 115.625)
(39.81484877368972, 101.8)
(32.2200857627755, 111.004)
(78.88199157495399, 103.49)
(44.462415929447005, 106.035)
(43.71006861078982, 106.035)
(44.46683524939147, 112.574)
(36.58272755343581, 104.879)
(36.71491210315229, 117.294)
(52.34952653170579, 110.946)
(98.18551156735775, 105.588)
(83.43867970007955, 117.294)
(105.30819849443994, 104.879)
(93.15406706450216, 100.444)
(99.84772996908671, 100.315)
(96.34057372251232, 103.678)
(82.37038506401588, 100.892)
(93.20779912570725, 103.678)
(93.840842329882, 112.574)
(95.04344484273439, 121.278)
(102.37720412207398, 100.444)
(97.77735416926505, 100.444)
(74.01292874288465, 100.892)
(90.66027619353719, 103.959)
(88.32162410157831, 105.588)
(90.30421895088917, 101.37)
(95.21040146373852, 103.49)
(105.7049146788645, 107.284)
(94.81301075715155, 103.959)
(92.6910447489211, 107.551)
(99.3203176858081, 106.038)
(95.84222994500622, 101.37)
(91.66484102959286, 103.959)
(117.5217812180605, 103.49)
(103.51789153315751, 103.678)
(84.53573940812356, 115.625)
(106.03303534223068, 101.8)
(88.27595965755852, 106.038)
(90.97030815696311, 101.37)
(85.6278298821929, 103.49)
(62.44808124763866, 101.8)
(40.97164023953515, 106.038)
(44.469262391017054, 106.038)
(33.068903050216804, 101.8)
(39.84040658461873, 103.49)
(51.42305070910137, 107.284)
(61.745927069198444, 103.49)
(64.29359223709507, 100.444)
(77.44532065076612, 103.959)
(31.146048169372943, 103.49)
(42.260804419574264, 101.37)
(48.72489992181559, 101.8)
(35.216424394066415, 106.038)
(29.747341421569963, 107.551)
(40.83054460370875, 103.49)
(38.049372793255884, 105.778)
(97.53313067283489, 100.444)
(33.59553861823529, 100.315)
(45.70794688274747, 106.038)
(60.679282845777394, 101.8)
(31.654923959840918, 103.959)
(29.307065888478068, 101.754)
(81.84871482672435, 100.892)
(66.16884694828846, 108.019)
(35.89861908398802, 109.527)
(33.501320457007125, 103.959)
(41.63964857594351, 101.37)
(61.434434836069684, 103.678)
(31.402960986590045, 105.778)
(36.25633383847583, 109.527)
(44.9846503387678, 103.959)
(38.19501796230798, 103.678)
(30.78123572652485, 108.019)
(42.39410604353156, 109.527)
(33.09915704105734, 110.04)
(40.8765281291678, 103.959)
(40.475132491263075, 103.49)
(43.16876385309617, 101.37)
(41.99032312491829, 108.019)
(61.58475149528897, 103.49)
(30.90763563875423, 103.49)
(45.96179842432197, 103.49)
(34.424139223518054, 104.879)
(39.336074092492865, 103.678)
(31.409379784216227, 110.04)
(66.16713797626585, 105.778)
(35.93237566957197, 103.959)
(37.180387956298624, 103.959)
(41.11175268554484, 103.49)
(28.07785955172797, 100.795)
(29.102752715214862, 101.8)
(54.76291180989211, 109.527)
(43.53521651680482, 105.588)
(32.332088576841294, 121.714)
(50.44815882834812, 137.169)
(26.57624507946431, 101.754)
(40.965949265505685, 127.501)
(26.887813464020887, 106.038)
 };

       \end{axis}
    \end{tikzpicture}

%% file: background.tex
\section{Preliminaries and Background}
\label{sec:background}

In this section, we discuss the necessary preliminaries and backgrounds that underpin our empirical study.

\begin{figure*}[t!]
\centering
\subfloat[Model-free tuners]{
\includegraphics[width=.325\textwidth]{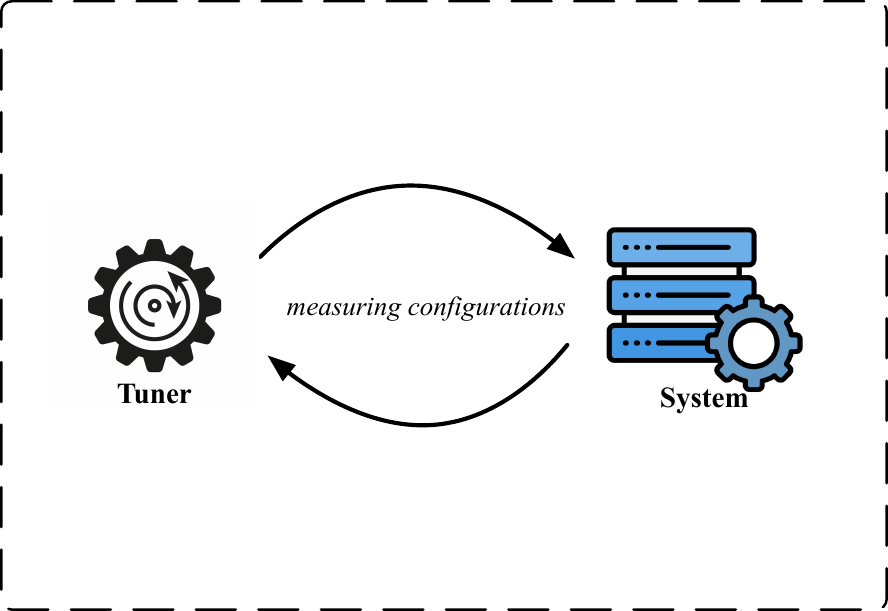}
\label{fig:tuners-a}
}
\hspace{-2mm}
\subfloat[Batch model-based tuners]{
\includegraphics[width=.325\textwidth]{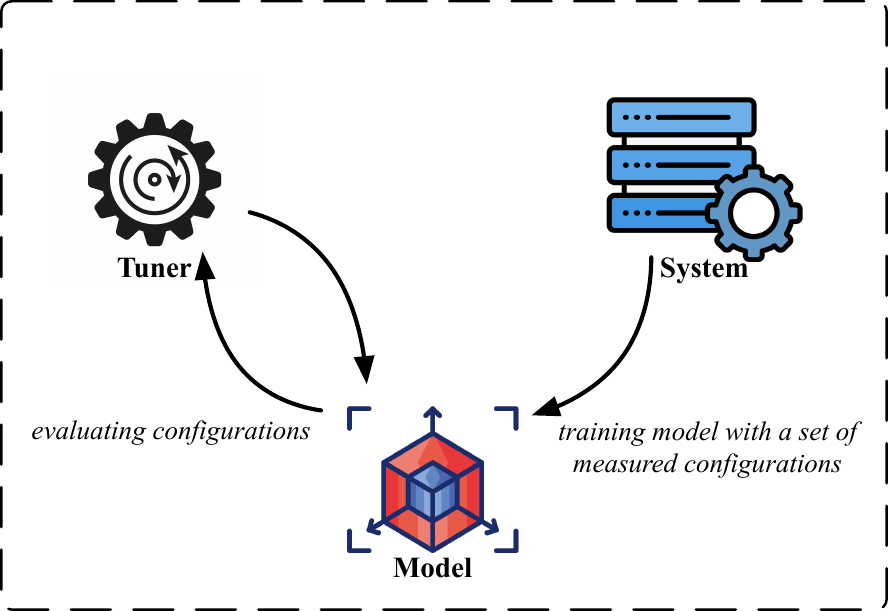}
\label{fig:tuners-b}
}
\hspace{-2mm}
\subfloat[Sequential model-based tuners]{
\includegraphics[width=.325\textwidth]{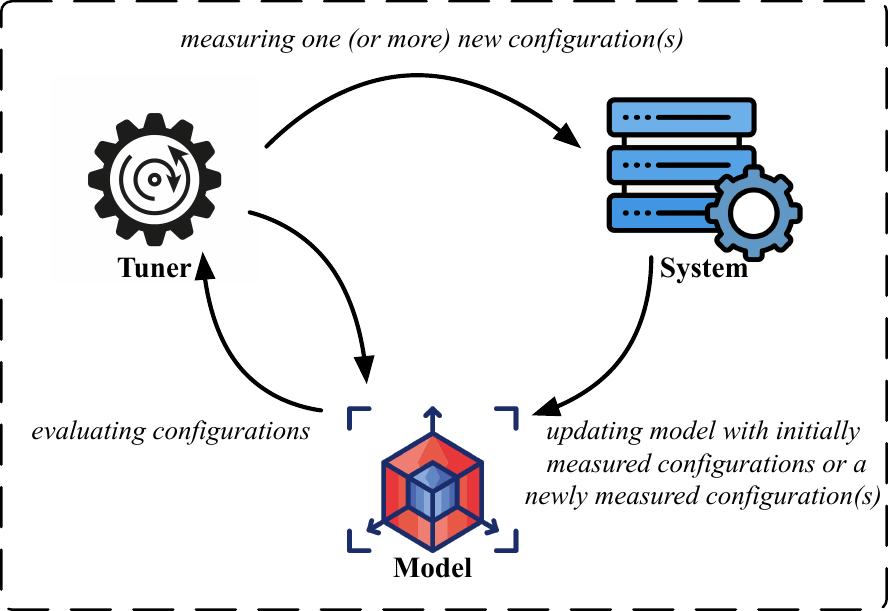}
\label{fig:tuners-c}
}
\caption{Illustrating the different types of tuners for configuration tuning in one tuning run.}
\label{fig:tuners}
\end{figure*}

\subsection{Software Configuration Tuning}

A configurable software system often comes with a set of critical configuration options to tune, for example, \textsc{Storm} allows one to change the \texttt{num\_splitters} and \texttt{num\_counters} for better latency or throughput~\cite{flash,DBLP:conf/sigsoft/0001L21}. $c_i$ denotes the $i$th option, which can be either a binary or integer variable, among $n$ options for a software system. The goal of configuration tuning is to search for better software configurations, from the space of $\mathbfcal{C}$ and with a given budget, that optimizes the performance objective $f$ under the given benchmark condition\footnote{Without loss of generality, we assume minimizing scenarios which can be converted to maximizing via additive inverse.}:
\begin{equation}
\argmin~f(\vect{c}),~~\vect{c} \in \mathbfcal{C}
\label{Eq:SOP}
\end{equation}
where $\vect{c} = (c_1, c_2, ..., c_n)$. The measurement of $f$ depends on the target system and the performance attribute, for which we make no assumption about the characteristics in this work.

A key difficulty of tuning configuration is the expensive measurement: Zuluaga et al.~\cite{DBLP:conf/icml/ZuluagaSKP13} report that it takes hours to merely measure a fraction of the configuration space. As a result, exhaustively profiling the system when tuning the configuration is often unrealistic, if not impossible.

\subsection{Model-free Tuners}

Model-free tuners represent the most straightforward way to address the problem of software configuration tuning as shown in Figure~\ref{fig:tuners-a}: designing intelligent heuristics, guided by direct measurement from the systems, to explore the configuration space and exploit the information to generate new directions of search~\cite{bestconfig,lopez2016irace,Sway,conex}. The configuration with the best-measured performance is used. The key focus has been on controlling the behavior of the tuners in exploring and exploiting the search space without having to traverse all of it. For example, works exist that leverage local search that discovers configurations around the best ones found so far~\cite{bestconfig}. Other tuners~\cite{conex} emphasize jumping out of local optima---an area in the configuration space where there are one (or more) configurations being optimal around the neighboring ones, but are sub-optimal with respect to the globally best configuration. 

Since the model-free tuners have relied on direct measurement of the systems, which provides the most accurate guidance in the tuning, it has been reported that it needs a substantial amount of trial-and-error before converging to some promising configurations~\cite{BOCA,Sway}. This can be undesirable when the measurement of configurations is rather expensive.

\subsection{Model-based Tuners}

Given the fact that measuring even a single configuration can be expensive, a natural resolution is to use a surrogate model that delegates the systems in the tuning process. In the past decade, there have been numerous studies that propose more and more accurate configuration performance models, primarily leveraging deep/machine learning algorithms, such as decision tree~\cite{DECART}, random forest~\cite{RF}, deep neural network~\cite{DeepPerf,DaL}, and hierarchical interaction neural network~\cite{HINNPerf}. Those models, if used appropriately, can be paired with different tuners, thereby significantly reducing the cost of evaluating a configuration as the better or worse between configurations can be compared directly using model prediction.

Broadly, there are two types of model-based tuners, namely batch model-based tuners and sequential model-based tuners, for which we elaborate below.



\subsubsection{Batch Model-based Tuners}

The batch model-based tuners~\cite{shi2024efficient,DBLP:journals/tosem/ChenLBY18,DBLP:journals/jss/ChengYW21,DBLP:conf/kbse/BaoLXF18,DBLP:journals/taco/LiL22,DBLP:conf/ijcai/AnsoteguiMSST15,DBLP:conf/asplos/YuBQ18} are inherent extensions from the model-free tuners, where the direct system measurements are replaced by model evaluations; the tuner itself remains unchanged. The optimal configuration with the best performance predicted by the model is then returned. This provides several advantages, for example, the newly proposed surrogate model can be paired with arbitrarily model-free tuners, in which the designed behaviors of the tuning algorithms can be preserved while an accurate model can be directly exploited. There is often no necessary inter-dependency between the model and tuner, since naturally model-free tuners rarely make assumptions about the internal structure of the system to be measured, providing a perfect foundation for it to be replaced by a model. However, the only additional step is that a high-quality model needs to be trained/built in advance with a good amount of measured configuration samples in order to effectively guide the tuning.

Figure~\ref{fig:tuners-b} shows the pipeline of the batch model-based tuners. The term ``batch'' refers to the fact that the model therein is trained/built with a set of training samples beforehand, but it would never be changed throughout the entire tuning run. Of course, the surrogate model can be updated when a newly measured set of data samples becomes available, but the updated model would only influence the next newly started tuning run.

Model evaluation is almost certainly cheaper than system measurement, the actual saving depends on the training data size, the model, and the system though. For example, even in some of the worst cases, a model evaluation merely takes half a minute~\cite{DBLP:conf/dac/SinghGMOCSMC19} as opposed to hours that are needed to measure a configuration on systems like \textsc{MariaDB}~\cite{DBLP:conf/sigsoft/0001L24}. Albeit the evaluation of configuration becomes much cheaper in batch model-based tuners, it might still be unrealistic to exhaustively traverse all configurations. As we will show in Section~\ref{sec:sys}, the search space of a system can go beyond a million, rendering the problem intractable even in the presence of a surrogate model. Yet, batch model-based tuners can still benefit from the sophistically designed heuristics of the model-free tuners during the tuning process, hence allowing them to deal with the intractability. Indeed, even with the reduced evaluation overhead, says 30 seconds per evaluation, heuristics that require 100 evaluations to obtain a promising configuration would remain much more preferred than the others that need a double evaluation to achieve the same, as this still leads to a saving of almost an hour for the entire tuning.


\subsubsection{Sequential Model-based Tuners}

The sequential model-based tuners~\cite{DBLP:conf/cloud/WangYYWLSHZ21,BOCA,flash,Ottertune,restune,SMAC}, in contrast, also rely on a surrogate model to determine the better or worse configurations while enabling a cheap exploration of the configuration space. However, after an initial model is trained/built with limited samples, it additionally permits new measurements of configuration on the system and updates the model using every new sample as the tuning proceeds, creating influence in guiding the current tuning run. The optimal configuration returned in the end is the one with the best-measured performance. In general, work on sequential model-based tuners often relies on a variant of Bayesian Optimization~\cite{garnett2023bayesian}, where the heuristic is guided by an acquisition function (e.g., Expected Improvement) that leverages the model prediction to identify the configuration to be measured next, which is most likely to improve performance while being uncertain enough to train and improve the model accuracy.

Indeed, the nature of sequential model-based tuners (and the acquisition function) might introduce a dependency between the model and the tuner, i.e., not all the models can quantify the uncertainty of the configurations, thereby models might not be compatible without some amendments. However, existing work has successfully exploited different surrogate models in sequential model-based tuners beyond the defaulted Gaussian Process Regression. For example \texttt{Flash}~\cite{flash} has been using a decision tree while \texttt{BOCA}~\cite{BOCA} has relied on random forest.

Figure~\ref{fig:tuners-c} shows the workflow of sequential model-based tuners. Clearly, the tuning progress does not only influence the configuration found but also the accuracy of the surrogate model. The term ``sequential'' implies the fact that the model is updated by sequentially obtained new measurements and then guides the search direction of the next configuration to measure.


\subsection{The ``Accuracy is All'' Belief}

Since the surrogate model serves as the delegate of the actual system that guides the tuner, it is natural to believe that better accuracy emulates the real system better, and hence should lead to superior tuning results~\cite{DBLP:conf/cloud/WangYYWLSHZ21,DBLP:conf/asplos/YuBQ18,bestconfig}. In addition to the examples and quotations mentioned in Section~\ref{sec:introduction}, there is an increasingly active research field, namely configuration performance learning~\cite{DBLP:journals/jss/PereiraAMJBV21,DBLP:journals/corr/abs-2403-03322}, that proposes sophisticated models with the main purpose of improving their accuracy. For example, to date, when using the common MAPE as the metric\footnote{Note that there are other accuracy metrics apart from MAPE.}, the prediction accuracy for system \textsc{VP9} has been pushed up to 0.44\% by a state-of-the-art model \texttt{HINNPerf}~\cite{HINNPerf} published at TOSEM'23, which is statistically better than the 0.86\% MAPE of the other already rather accurate model \texttt{DECART}~\cite{DECART}. Yu et al.~\cite{DBLP:conf/asplos/YuBQ18} also claim that since the proposed model can improve the accuracy by up to 22.4\%, hence it should be more useful than the others for tuning configuration. Therefore, all of those imply a general belief:


\input{Tables/systems}

\begin{tcbitemize}[%
    raster columns=1, 
    raster rows=1
    ]
  \tcbitem[myhbox={}{``Accuracy is All'' Belief}]  \textit{``Regardless of how the model is applied for configuration tuning, the higher the model accuracy, the more useful it becomes for improving the tuning quality and vice versa.''}
\end{tcbitemize}

As such, it is not hard to anticipate that future research and design choices on configuration tuning (and configuration performance modeling) will still be strongly driven by the model accuracy when a surrogate model is involved. Yet, our experience and preliminary results (as discussed in Section~\ref{sec:introduction}) question this practice, e.g., is the \% MAPE improvement in the aforementioned case really that important for configuration tuning? In what follows, through a large-scale empirical study, this work challenges the above belief and provides an in-depth understanding of the role of the surrogate model and its accuracy for configuration tuning.

%% file: Tables/systems.tex
\begin{table*}[t!]
\centering
\footnotesize
\caption{Details of the subject systems ordered by their sizes of configuration/search space $\mathbfcal{S}_{space}$. ($|\mathbfcal{B}|$/$|\mathbfcal{N}|$) denotes the number of binary/numeric options, $\mathbfcal{S}_{train}$ and $\mathbfcal{S}_{test}$ are the training size and testing size for batch model-based tuners, respectively.}
\label{tb:systems}

\begin{adjustbox}{width=\linewidth,center}

\begin{tabular}{lllllllll|lll}
\toprule
\textbf{System} & \textbf{Version} & \textbf{Workload} & \textbf{Domain} & \textbf{Performance} & \textbf{$|\mathbfcal{B}|$/$|\mathbfcal{N}|$} & \textbf{Scale} & \textbf{$\mathbfcal{S}_{space}$} & \textbf{Ref.} & \textbf{Budget} & \textbf{$\mathbfcal{S}_{train}$} &  \textbf{$\mathbfcal{S}_{test}$}\\ \hline
\textsc{Brotli} & 1.0.7 & Compressing a 1 GB file & File Compressor & Runtime (s) & 1/2 & Small & $3.08 \times 10^2$ & \cite{weber2023twins} & 203 & 55 & 253\\ 
\rowcolor{steel!10}\textsc{LLVM} & 3.0 & LLVM’s test suite & Compiler & Runtime (ms) & 10/0 & Small & $1.02 \times 10^3$ & \cite{DeepPerf} & 182 & 50 & 974\\ 
\textsc{Lrzip} & 0.10 & Community compression benchmark & File Compressor & Runtime (s) & 11/0 & Small & $2.04 \times 10^3$ & \cite{muhlbauer2023analyzing} & 184 & 55 & 136\\ 
\rowcolor{steel!10}\textsc{XGBoost} & 12.0 & Two standard datasets & Machine Learning Tool & Runtime (min) & 11/0 & Small & $2.04 \times 10^3$ & \cite{DBLP:conf/sigsoft/JamshidiVKS18} & 278 & 55 & 1000\\ 
\textsc{noc-CM-log} & 1.0 & Coremark benchmark workload & Database & Runtime (s) & 1/3 & Small & $2.38 \times 10^3$ & \cite{flash} & 129 & 55 &204\\ 
\rowcolor{steel!10}\textsc{DeepArch} & 2.2.4 & UCR Archive time series dataset & Deep Learning Tool & Runtime (min) & 12/0 & Small & $4.10 \times 10^3$ & \cite{jamshidi2018learning} & 207 & 60 & 1000\\ 
\textsc{BDB\_C} & 18.0 & Benchmark provided by vendor  & Database & Latency (s) & 16/0 & Small & $6.55 \times 10^4$ & \cite{DeepPerf} & 259 & 80 & 1000\\ 
\rowcolor{steel!10}\textsc{HSQLDB} & 19.0 & PolePosition 0.6.0 & Database & Runtime (ms) & 18/0 & Small & $2.62 \times 10^5$ & \cite{weber2023twins} & 149 & 90 & 774\\ 
\textsc{DConvert} & 3.0 & transform resources at different scales & Image Scaling & Runtime (s) & 17/1 & Large & $1.05 \times 10^7$ & \cite{muhlbauer2023analyzing} & 335 & 289 & 1000\\ 
\rowcolor{steel!10}\textsc{7z} & 9.20 & Compressing a 3 GB directory & File Compressor & Runtime (ms) & 11/3 & Large & $1.68 \times 10^8$ & \cite{weber2023twins} & 382 & 363 & 6827\\ 
\textsc{Apache} & 21.0 & ApacheBench 2.3 & Web Server & Maximum load ($\#$ users) & 14/2 & Large & $2.35 \times 10^8$ & \cite{weber2023twins} & 271 & 392 & 1304\\ 
\rowcolor{steel!10}\textsc{HSMGP} & 1.0 & Perform
one V-cycle & Stencil-Grid Solver & Latency (ms) & 11/3 & Large & $2.97 \times 10^8$ & \cite{DeepPerf} & 218 & 363 & 1000\\ 
\textsc{MongoDB} & 4.4 & Yahoo! cloud serving benchmark & Database & Runtime (ms) & 14/2 & Large & $3.77 \times 10^{8}$ & \cite{peng2021veer} & 278 & 392 & 1000\\ 
\rowcolor{steel!10}\textsc{PostgreSQL} & 22.0 & PolePosition 0.6.0 & Database & Runtime (ms) & 6/3 & Large & $1.42 \times 10^9$ & \cite{weber2023twins} & 298 & 108 & 1890\\ 
\textsc{ExaStencils} & 1.2 & Three default benchmarks & Code Generator & Runtime (ms) & 7/5 & Large & $1.61 \times 10^9$ & \cite{weber2023twins} & 416 & 224 & 8583\\ 
\rowcolor{steel!10}\textsc{Kanzi} & 5.0 & All cpp files of TMV & File Compressor & Runtime (ms) & 31/0 & Large & $2.14 \times 10^9$ & \cite{weber2023twins} & 237 & 160 & 1000\\ 
\textsc{Jump3r} & 1.0 & All cpp files of TMV & Audio Encoder & Runtime (s) & 37/0 & Large & $1.37 \times 10^{11}$ & \cite{weber2023twins} & 232 & 185 & 1000\\ 
\rowcolor{steel!10}\textsc{MariaDB} & 10.5 & Sysbench & Database & Runtime (ms) & 8/3 & Large & $5.31 \times 10^{11}$ & \cite{peng2021veer} & 226 & 192 & 780\\ 
\textsc{Polly} & 3.9 & The gemm program from polybench & Code Optimizer & Runtime (s) & 39/0 & Large & $5.50 \times 10^{11}$ & \cite{HINNPerf} & 285 & 195 & 5980\\ 
\rowcolor{steel!10}\textsc{SQLite} & 3.0 & Benchmark provided by vendor & Database & Runtime (s) & 39/0 & Large & $5.50 \times 10^{11}$ & \cite{DeepPerf} & 206 & 195 & 1000\\

\textsc{VP9} & 1.0 & 2 encoding Big Buck Bunny trailer (s) & Video Encoder & Runtime (s) & 41/0 & Large & $2.20 \times 10^{12}$ & \cite{HINNPerf} & 271 & 205 & 21579\\ 
\rowcolor{steel!10}\textsc{Spark} & 3.0 & HiBench & Data Analytics & Throughput (MB/s) & 5/8 & Large & $2.55 \times 10^{12}$ & \cite{cao2023cm} & 326 & 200 & 1000\\ 
\textsc{HIPAcc} & 0.8 & A set of partial
differential equations & Image Processing & Latency (ms) & 31/2 & Large & $3.30 \times 10^{12}$ & \cite{DeepPerf} & 371 & 1008 & 1247\\ 
\rowcolor{steel!10}\textsc{Redis} & 6.0 & Sysbench & Database & Requests per second & 1/8 & Large & $5.78 \times 10^{16}$ & \cite{cao2023cm} & 298 & 192 & 1000\\ 
\textsc{Storm} & 0.9.5 & Randomly generated benchmark & Data Analytics & Latency (s) & 0/12 & Large & $2.83 \times 10^{23}$ & \cite{krishna2020whence} & 263 & 288 & 1000\\ 
\rowcolor{steel!10}\textsc{SaC} & 1.0 & An n-body simulation & Compiler & Runtime (s) & 52/7 & Large & $3.14 \times 10^{28}$ & \cite{SPL} & 316 & 2704 & 6247\\ 
\textsc{Hadoop} & 3.0 & HiBench & Data Analytics & Throughput (MB/s) & 2/7 & Large & $1.27 \times 10^{29}$ & \cite{cao2023cm} & 297 & 192  & 1000\\ 
\rowcolor{steel!10}\textsc{Tomcat} & 8.0 & Sysbench & Web Server & Requests per second & 0/12 & Large & $7.93 \times 10^{34}$ & \cite{cao2023cm} & 282 & 288 & 1000\\ 
\textsc{JavaGC} & 7.0 & DaCapo benchmark suite & Java Runtime& Runtime (ms) & 12/23 & Large & $2.67 \times 10^{41}$ & \cite{DeepPerf} & 289 & 1922 & 1000\\ 
\bottomrule
\end{tabular}

\end{adjustbox}

\end{table*}

%% file: methodology.tex
\section{Research Methodology}
\label{sec:methodology}

We now delineate the methodology of our empirical study.

\subsection{Research Questions}

To provide a more in-depth understanding of the aforementioned ``accuracy is all'' belief, in this paper, we answer several important research questions (RQs). In particular, we seek to first confirm the benefit and usefulness of using a model for configuration tuning against the model-free counterparts by examining:

\keybox{
\textbf{RQ1:} How useful is the model for tuning quality?
}

Indeed, some work did compare a small set of model-based tuners with selected model-free counterparts to showcase the benefits of models~\cite{BOCA}, but their scale is rather limited and they are often based on a biased budget, e.g., only tens of measurements are considered, which would be more beneficial to model-based tuners. 

Since the sequential model-based tuners are commonly paired with a specifically chosen/designed model, \textbf{\textit{one would expect that such a model should help the most in tuning quality compared with the alternatives for the majority of the cases}}. Subsequently, it is natural to ask:

\keybox{
\textbf{RQ2:}  Do the chosen models work the best on tuning quality?
}

Both \textbf{RQ1} and \textbf{RQ2} provide a more thorough high-level understanding of the necessity and benefit of using a model to tune configuration. However, it remains unclear what role the model accuracy plays in terms of the tuning quality. Often, one would be interested in whether the most (least) accurate model can lead to the best (worst) tuning results, hence the next question we seek to answer is:

\keybox{
\textbf{RQ3:} Dose the goodness of the model consistent with the resulted tuning quality?
}

\textbf{\textit{If the ``accuracy is all'' belief is correct, the normal intuition is that the model accuracy should be a good indication of the tuning quality for most cases}}. 

As a next step, an extended question that encourages finer-grained investigations therein would be:

\keybox{
\textbf{RQ4:}  How do the model accuracy and tuning quality correlated?
}

\textbf{\textit{Again, following the belief, we would expect that there should be an overwhelmingly strong and positive correlation between them for most cases}}. 

A crucial issue with the ``accuracy is all'' belief is that research on newly proposed models is solely driven by accuracy comparison~\cite{DeepPerf,DaL,HINNPerf,DECART,SPL,DBLP:conf/asplos/YuBQ18,DBLP:conf/cloud/WangYYWLSHZ21}, but whether a certain extent of the accuracy change is significant or not for configuration tuning is often unclear. Therefore, the last question we aim to answer is:

\keybox{
\textbf{RQ5:} How much accuracy change do we need?
}

From this, we hope to provide some figures that can serve as references for future research on surrogate models for configuration tuning.

\subsection{Configurable Systems}
\label{sec:sys}

\subsubsection{System Selection}

We select the software systems and their benchmarks used by the most notable works from the key venues in software and system engineering. After extensively surveying the systems, we conduct a screening of them following two criteria:

\begin{itemize}
    \item To balance the generality of our study and the realism of the efforts required, we select the system with the most number of options as the representative for systems from the same origin, e.g., \textsc{BDB\_C} and \textsc{BDB\_J} are both the variants of the Berkeley database system written in C and Java, respectively. In that case, only \textsc{BDB\_C} is used as it has more configuration options.  
    \item For the same systems where different sets of relevant configuration options have been used, we select the one with the highest number of options. For example, \textsc{STORM} is a configurable system that has been studied in many prior studies, and we use the instance with 12 options to tune which is the most complicated case.
\end{itemize}




\input{Tables/surrogate}

Table~\ref{tb:systems} shows the final set of 29 configurable systems considered in this study. It is clear that those systems come from various domains with diverse performance attributes that are of concern; can exhibit only 3 options in the tuning (\textsc{Brotli}) or can be as complex as having up to 59 options to tune (\textsc{SaC}). The search space also varies, ranging from $308$ to $2.67\times 10^{41}$. It is worth noting that, even for those with relatively small search space, the measurement of a single configuration can still be time-consuming. For instance, it can take up to 166 minutes to measure one configuration on \textsc{MariaDB}~\cite{DBLP:conf/sigsoft/0001L24}, and a single run of measurement for \textsc{DeepArch} might even be a couple of hours \cite{jamshidi2018learning}.


When tuning under a surrogate model, we pragmatically consider those systems with less than a million configurations to search as \textit{``small''} systems, because the space can be reasonably covered when the measurements are replaced by model evaluations. Otherwise, we treat the systems as \textit{``large''} ones, i.e., they tend to be intractable as exhaustively traversing the entire space is expensive even with a cheaply evaluated model.

To the best of our knowledge, such a comprehensive set of configurable systems results in one of the largest-scale studies in this area\footnote{Among the related empirical studies, Cao et al.~\cite{DBLP:journals/ase/CaoBZWWSLZ24} include six systems; Zhang et al.~\cite{DBLP:journals/pvldb/ZhangCLWTLC22} study \textsc{MySQL} and \textsc{PostgreSQL} only. In contrast, we cover 29 systems. MCBO~\cite{DBLP:conf/nips/DreczkowskiGB23} and Grosnit et al.~\cite{DBLP:journals/jmlr/GrosnitCTGWB21} 4,000 and 3,958 cases, respectively, while we investigate 13,612 cases in our work. Note that apart from the scale, none of the above seek to provide a thorough understanding of how the models (and their accuracy) can impact configuration tuning.}.

\subsubsection{Option, Benchmark and Performance Selection}

The configuration options, their possible values, workloads/benchmarks, and target performance metric have been carefully selected as exactly the same as those studied in previous works of the corresponding systems~\cite{DBLP:conf/icse/WeberKSAS23,DBLP:conf/icse/HaZ19,DBLP:journals/tse/Nair0MSA20, muhlbauer2023analyzing, DBLP:journals/corr/abs-2106-02716, DBLP:conf/sigsoft/JamshidiVKS18, DBLP:journals/tse/KrishnaNJM21,DBLP:journals/jss/CaoBWZLZ23}. As such, in this regard, we did not use our own selection criteria but followed what has been widely acknowledged in the community.

Notably, it is well-known that not all options might significantly contribute to the performance---this is known as the sparsity problem in configuration performance learning~\cite{DeepPerf,DaL}. Practically, it would be difficult to obtain a thorough understanding beforehand regarding which options are more significant to the performance (especially given the expensive measurements). In fact, some configuration performance models have been proposed to tackle exactly such (e.g., \texttt{DeepPerf}~\cite{DeepPerf} and \texttt{DaL}~\cite{DaL}), such that they are capable of learning a model where only those more influential options would contribute to the performance prediction while those less relevant ones would have minimal impact (via, e.g., regularization or data division). Some other models, such as \texttt{SPLConqueror}~\cite{SPL}, would be more vulnerable to the sparsity problem but we regard this as a natural limitation of the proposed model rather than a problem that can be addressed fundamentally.

The strategy of measuring the performance follows exactly the same setting as previous work\cite{DBLP:conf/icse/WeberKSAS23,DBLP:conf/icse/HaZ19,DBLP:journals/tse/Nair0MSA20, muhlbauer2023analyzing, DBLP:journals/corr/abs-2106-02716, DBLP:conf/sigsoft/JamshidiVKS18, DBLP:journals/tse/KrishnaNJM21,DBLP:journals/jss/CaoBWZLZ23}. For example, when tuning \textsc{SQLite}, the workloads are generated by \textsc{Sysbench}, which is a standard benchmark for testing performance. Similarly, the performance is also measured by \textsc{Sysbench} as a final overall result after processing some workloads. Further, the system is rebooted when measuring different configurations, ensuring that none of them would benefit from the cache. The measurements are also taken as the average/median of 3 to 23 repeats depending on the specific system as used in prior work~\cite{DBLP:conf/icse/WeberKSAS23,DBLP:conf/icse/HaZ19,DBLP:journals/tse/Nair0MSA20, muhlbauer2023analyzing, DBLP:journals/corr/abs-2106-02716, DBLP:conf/sigsoft/JamshidiVKS18, DBLP:journals/tse/KrishnaNJM21,DBLP:journals/jss/CaoBWZLZ23}. To expedite the experiments and verify accuracy, we also use the same configuration datasets measured in previous research.

\input{Tables/tuners}

\subsection{Surrogate Models and Configuration Tuners}

\subsubsection{Models}

Table~\ref{tb:models} shows the 10 surrogate models we considered in this work. Those models are either commonly used by model-based tuners~\cite{DBLP:conf/cloud/WangYYWLSHZ21,BOCA,flash,Ottertune,restune,SMAC,DBLP:journals/tosem/ChenLBY18,DBLP:journals/taco/LiL22}, e.g., \texttt{GP} and \texttt{DT}, or are state-of-the-art ones from the field of configuration performance learning~\cite{DeepPerf,DaL,HINNPerf,DECART,SPL}, e.g., \texttt{DeepPerf} and \texttt{HINNPerf}, covering the key venues from both software engineering and system engineering communities. We also cater to the diverse nature of the models, as they either belong to statistical machine learning or deep learning; and distinct application domains, since they could be designed for general purpose, e.g., \texttt{RF}, or are specifically tailored to handle the characteristics of configuration data, e.g., \texttt{DaL}~\cite{DaL}.

While we aim to cover the most common and state-of-the-art models, some of them are intentionally omitted due to their various restrictions and inflexibility:

\begin{itemize}
    \item We rule out \texttt{$k$NN} since it is a lazy model, i.e., the overhead occurs at the prediction rather than training, which causes unacceptable long running time when paired with some tuners that leverage excessive model evaluations, e.g., \texttt{FLASH}. This is because, in our experiments, the prediction happens much more often than the training.

    \item \texttt{Perf-AL}, a deep learning model specifically designed to learn configuration performance, is also omitted due to its prohibitive training time required for training the adversarial neural network. Besides, hyperparameter tuning further complicates the process and the default values suggested by the authors did not lead to any useful models on our systems.

    \item We remove \texttt{KRR} due to its lengthy training process and high similarity to \texttt{SVR}, which is part of the model considered.
\end{itemize}

It is, however, worth noting that our set of models is comparable to those considered in recent empirical studies on modeling configuration performance~\cite{DBLP:conf/icse/Chen19b,DBLP:conf/msr/GongC22,DBLP:conf/kbse/JamshidiSVKPA17,DBLP:conf/splc/0003APJ21}.



\subsubsection{Tuners}

We consider up to 17 tuners in our study, including 8 sequential model-based ones, 8 model-free/batch model-based ones, and one that is specifically used as a batch model-based tuner (\texttt{Brute-force}) when the configuration space is tractable, as shown in Table~\ref{tb:tuners}.

To ensure good coverage, we select the most notable work from software engineering (e.g., \texttt{FLASH}~\cite{flash} and \texttt{SWAY}~\cite{Sway}), system engineering (e.g., \texttt{OtterTune}~\cite{Ottertune} and \texttt{Tuneful}~\cite{tuneful}), general parameter optimization (e.g., \texttt{SMAC}~\cite{SMAC} and \texttt{ParamILS}~\cite{paramils}), and commonly used baselines, such as \texttt{Random} and \texttt{Brute-force}. Those tuners could aim for any configurable systems/parameter optimization in general or they might be originally designed for a particular type of system, such as database systems or big data systems, but are generic enough to work on a configurable system of any type without substantial extension. They also come with a diverse set of underlying designs, containing different option reduction methods, search heuristics, and acquisition functions (for sequential model-based tuners). While most batch model-based tuners can find a model-free counterpart, \texttt{Brute-force} is the only exception. This is because even though the system might be tractable, the expensive measurements prevent any tuner from covering the entire space. However, with the help of a surrogate model, traversing the space might become plausible. As a result, for those ``small'' systems (recall Section~\ref{sec:sys}), we do not need sophisticated heuristics but only the \texttt{Brute-force} tuner is considered.

However, like the selection of surrogate models, we have to omit certain tuners for different reasons:

\begin{itemize}
    \item We rule out \texttt{SCOPE}~\cite{kim2022scope} and \texttt{OnlineTune}~\cite{onlinetune} because they are specifically designed to ensure the safety of a configurable system---a property that is beyond the scope of this work.

    \item We do not consider \texttt{HyperBand}~\cite{Hyperband} and its extension \texttt{BOHB}~\cite{falkner2018bohb}. This is because they assume significantly different fidelity/cost when measuring a hyperparameter, which is a property that not all configurable systems would have.

    \item Tuners that assume multiple performance objectives~\cite{DBLP:journals/jss/ChengYW21} or leverage on multi-objectivization for a single performance objective~\cite{DBLP:conf/sigsoft/0001L21,DBLP:conf/sigsoft/0001L24,DBLP:journals/corr/abs-2112-07303,DBLP:conf/wcre/Chen22} are also omitted since not all systems studied have more than one meaningful performance objective.

    \item There are omissions due to the lack of publicly available artifacts and the difficulty of re-implementing the tuners. For example, we do not utilize \texttt{LOCAT}~\cite{xin2022locat} due to the absence of public source code and we rule out \texttt{LlamaTune}~\cite{llamatune} since its source code is out of date, which compromises the usability. 
\end{itemize}

Despite the above omissions, to the best of our knowledge, the selected ones remain leading to the largest set of tuners considered in the field of configuration tuning. For all parameters of the tuners, we set those exactly the same as used in their corresponding work.

We also have to slightly improve a few tuners to enable them to work effectively on all systems considered: since \textsc{Flash} uses exhaustive search on the model to find the next configuration to measure without any option/space reduction, it might not be scalable to all ``large'' systems. To tackle this, for each system, we cap the exhaustive search therein with the maximum number of configurations that is commonly explored in the literature~\cite{DBLP:journals/corr/abs-2112-07303}. Certain tuners, e.g., \texttt{ResTune}, \texttt{ATConf}, \texttt{ROBOTune} and \texttt{Tuneful}, are designed for continuous variables only while most of the configurable systems have discrete options. To deal with this incompatibility, whenever the tuner needs to measure an invalid configuration that does not meet the possible values of the options, we measure the performance of the most similar valid configuration, which is quantified by using normalized Euclidean distance, as its performance value. In other words, we make the shape that covers a valid configuration and its neighboring invalid configuration flat in the configuration landscape (as they have the same performance value), hence helping to prevent the tuner from being trapped at local optima. Yet, if one of those invalid configurations happens to be the best one returned by the end of tuning, then they are converted back into its most similar valid configuration. In this way, we do not need to largely change those tuners while allowing them to work effectively on the systems considered.

\subsubsection{Pairing Model-based Tuners with Models}
\label{sec:pair}

Due to the nature of model-free tuners, they can be naturally paired with arbitrary surrogate models, as long as those models are trained with a batch of configuration samples in advance. As mentioned, this forms the foundation of batch model-based tuners, where the tuning is guided by the model rather than real measurement of the systems as in the case of its model-free counterpart, but the tuner and search itself are the same. Since a model is trained in advance, all batch model-based tuners share the said model under each system.

We can do the same arbitrary model pairing for the sequential model-based tuners, but the setting becomes slightly more complicated because they also need to update the model sequentially as new measurement(s) of configurations, which are decided by the tuner, become available. In particular, a key part of this type of model-based tuner is the acquisition function that determines which next configuration to measure. In essence, most sequential model-based tuners (e.g., \texttt{BOCA}~\cite{BOCA} and \texttt{OtterTune}~\cite{Ottertune}) leverage acquisition function to balance the exploitation and exploration of the tuners via the performance and uncertainty of the configurations predicted by a surrogate model, respectively. Exploitation focuses on the better configuration as predicted while exploration emphasizes the most unknown configurations inferred, which might not be better but could improve the surrogate model and allow the tuner to jump into new regions in the configuration landscape. All surrogate models would be able to predict the performance of a configuration, but not every one of them can quantify its uncertainty, except \texttt{GP} and \texttt{RF}.

To mitigate the above and allow sequential model-based tuners to be paired with any surrogate model as its batch model-based counterpart, we take inspiration from \texttt{RF} which can naturally quantify uncertainty. \texttt{RF}, as a bagging-based ensemble model, has been used to quantify the uncertainty of a configuration with the variance of the prediction made by all sub-models of decision trees~\cite{BOCA}. Therefore, when pairing with a tuner that needs to quantify the uncertainty of a configuration, we train 10 models (except \texttt{GP} and \texttt{RF}) based on the data partitioned by bagging, and measuring their prediction variance as the uncertainty for exploration. The actual prediction of a configuration for exploitation is still the model trained with all the data available. 


In summary, recall that Figure~\ref{fig:tuners} illustrates the patterns of different relationships between the key components in configuration tuning, particularly for the tuner and model. Tables~\ref{tb:models} and~\ref{tb:tuners} basically illustrate the concrete techniques/algorithms that one can use as the model and tuner, respectively, which can be arbitrarily replaced according to one of the patterns in Figure~\ref{fig:tuners} (if applicable). As such, with the model-based tuners, we have 10 models $\times$ (8 sequential model-based tuners $+$ 9 batch model-based tuners) $=170$ tuner-model pairs, together with the 8 model-free tuners, leading to $178$ instances to be considered in our study.


\subsection{Sample Sizes and Budgets}
\label{sec:size}


\subsubsection{Hot-start Sample Size for Sequential Model-based Tuners}

While the sequential model-based tuners can conduct a \textit{cold start} with a surrogate model learned under the one-shot setting, i.e., the surrogate model is directly used by a tuner without pre-training, they impose a high risk of measuring unreliable configurations at the beginning, which negatively affects the outcome. Therefore, those tuners often follow a \textit{hot-start} manner where the surrogate model is well-trained with a certain number of measured configuration samples~\cite{flash,BOCA,SMAC}. A larger hot-start sample size means that more configurations are randomly measured initially, allowing for a more precise initialization of the model and, consequently, being more favorable for finding new configurations. However, an increased sample size also comes with higher measurement costs. In this work, we turn to the literature of the eight sequential model-based tuners considered for identifying a common practice of setting the hot-start sample size, we found that existing work generally uses a fixed size across the systems, but the number can vary, e.g., two samples for \texttt{BOCA}~\cite{BOCA} and five samples for \texttt{ResTune}~\cite{restune}. To ensure a reasonable balance between building a reliable initial model and the additional measurement cost, we set a hot-start size of 20 as suggested previously, which is also the most commonly used setting we found among the sequential model-based tuners \cite{tuneful,ROBOTune,flash}. The data is then collected via conducting random sampling\footnote{Admittedly, random sampling is only one choice among the other sampling methods, but it is widely accepted as a default due to its simplicity and parameter-free nature \cite{conex, DBLP:journals/ase/CaoBZWWSLZ24, DBLP:journals/pvldb/CeredaVCD21, BOCA, HINNPerf, DBLP:conf/icse/HaZ19, DBLP:conf/kbse/BaoLWF19}. The experiment is repeated 30 runs with random seeds, hence prevnting the samples from being concentrated in certain parts.} on the measured dataset.

Noteworthily, we found that the budget size is much more important and influential to the tuning results compared with the hot start size, and hence we decided to adapt the budget according to system complexity but leave the hot start size fixed for all systems.

\subsubsection{Training Sample Size for Batch Model-based Tuners}

As mentioned, the batch model-based tuners require the surrogate model to be trained in advance with measured configurations. In particular, since the model would not be updated in the same configuration cycle again, the size required is often larger than that of the sequential model-based tuner. To identify the training set from the measured dataset, we follow random sampling to obtain the training samples, together with the most common method for determining the corresponding size as below~\cite{DeepPerf,DaL}:


\begin{itemize}
    \item \textbf{Binary systems:} for binary system, we randomly sample $5n$ configurations, where $n$ is the dimension of options. Note that $5n$ is often the largest size used from prior work~\cite{DeepPerf,DaL}.
    \item \textbf{Mixed systems:} for systems that come with both binary and numeric options, we use the sampling strategy from \texttt{SPLConqueror}~\cite{SPL} to determine the size with the parameters that lead to the largest number as considered in existing work~\cite{DeepPerf,DaL}.
\end{itemize}


The system-specific training sample sizes are shown in Table~\ref{tb:systems}. The above serves as the standard method to provide sufficient training data for building a sufficiently good model beforehand. 


\subsubsection{Tuning Budget}


In this work, we measure the tuning budget using the number of measurements as suggested by many prior works~\cite{flash,DBLP:conf/sigsoft/0001L21,DBLP:conf/sigsoft/0001L24,DBLP:conf/mascots/JamshidiC16}---over 60\% of the tuners studied have suggested such. In particular, the number of measurements comes with several advantages:

\begin{itemize}
    \item It eliminates the interference of clock time caused by the running software system under tuning at the same machine.
    \item The measurements of configuration are often the most expensive part throughout configuration tuning.
    \item It is independent of the underlying implementation details such as programming language and the version of libraries. 
\end{itemize}

Since the ideal budget can differ depending on the tuner, model, and system, we seek to find a budget that ensures our study is conducted fairly where all tuners would have the chance to reach their best state. To that end, under each system, we perform pilot experiments as follows:

\begin{enumerate}
    \item Run all sequential model-based tuners (which are equipped with their original surrogate models) and model-free tuners. We do not pair the sequential model-based tuners with other models since the heuristics in tuning the acquisition function remain unchanged regardless of the model. We also omit the batch model-based ones as their search/optimization behavior is identical to those of model-free tuners or only seeks to traverse the entire space, i.e., \texttt{Brute-force}.
    \item We set a maximum consecutive count, i.e., 100 consecutive measurements. We say a tuner-model pair/model-free tuner has reasonably converged if no better configuration can be found after 100 new and consecutively measured configurations have been explored and force the tuner-model pair/model-free tuner to terminate.
    \item We find the biggest required budget upon which a tuner-model pair/model-free tuner can be terminated among the others.
    \item Repeat the above five times with different random seeds and use the maximum budget therein.
\end{enumerate}

The above setting guarantees that all the tuner-model pairs and model-free tuners can reasonably converge, achieving their best possible state on a system, which is a common way to fairly compare optimization processes in software engnieering~\cite{DBLP:journals/tosem/ChenL23a,DBLP:journals/tosem/ChenL23,DBLP:journals/ase/GerasimouCT18}. The budgets found are shown in Table~\ref{tb:systems}. For each system, all the sequential model-based tuners and model-free tuners would have the same budget for configuration measurements. In particular, the budget needs to deduct the amount used for any hot-start sampling. The budget is also used as the number of model evaluations for the batch model-based tuners since it is the only factor that influences their convergence and search behaviors. Particularly, under each tuning run, we follow the common practice that only the unique configurations would consume the budget~\cite{flash,DBLP:conf/sigsoft/0001L21,DBLP:conf/sigsoft/0001L24,DBLP:conf/mascots/JamshidiC16}. 

Note that, to expedite the experiments, we store and reuse the measurements of configurations that have already been measured across different experiment runs as adopted in prior work~\cite{DBLP:conf/sigsoft/0001L21,DBLP:conf/sigsoft/0001L24}.

\subsection{Metrics}
\label{sec:metrics}

To answer our RQs, we define two types of metrics: one for measuring the accuracy of the surrogate model and the other for quantifying the quality of configuration tuning.

\subsubsection{Accuracy of Surrogate Models}

Accuracy is undoubtedly the most common metric that reflects the goodness of models. Measuring the accuracy of models for the batch model-based tuners is straightforward but it becomes slightly more complicated for the sequential model-based ones, in which the model evolves. In this work, we measure the model accuracy in sequential model-based tuners every 10 new measurements and use the average of those accuracy values as the final accuracy for the model. 



To choose the accuracy metrics, we turn to the literature on configuration performance learning and model-based tuning, from which we found the two metrics, namely mean absolute percentage error (MAPE)~\cite{bestconfig,DeepPerf,DaL,DBLP:conf/asplos/YuBQ18} and rank difference ($\mu$RD)~\cite{flash,DBLP:conf/sigsoft/NairMSA17}, are predominately used from their categories (residual and ranked) when measuring the accuracy of the surrogate model for configuration performance. 

\textbf{MAPE}: As a popular residual metric for measuring accuracy, MAPE (a.k.a. MRE or MMRE) computes the percentage of absolute errors relative to the actual values. Formally, MAPE is calculated as:


\begin{equation}
\text{MAPE} = \frac{1}{n} \sum_{i=1}^{n} \left|\frac{y_i - \hat{y}_i}{y_i}\right| \cdot 100\% 
\label{eq:MAPE}
\end{equation}
where $n$ is the number of testing samples; $\hat{y}_i$ and $y_i$ correspond to the model-predicted and true performance value of a configuration, respectively. Since MAPE measures the error, the smaller the MAPE, the better the accuracy. 



It is not hard to understand that MAPE is independent of the performance scale, therefore enabling cross-comparisons of models built for different systems and performance attributes. In contrast to other common metrics such as root means square error (RMSE) and mean absolute error (MAE), MPAE has its unique advantage: it is more scale-robust compared with MAE while it is more resilient to outliers and is more interpretable than RMSE.

\textbf{$\mathbf{\mu}$RD}: Since the residual nature of MAPE might impose rather fine-grained discrimination between the model accuracies, Vivek et al.~\cite{flash} propose a relaxed version of the accuracy metric, namely rank difference $\mu$RD, seeking to provide a more coarse-grained quantification of the accuracy. 

In a nutshell, $\mu$RD computes the number of pairs in the test data where the performance model ranks them incorrectly and measures the average rank difference. It can be calculated using equation~(\ref{eq:MURD}) :

\begin{equation}
\mu RD = \frac{1}{n} \sum_{i=1}^{n} |rank(y_i) - rank(\hat{y}_i)|
\label{eq:MURD}
\end{equation}
in which $rank$ denotes the function that produces the rank value of a performance value among the others in the testing data; $n$ signifies the number of testing samples. $y_i$ corresponds to the true performance value of the configuration, and $\hat{y}_i$ stands for the predicted performance value for the same configuration. 

\begin{figure}[t!]
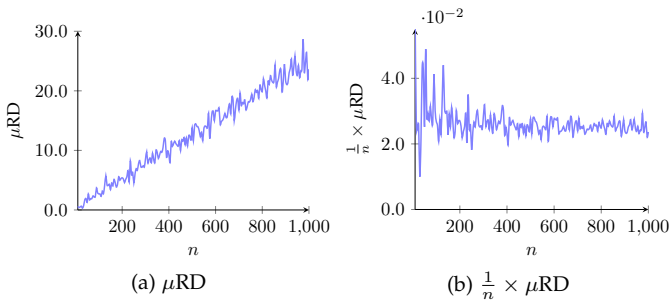

\centering
\subfloat[$\mu$RD]{\includestandalone[width=0.5\columnwidth]{Figures/rank_diff_1}}
~\hfill
\subfloat[${1\over n} \times \mu$RD]{\includestandalone[width=0.5\columnwidth]{Figures/rank_diff_2}}

\caption{Comparing $\mu$RD and its better standardized version under Decision Tree for system \textsc{7z}.}
\label{fig:rank-diff}
\end{figure}

However, we observed an issue with $\mu$RD: both the ranks of the predicted and true performance are highly dependent on the testing sample size $n$, but the original $\mu$RD only factors in a single $n$. As illustrated in Figure~\ref{fig:rank-diff}a, this still causes the original $\mu$RD metric to be highly sensitive to $n$, thereby preventing it from fairly comparing and averaging the accuracy across different systems (which is important for \textbf{RQ4}). To mitigate such, we additionally divide $\mu$RD by a further $n$ in this work, i.e., $\frac{1}{n} \times \mu$RD, thereby better standardizing it by further eliminating the influence of sample sizes brought by the ranks of both $y_i$ and $\hat{y}_i$. Clearly, from Figure~\ref{fig:rank-diff}b, $\frac{1}{n} \times \mu$RD is more robust to $n$ than its origin. 


\subsubsection{Testing Data for Accuracy Metrics}

To determine the testing samples of the accuracy metrics, we follow the typical procedure for configuration performance learning~\cite{DBLP:journals/corr/abs-2403-03322,DBLP:journals/jss/PereiraAMJBV21,DeepPerf,DaL,HINNPerf,DECART,SPL}, excluding those used for training/updating the model. Specifically, for each system, we prepare a dataset that contains the most measured valid configurations as consistent with prior work~\cite{DeepPerf,DaL,HINNPerf,DECART,SPL}; this avoids the need to cover the entire search space which can be intractable. After ruling out the samples used for training, we then randomly test 10\%  configurations from all those measured ones if that 10\% of which is greater than $1,000$. Otherwise, we test all the remaining configurations measured or $1,000$ randomly sampled ones whichever is smaller. The testing sample sizes have been illustrated in Table~\ref{tb:systems}. To verify that, we conduct significance tests on the model accuracy tested from our setting against that tested under all remaining samples of the five largest systems over 30 runs, as shown in Table~\ref{p_res} (for MAPE) and Table~\ref{p_res_RD} (for $\mu$RD) of this letter. As can be seen, under all models and systems, we have $p>0.05$, meaning that there are no statistically significant differences between the results using the above two strategies that set the testing sample sizes in the testing.

\input{Tables/test1}
\input{Tables/test2}

\subsubsection{Tuning Quality}

In this work, we are concerned with two aspects of the tuning quality, namely effectiveness and efficiency.

\textbf{Performance}: Naturally, we use the measured performance of the best configuration returned by a tuner-model pair/model-free tuner to quantify the effectiveness. Notably, according to Table~\ref{tb:systems}, some of those performance attributes are to be minimized (e.g., the latency for \textsc{MariaDB}) while others are to be maximized (e.g., the throughput for \textsc{Hadoop}).

\textbf{Efficiency}: While a tuner-model pair/model-free tuner can perform well in terms of effectiveness, it remains undesirable if it needs to consume a large budget to do so. Efficiency hence measures the speed of convergence of a tuner-model pair/model-free tuner towards a promising configuration. Following the same metrics as used in prior work~\cite{DBLP:conf/sigsoft/0001L21,DBLP:conf/icse/GaoZ0LY21}, we compute efficiency, $r$, as below: 
\begin{equation}
r = \frac{m}{b} 
\label{eq:r}
\end{equation}
Specifically, the procedure for calculating $r$ is:
\begin{enumerate}

\item For each set of the tuner-model pairs/model-free tuners to be compared\footnote{This excludes \texttt{Brute-force} (a batch model-based tuner) as it will certainly find the optimal configuration.}, find a baseline, $b$, as the smallest number of measurements (up to the budget) that the worst tuner-model pairs/model-free tuners\footnote{The worst tuner-model pair/model-free tuner is determined by using the Scott-Knott ESD test~\cite{DBLP:journals/tse/Tantithamthavorn19} and the average performance value (if there are multiple tuner-model pairs/model-free tuners in the worst rank).} (among all tuner-model pairs/model-free tuners considered) need to reach its best average (over 30 runs) of the performance (says $T$).

\item For every tuner-model pair/model-free tuner in the set, find the smallest number of measurements, $m$, at which the concerned performance is at least the same as $T$.

\item The efficiency $r$ of a tuner-model pair/model-free tuner, or in other words the speedup of a tuner-model pair/model-free tuner over the worst counterpart, is reported.


\end{enumerate}

If a tuner-model pair/model-free tuner has $r<1$, then we say the tuner-model pair/model-free tuner has better budget efficiency than the worst-performed tuner-model pair/model-free tuner in the set; if we have $r>1$ then it is less efficient; $r=1$ simply means they have the same efficiency. The value of $r$ naturally quantifies the extent of efficiency improvement/degradation against the worst-performed tuner-model pair/model-free tuner.

Note that, since the tuning for batch model-based tuners does not involve additional measurements but model evaluations, we calculate its efficiency based on the model evaluation throughout the tuning. However, the best performance would still be a measured value of the best-predicted configuration, yet this is for evaluation purposes only (e.g., when comparing and obtaining $T$).

\begin{figure*}[t!]
\centering
\includegraphics[width=\textwidth]{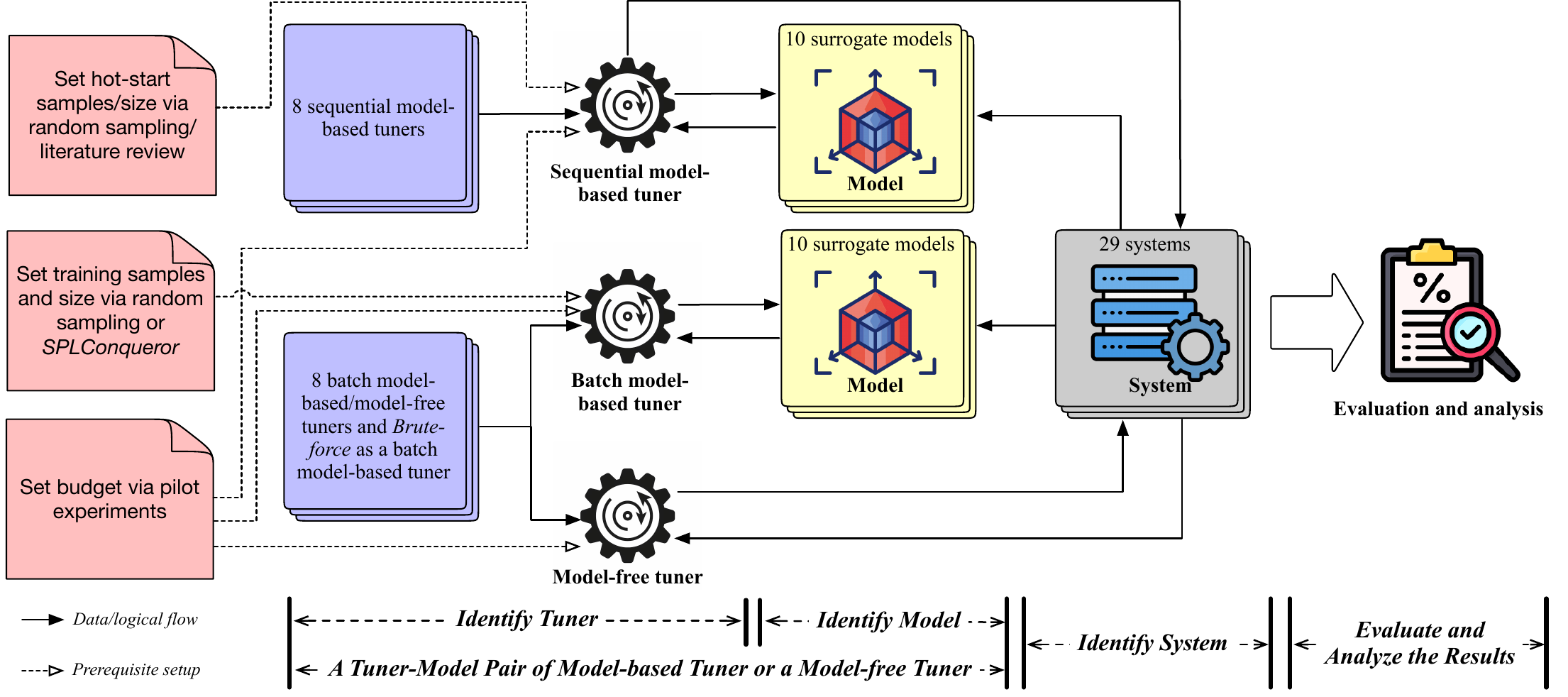}
\caption{The architecture and workflow of our empirical study.}
\label{fig:overview}
\end{figure*}

\subsection{Statistical Validation}
\label{sec:sta}

While the metrics provide a convenient means to answer our RQs, it remains unclear whether the gap between the results is of significance over repeated runs. To that end, we apply different statistical tests and correlation analysis indicators to validate our results.

\subsubsection{Mann-Whitney U-Test}

To verify pairwise comparisons of the metrics for accuracy and tuning quality, we use Mann-Whitney U-Test---a non-parametric and non-paired test that has been recommended for software engineering research~\cite{DBLP:conf/icse/WangAYLL16}. We set a confidence level as $\alpha=95\%$, meaning that if the comparison results in $p<0.05$ then the difference is statistically significant. Noteworhily, as a non-paired test, the Mann-Whitney U-Test bears weak statistical power, i.e., the statistical significance needs to be strong in order to be detected.

\subsubsection{Scott-Knott Effect Size Difference (ESD) Test}

When comparing multiple tuner-model pairs/model-free tuners or models, we leverage the non-parametric version of the Scott-Knott Effect Size Difference (ESD) test~\cite{tantithamthavorn2017mvt}. In a nutshell, Scott-Knott ESD test sorts the list of treatments (the tuner-model pairs/model-free tuners or models) by their median values of the metric. Next, it splits the list into two sub-lists with the largest expected difference~\cite{xia2018hyperparameter}. For example, suppose that we compare $A$, $B$, and $C$, a possible split could be $\{A, B\}$, $\{C\}$, with the rank ($r$) of 1 and 2, respectively. This means that, in the statistical sense, $A$ and $B$ perform similarly, but they are significantly better than $C$. Formally, Scott-Knott ESD test aims to find the best split by maximizing the difference $\Delta$ in the median before and after each split:
\begin{equation}
    \Delta = \frac{|l_1|}{|l|}(\overline{l_1} - \overline{l})^2 + \frac{|l_2|}{|l|}(\overline{l_2} - \overline{l})^2
\end{equation}
whereby $|l_1|$ and $|l_2|$ are the sizes of two sub-lists ($l_1$ and $l_2$) from list $l$ with a size $|l|$. $\overline{l_1}$, $\overline{l_2}$, and $\overline{l}$ denote their median metric value.

During the splitting, we apply a statistical hypothesis test $H$ to check if $l_1$ and $l_2$ are significantly different. This is done by using bootstrapping and Cliff's delta effect size~\cite{cliff1993dominance}. If that is the case, Scott-Knott ESD test recurses on the splits. In other words, we divide the approaches into different sub-lists if both bootstrap sampling and effect size test suggest that a split is statistically significant (with a confidence level of 99\%) and with a good effect measured by Cliff effect size. The sub-lists are then ranked based on their median metric.

In contrast to other non-parametric statistical tests on multiple comparisons (e.g., Kruskal-Wallis test~\cite{mckight2010kruskal}), Scott-Knott ESD test offers the following unique advantages:

\begin{itemize}
    \item It does not require posterior correction, as the comparisons are essentially conducted in a pairwise manner.
    \item It does not only show whether some treatments are statistically different or not, but also indicates which one is better than another, i.e., by means of ranking, while avoiding overlapping groups.
\end{itemize}

\subsubsection{Spearman's Rank Correlation}

We leverage Spearman's rank correlation ($\rho$)~\cite{myers2004spearman}, which is a widely used indicator in software engineering~\cite{DBLP:conf/icse/Chen19b,DBLP:journals/tse/WattanakriengkraiWKTTIM23}, to quantify the relationship between two metrics. Specifically, Spearman's rank correlation measures the nonlinear monotonic relation between two random variables and we have $-1\leq \rho \leq1$. $|\rho|$ represents the strength of monotonic correlation and $\rho=0$ means that the two metrics do not correlate with each other in any way; $-1\leq \rho<0$ and $0< \rho \leq 1$ denote that the monotonic correlation is negative and positive, respectively. 

To interpret the strength of Spearman's rank correlation, we follow the common patterns below, which have also been widely used in software engineering~\cite{DBLP:journals/tse/WattanakriengkraiWKTTIM23}: 

\begin{itemize}
    \item \textbf{negligible:} $0 \leq |\rho| \leq 0.09$
    \item \textbf{weak:} $0.09 < |\rho| \leq 0.39$
    \item \textbf{moderate:} $0.39 < |\rho| \leq 0.69$
    \item \textbf{strong:} $0.69 < |\rho| \leq 1$
\end{itemize}

We also assess the statistical significance of the $\rho$ value using z-score under a confidence level of $95\%$, i.e., the correlation is significant only when $p<0.05$.

\subsection{Procedure of Empirical Study}

The workflow and procedure of our empirical study have been depicted in Figure~\ref{fig:overview}. As can be seen, a tuner-model pair for model-based tuners or a model-free tuner is used on all systems considered. The best configuration returned by the tuner-model pairs/model-free tuners would be then evaluated/analyzed using the metrics and statistical validation presented in Sections~\ref{sec:metrics} and~\ref{sec:sta}. In particular, depending on the type of tuners, there can be different considerations and prerequisites:

\begin{itemize}
    \item \textbf{Sequential model-based tuners}: Each tuner of this type is paired with all 10 surrogate models, including the one that was originally chosen for the tuner and other replacement models. We set two key values beforehand: the hot-start sample size and the budget of measurements, as specified in Section~\ref{sec:size}.
    \item \textbf{Batch model-based tuners}: Each batch model-based tuner is also paired with all 10 surrogate models. We fixed the training sample size and budget as determined in Section~\ref{sec:size}.
    \item \textbf{Model-free tuners}: Since the model-free tuners do not rely on models, they are run directly by measuring the system given a budget. Yet, the budget can vary based on the \textbf{RQ} and which tuner type they are compared with (see Section~\ref{sec:rq1}).
\end{itemize}

As such, our study covers 2 accuracy metrics $\times$ 29 systems $\times$ (10 models $\times$ 8 sequential model-based tuners $+$ 10 models paired with the batch model-based tuners)\footnote{Since the model evolves depending on the sequential model-based tuners, a model's accuracy need to be evaluated on each specific tuner. For the batch model-based tuners, we only evaluate the accuracy of 10 models independent to the tuners as they essentially share the same trained model.} $=5,220$ cases for accuracy. For the tuning quality, we have 2 tuning quality metrics $\times$ [(21 large systems $\times$ 80 instances $+$ 8 small systems $\times$ 10 instances for batch model-based tuner-model pairs) $+$ (29 systems $\times$ 80 instances for sequential model-based tuner-model pairs) $+$ 29 systems $\times$ 8 instances] $=8392$ cases. In total, the above leads to $13,612$ cases of investigation---to the best of our knowledge, this is the largest-scale study to date on this topic for software configuration tuning. To mitigate stochastic noise, we repeat each experiment 30 times with varying seeds.

Notably, when comparing two tuners (or tuner-model paris), we say one wins (or loses) only if its results are better (or worse) with statistical significance; otherwise, we say the comparison is a tie.

All experiments were run in parallel on two high-performance servers each with Ubuntu 20.04.6 LTS, two Intel(R) Xeon(R) Platinum 50 cores CPU @ 2.30GHz (with two Nvidia A100 80GB GPU support for training deep learning-based models), and 500GB memory over the period of 13 months, $24 \times 7$---roughly $9,490.01$ GPU/CPU hours.


%% file: Tables/surrogate.tex
\begin{table*}[t!]
\centering
\footnotesize
\caption{Specification of the surrogate models considered in our study. \texttt{SPLConqueror} is basically \texttt{LR} while \texttt{DT} is the same as \texttt{DECART}. However, \texttt{SPLConqueror} considers the interaction between options as the terms while \texttt{LR} does not; \texttt{DECART} involves hyperparameter tuning while \texttt{DT} uses the defaults.}
\begin{adjustbox}{width=\linewidth,center}
\begin{tabular}{llllll}
\toprule
\textbf{Model} & \textbf{Type} & \textbf{Domain} & \textbf{Characteristics} & \textbf{Ref.} & \textbf{Used by} \\ \midrule

Support Vector Regression (\texttt{SVR})&Machine learning&General&Good for high dimensionality but can be sensitive to its parameters&\cite{SVM}&\cite{DaL}\\
\rowcolor{steel!10}Linear Regression (\texttt{LR})&Machine learning&General&Fast execution but only suitable for independent linear relationships&\cite{montgomery2021LR}&\cite{DaL}\\
Gaussian Process (\texttt{GP})&Machine learning&General&Providing uncertainty estimates but only suitable for low dimension&\cite{GPR}&\cite{llamatune}\\
\rowcolor{steel!10}Decision Tree (\texttt{DT})&Machine learning&General&Easy to interpret but can be overfitting and instable&\cite{DT}&\cite{flash}\\
Random Forests (\texttt{RF})&Machine learning&General&High accuracy and stable but can be complex and difficult to explain&\cite{RF}&\cite{BOCA}\\
\rowcolor{steel!10}\texttt{DECART} (\texttt{DCT})&Machine learning&Configurable systems&Handling non-linear relationships with tuned parameters but can overfit&\cite{DECART}&\cite{DeepPerf}\\
\texttt{SPLConqueror} (\texttt{SPL})&Machine learning&Configurable systems&Handling option interactions but mainly work linearly&\cite{SPL}&\cite{HINNPerf}\\
\rowcolor{steel!10}\texttt{DaL}&Machine/Deep learning&Configurable systems&Efficiently addressing sample
sparsity but can be expensive to train&\cite{DaL}&\cite{DaL}\\
\texttt{DeepPerf} (\texttt{DeP})&Deep learning&Configurable systems&Accurate for sparse options but sensitive to hyperparameters&\cite{DeepPerf}&\cite{HINNPerf}\\
\rowcolor{steel!10}\texttt{HINNPerf} (\texttt{HIP})&Deep learning&Configurable systems&High accuracy through hierarchical modeling but hard to tune&\cite{HINNPerf}&\cite{DaL}\\

\bottomrule

%
\end{tabular}
\end{adjustbox}
\label{tb:models}
\end{table*}

%% file: Tables/tuners.tex
\begin{table*}[t!]
\centering
\footnotesize
\caption{Specification of the tuners considered in our study.}
\label{tb:tuners}

\begin{adjustbox}{width=\linewidth,center}

\begin{tabular}{llllllll} 
\toprule
\textbf{Tuner} & \textbf{Type} & \textbf{Domain} & \textbf{Option Reduction} & \textbf{Surrogate Model} & \textbf{Acquisition} & \textbf{Search Heuristic}&\textbf{Ref.} \\ 
\midrule

\texttt{BOCA}&SMBT&Compilers&Random Forest&Random Forest&Maximum Expected Improvement&Local search that focus on important options& \cite{BOCA}\\
\rowcolor{steel!10}\texttt{ATconf}&SMBT&Big data systems&N/A&Gaussian Process&Minimize Lower Confidence Bound&Dropping options and exhaustive search& \cite{atconf}\\
\texttt{FLASH}&SMBT&Configurable systems&N/A&Decision Tree&Maximum Mean&Trying all configurations& \cite{flash}\\
\rowcolor{steel!10}\texttt{OtterTune}&SMBT&Database systems&Lasso&Gaussian Process&Maximum Upper Confidence Bound&Random sampling and exploration with gradient descent& \cite{Ottertune}\\
\texttt{ResTune}&SMBT&Database systems&Pre-selection&Gaussian Process&Maximum Expected Improvement&Exhaustive search& \cite{restune}\\
\rowcolor{steel!10}\texttt{ROBOTune}&SMBT&Data analytics&Random Forest&Gaussian Process&Maximum Hedge&Exhaustive search& \cite{ROBOTune}\\
\texttt{SMAC}&SMBT&Parameter optimization&N/A&Random Forest&Maximum Expected Improvement&Random sampling and exploration around good configurations& \cite{SMAC}\\
\rowcolor{steel!10}\texttt{Tuneful}&SMBT&Data analytics&Sensitivity analysis&Gaussian Process&Maximum Expected Improvement&Exhaustive search& \cite{tuneful}\\

\texttt{BestConfig}&MFT/BMBT&Database systems&N/A&N/A&N/A&Local search with recursive bound& \cite{bestconfig}\\
\rowcolor{steel!10}\texttt{Irace}&MFT/BMBT&Parameter optimization&N/A&N/A&N/A&Iterated race and focusing on good options& \cite{lopez2016irace}\\
\texttt{GGA}&MFT/BMBT&Configurable solvers&N/A&N/A&N/A&Gender separation with genetic algorithm& \cite{gga}\\
\rowcolor{steel!10}\texttt{ParamILS}&MFT/BMBT&Parameter optimization&N/A&N/A&N/A& Iterated recursive bound with parameter values& \cite{paramils}\\
\texttt{Random}&MFT/BMBT&General&N/A&N/A&N/A&Random search& -\\
\rowcolor{steel!10}\texttt{GA}&MFT/BMBT&General&N/A&N/A&N/A&Genetic algorithm& \cite{k2vtune}\\
\texttt{SWAY}&MFT/BMBT&Parameter optimization&N/A&N/A&N/A&Prune half of candidate based on the representatives& \cite{Sway}\\
\rowcolor{steel!10}\texttt{ConEx}&MFT/BMBT&Big data systems&N/A&N/A&N/A&Evolutionary Markov
Chain Monte Carlo sampling strategy& \cite{conex}\\
\texttt{Brute-force}&BMBT&General&N/A&N/A&N/A&Trying all configurations&-\\

\bottomrule

\end{tabular}
\end{adjustbox}

\end{table*}

%% file: Figures/rank_diff_1.tex
   \begin{tikzpicture}
  \begin{axis}[
  width = 6cm,
  height = 5cm,
  ymin=0,
  ymax=30,
  xmax=1000,
    xmin=10,
    xlabel=$n$,
     ylabel={$\mu$RD}, 
    y tick label style={
        /pgf/number format/.cd,
        fixed,
        fixed zerofill,
        precision=1,
        /tikz/.cd
    },
      axis x line  = bottom,
    axis y line  = left  ,
     xticklabel style={rotate=0},
    legend cell align={left},
    legend columns=2,
    legend style={font=\footnotesize,draw=none,fill=none,at={(1.1,1.15)}},                 
  ]

   \addplot [smooth,blue!50,thick] plot coordinates {

(10, 0.6)
(15, 0.4)
(20, 0.5)
(25, 0.64)
(30, 0.3)
(35, 0.7428571428571429)
(40, 1.725)
(45, 1.9111111111111112)
(50, 1.26)
(55, 2.690909090909091)
(60, 1.7)
(65, 1.9076923076923078)
(70, 1.8857142857142857)
(75, 2.36)
(80, 2.175)
(85, 2.2470588235294118)
(90, 3.7111111111111112)
(95, 2.873684210526316)
(100, 2.69)
(105, 2.6857142857142855)
(110, 2.2818181818181817)
(115, 3.1652173913043478)
(120, 2.7333333333333334)
(125, 4.152)
(130, 5.723076923076923)
(135, 4.207407407407407)
(140, 4.107142857142857)
(145, 4.310344827586207)
(150, 3.4133333333333336)
(155, 4.529032258064516)
(160, 4.03125)
(165, 5.133333333333334)
(170, 4.6)
(175, 3.84)
(180, 4.277777777777778)
(185, 5.589189189189189)
(190, 4.989473684210527)
(195, 5.282051282051282)
(200, 5.4)
(205, 5.204878048780488)
(210, 4.595238095238095)
(215, 5.130232558139535)
(220, 5.5636363636363635)
(225, 6.613333333333333)
(230, 4.665217391304348)
(235, 8.238297872340425)
(240, 5.895833333333333)
(245, 6.3061224489795915)
(250, 4.544)
(255, 6.023529411764706)
(260, 6.946153846153846)
(265, 6.660377358490566)
(270, 7.766666666666667)
(275, 8.298181818181819)
(280, 7.267857142857143)
(285, 7.171929824561404)
(290, 7.410344827586207)
(295, 7.206779661016949)
(300, 8.583333333333334)
(305, 9.754098360655737)
(310, 7.867741935483871)
(315, 8.431746031746032)
(320, 8.721875)
(325, 7.92)
(330, 9.512121212121212)
(335, 8.450746268656717)
(340, 7.923529411764706)
(345, 10.023188405797102)
(350, 9.665714285714285)
(355, 8.650704225352113)
(360, 9.188888888888888)
(365, 9.832876712328767)
(370, 9.68108108108108)
(375, 10.488)
(380, 12.007894736842106)
(385, 11.65974025974026)
(390, 8.830769230769231)
(395, 9.931645569620253)
(400, 9.275)
(405, 7.970370370370371)
(410, 11.446341463414635)
(415, 10.69156626506024)
(420, 11.119047619047619)
(425, 11.543529411764705)
(430, 11.148837209302325)
(435, 10.885057471264368)
(440, 10.709090909090909)
(445, 11.649438202247191)
(450, 9.977777777777778)
(455, 11.753846153846155)
(460, 11.523913043478261)
(465, 12.180645161290322)
(470, 10.938297872340426)
(475, 12.12)
(480, 10.747916666666667)
(485, 12.472164948453608)
(490, 12.369387755102041)
(495, 11.745454545454546)
(500, 13.824)
(505, 11.615841584158416)
(510, 14.107843137254902)
(515, 14.464077669902913)
(520, 14.023076923076923)
(525, 13.944761904761904)
(530, 13.29622641509434)
(535, 12.300934579439252)
(540, 13.05)
(545, 11.616513761467889)
(550, 13.658181818181818)
(555, 14.106306306306307)
(560, 13.948214285714286)
(565, 13.828318584070797)
(570, 16.235087719298246)
(575, 11.963478260869564)
(580, 14.370689655172415)
(585, 16.394871794871793)
(590, 16.515254237288136)
(595, 14.99327731092437)
(600, 16.623333333333335)
(605, 16.647933884297522)
(610, 16.640983606557377)
(615, 17.089430894308943)
(620, 14.370967741935484)
(625, 14.8672)
(630, 15.046031746031746)
(635, 15.522834645669292)
(640, 16.1984375)
(645, 15.516279069767442)
(650, 15.64)
(655, 17.061068702290076)
(660, 17.792424242424243)
(665, 16.13684210526316)
(670, 18.96716417910448)
(675, 18.423703703703705)
(680, 14.988235294117647)
(685, 16.814598540145987)
(690, 18.33768115942029)
(695, 18.687769784172662)
(700, 16.748571428571427)
(705, 19.64113475177305)
(710, 17.52394366197183)
(715, 18.041958041958043)
(720, 17.8125)
(725, 18.489655172413794)
(730, 16.806849315068494)
(735, 17.268027210884355)
(740, 18.124324324324323)
(745, 18.1248322147651)
(750, 18.556)
(755, 19.429139072847683)
(760, 18.17105263157895)
(765, 18.894117647058824)
(770, 20.203896103896103)
(775, 19.753548387096775)
(780, 20.02179487179487)
(785, 18.729936305732483)
(790, 20.969620253164557)
(795, 20.257861635220127)
(800, 19.90375)
(805, 22.026086956521738)
(810, 20.925925925925927)
(815, 19.97791411042945)
(820, 18.789024390243902)
(825, 21.612121212121213)
(830, 21.06867469879518)
(835, 22.55449101796407)
(840, 21.80952380952381)
(845, 21.418934911242605)
(850, 21.574117647058824)
(855, 24.369590643274854)
(860, 20.709302325581394)
(865, 21.44277456647399)
(870, 22.088505747126437)
(875, 23.456)
(880, 23.473863636363635)
(885, 19.78870056497175)
(890, 21.67191011235955)
(895, 24.481564245810056)
(900, 21.72888888888889)
(905, 21.044198895027623)
(910, 21.38791208791209)
(915, 23.01530054644809)
(920, 23.77282608695652)
(925, 25.34918918918919)
(930, 22.61182795698925)
(935, 23.95080213903743)
(940, 25.859574468085107)
(945, 23.36931216931217)
(950, 22.591578947368422)
(955, 24.26282722513089)
(960, 22.195833333333333)
(965, 23.323316062176165)
(970, 23.671134020618556)
(975, 28.657435897435896)
(980, 23.131632653061224)
(985, 24.778680203045685)
(990, 26.393939393939394)
(995, 21.930653266331657)
(1000, 23.64)

 };

  \end{axis}

\end{tikzpicture}

%% file: Figures/rank_diff_2.tex
   \begin{tikzpicture}
  \begin{axis}[
  width = 6cm,
  height = 5cm,
  ymin=0,
  ymax=0.055,
  xmax=1000,
    xmin=10,
    xlabel=$n$,
       ylabel={${1\over n} \times \mu$RD},
    y tick label style={
        /pgf/number format/.cd,
        fixed,
        fixed zerofill,
        precision=1,
        /tikz/.cd
    },
      axis x line  = bottom,
    axis y line  = left  ,
     xticklabel style={rotate=0},
    legend cell align={left},
    legend columns=2,
    legend style={font=\footnotesize,draw=none,fill=none,at={(1.1,1.15)}},                 
  ]

   \addplot [smooth,blue!50,thick] plot coordinates {
(10, 0.06)
(15, 0.02666666666666667)
(20, 0.025)
(25, 0.0256)
(30, 0.01)
(35, 0.02122448979591837)
(40, 0.043125000000000004)
(45, 0.04246913580246914)
(50, 0.0252)
(55, 0.04892561983471074)
(60, 0.028333333333333332)
(65, 0.029349112426035506)
(70, 0.026938775510204082)
(75, 0.031466666666666664)
(80, 0.027187499999999996)
(85, 0.02643598615916955)
(90, 0.04123456790123457)
(95, 0.03024930747922438)
(100, 0.0269)
(105, 0.025578231292517004)
(110, 0.020743801652892562)
(115, 0.027523629489603026)
(120, 0.02277777777777778)
(125, 0.033216)
(130, 0.04402366863905326)
(135, 0.031165980795610424)
(140, 0.02933673469387755)
(145, 0.02972651605231867)
(150, 0.022755555555555557)
(155, 0.029219562955254946)
(160, 0.0251953125)
(165, 0.031111111111111114)
(170, 0.027058823529411764)
(175, 0.021942857142857142)
(180, 0.023765432098765433)
(185, 0.030211833455076696)
(190, 0.02626038781163435)
(195, 0.027087442472057854)
(200, 0.027000000000000003)
(205, 0.025389649018441405)
(210, 0.021882086167800453)
(215, 0.02386154678204435)
(220, 0.025289256198347106)
(225, 0.02939259259259259)
(230, 0.020283553875236296)
(235, 0.03505658669081032)
(240, 0.02456597222222222)
(245, 0.025739275301957517)
(250, 0.018175999999999998)
(255, 0.02362168396770473)
(260, 0.026715976331360948)
(265, 0.025133499466002136)
(270, 0.02876543209876543)
(275, 0.03017520661157025)
(280, 0.025956632653061227)
(285, 0.025164666051092643)
(290, 0.025552913198573127)
(295, 0.02442976156276932)
(300, 0.02861111111111111)
(305, 0.031980650362805695)
(310, 0.025379812695109263)
(315, 0.026767447719828672)
(320, 0.027255859375000004)
(325, 0.024369230769230768)
(330, 0.028824609733700643)
(335, 0.025226108264646916)
(340, 0.023304498269896196)
(345, 0.029052720016803195)
(350, 0.027616326530612245)
(355, 0.024368180916484825)
(360, 0.02552469135802469)
(365, 0.026939388252955526)
(370, 0.026165084002921838)
(375, 0.027968)
(380, 0.031599722991689755)
(385, 0.030285039635688987)
(390, 0.022642998027613413)
(395, 0.025143406505367727)
(400, 0.0231875)
(405, 0.019679926840420667)
(410, 0.02791790600832838)
(415, 0.025762810277253592)
(420, 0.02647392290249433)
(425, 0.027161245674740483)
(430, 0.025927528393726338)
(435, 0.025023120623596247)
(440, 0.024338842975206613)
(445, 0.026178512814038633)
(450, 0.02217283950617284)
(455, 0.02583262890955199)
(460, 0.025051984877126655)
(465, 0.026194935830731875)
(470, 0.023272974196468992)
(475, 0.02551578947368421)
(480, 0.022391493055555555)
(485, 0.02571580401743012)
(490, 0.025243648479800082)
(495, 0.023728191000918275)
(500, 0.027648)
(505, 0.02300166650328399)
(510, 0.027662437524029217)
(515, 0.028085587708549346)
(520, 0.026967455621301775)
(525, 0.02656145124716553)
(530, 0.025087219651121397)
(535, 0.022992401083064024)
(540, 0.02416666666666667)
(545, 0.021314704149482364)
(550, 0.02483305785123967)
(555, 0.02541676811947082)
(560, 0.024907525510204084)
(565, 0.02447490014879787)
(570, 0.028482610033856572)
(575, 0.02080604914933837)
(580, 0.02477705112960761)
(585, 0.028025421871575716)
(590, 0.027991956334386673)
(595, 0.025198785396511547)
(600, 0.02770555555555556)
(605, 0.02751724608974797)
(610, 0.027280300994356355)
(615, 0.027787692511071452)
(620, 0.0231789802289282)
(625, 0.02378752)
(630, 0.023882590073066264)
(635, 0.024445408890817782)
(640, 0.02531005859375)
(645, 0.024056246619794483)
(650, 0.024061538461538464)
(655, 0.02604743313326729)
(660, 0.02695821854912764)
(665, 0.024265927977839337)
(670, 0.028309200267320117)
(675, 0.02729437585733882)
(680, 0.02204152249134948)
(685, 0.02454685918269487)
(690, 0.026576349506406215)
(695, 0.026888877387298796)
(700, 0.023926530612244895)
(705, 0.027859765605351843)
(710, 0.02468161079150962)
(715, 0.025233507750990272)
(720, 0.024739583333333332)
(725, 0.025502972651605234)
(730, 0.023023081253518485)
(735, 0.023493914572631776)
(740, 0.02449233016800584)
(745, 0.02432863384532228)
(750, 0.024741333333333334)
(755, 0.025733959036884347)
(760, 0.023909279778393352)
(765, 0.024698193002691275)
(770, 0.026238826108955976)
(775, 0.025488449531737775)
(780, 0.0256689677843524)
(785, 0.02385979147227068)
(790, 0.02654382310527159)
(795, 0.02548158696254104)
(800, 0.024879687499999997)
(805, 0.027361598703753712)
(810, 0.025834476451760405)
(815, 0.024512778049606686)
(820, 0.022913444378346224)
(825, 0.026196510560146925)
(830, 0.025383945420235157)
(835, 0.027011366488579727)
(840, 0.025963718820861677)
(845, 0.02534785196596758)
(850, 0.025381314878892735)
(855, 0.028502445196812696)
(860, 0.02408058409951325)
(865, 0.024789334758929468)
(870, 0.02538908706566257)
(875, 0.026806857142857143)
(880, 0.026674845041322313)
(885, 0.022360113632736444)
(890, 0.02435046080040399)
(895, 0.027353703067944195)
(900, 0.024143209876543212)
(905, 0.023253258447544334)
(910, 0.02350320009660669)
(915, 0.02515333393054436)
(920, 0.025840028355387522)
(925, 0.027404528853177502)
(930, 0.024313793502138976)
(935, 0.025615831164745917)
(940, 0.027510185604345858)
(945, 0.024729430866997006)
(950, 0.02378060941828255)
(955, 0.025406101806419782)
(960, 0.02312065972222222)
(965, 0.0241692394426696)
(970, 0.024403230949091297)
(975, 0.0293922419460881)
(980, 0.023603706788837983)
(985, 0.02515602051070628)
(990, 0.026660544842363024)
(995, 0.022040857554102167)
(1000, 0.02364)
 };

  \end{axis}

\end{tikzpicture}

%% file: Tables/test1.tex
\begin{table}[t!]
\centering
\caption{The $p$-values from the U-Test when comparing the MAPE (of all models) under our method of setting the testing sample size and that under using all remaining samples on the five largest systems over 30 runs.}
\begin{adjustbox}{width=\linewidth,center}
\begin{tabular}{lcccccccccc}
\toprule
{} & \multicolumn{10}{c}{Model} \\
\cmidrule(lr){2-11}
System & \texttt{DT} & \texttt{RF} &  \texttt{LR} &  \texttt{SVR} &  \texttt{GP} &  \texttt{SPL} &  \texttt{DCT} &  \texttt{DeP} &  \texttt{HIP} &  \texttt{DaL} \\
\midrule
\textsc{Storm}   & 1.00 & 0.85 & 0.95 & 0.54 & 0.90 & 0.97 & 0.74 & 0.85 & 0.97 & 0.95 \\
\textsc{SaC}    & 0.81 & 0.63 & 0.83 & 0.82 & 0.45 & 0.99 & 0.26 & 0.94 & 0.97 & 0.70 \\ 
\textsc{Hadoop}  & 0.46 & 0.58 & 0.36 & 0.86 & 0.25 & 0.31 & 0.90 & 0.77 & 0.67 & 0.67 \\
\textsc{Tomcat}  & 0.82 & 0.53 & 0.80 & 0.40 & 0.76 & 0.84 & 0.98 & 0.53 & 0.75 & 0.43 \\
\textsc{JavaGC}  & 0.46 & 0.60 & 0.60 & 0.62 & 0.70 & 0.61 & 0.74 & 0.95 & 0.94 & 0.84 \\
\bottomrule
\label{p_res}
\end{tabular}
\end{adjustbox}
\end{table}

%% file: Tables/test2.tex
\begin{table}[t!]
\centering
\caption{The $p$-values from the U-Test when comparing the $\mu$RD (of all models) under our method of setting the testing sample size and that under using all remaining samples on the five largest systems over 30 runs.}
\begin{adjustbox}{width=\linewidth,center}
\begin{tabular}{lcccccccccc}
\toprule
{} & \multicolumn{10}{c}{Model} \\
\cmidrule(lr){2-11}
System & \texttt{DT} & \texttt{RF} &  \texttt{LR} &  \texttt{SVR} &  \texttt{GP} &  \texttt{SPL} &  \texttt{DCT} &  \texttt{DeP} &  \texttt{HIP} &  \texttt{DaL} \\
\midrule
\textsc{Storm}   & 0.18 & 0.43 & 0.37 & 0.82 & 0.43 & 1.00 & 0.22 & 0.26 & 0.18 & 0.26 \\
\textsc{SaC}     & 0.50 & 0.22 & 0.18 & 0.15 & 0.82 & 0.91 & 0.76 & 0.50 & 0.74 & 0.43 \\
\textsc{Hadoop}  & 0.22 & 0.37 & 0.74 & 0.26 & 0.37 & 0.74 & 0.43 & 1.00 & 0.22 & 0.50 \\
\textsc{Tomcat}  & 0.65 & 0.31 & 0.18 & 0.22 & 0.18 & 0.18 & 0.31 & 0.65 & 0.18 & 0.82 \\
\textsc{JavaGC}  & 0.74 & 0.31 & 0.18 & 0.65 & 0.82 & 0.37 & 0.58 & 0.50 & 0.12 & 0.26 \\
\bottomrule
\label{p_res_RD}
\end{tabular}
\end{adjustbox}
\end{table}

%% file: results.tex
\section{Results and Analysis}
\label{sec:results}

In this section, we summarize and discuss the results obtained.

\subsection{RQ1: How Useful is the Model for Tuning Quality?}
\label{sec:rq1}




\input{Tables/RQ4/SMBO}

\subsubsection{Method}

To answer \textbf{RQ1}, we respectively compare the sequential and batch model-based tuners, paired with all possible models, against their model-free counterparts. Since there are many tuner-model pairs and model-free tuners to compare, for each system, we select the best representative from each tuner type. To that end, we run Scott-Knott ESD test on each tuning quality metric, and use the one that is ranked the first as the representative; if there is more than one pair/tuner in the best rank, we choose the one with the best average quality. It is worth noting that, when comparing model-free tuners with sequential model-based tuners, they are set with the budget identified in Section~\ref{sec:size}. In contrast, when comparing them with batch model-based tuners, their budget is set as the same as the training sample size used by the batch model-based ones, which is already a large figure compared with existing work. This is because only the measurement of systems is expensive in configuration tuning while the other factors are relatively marginal, hence matching them in that aspect ensures a fairer comparison. Since there are 29 systems, we have 29 cases under each metric.

The pairwise comparisons are verified by the Mann-Whitney U-test. All other settings are the same as discussed in Section~\ref{sec:methodology}. In particular, we summarize the result under a case as follows:

\begin{itemize}
    \item \textbf{Win:} This means a model-based tuner has better results than the model-free counterpart while the comparison is statistically significant.
    \item \textbf{Loss:} This means a model-based tuner has worse results than the model-free counterpart while the comparison is statistically significant.
    \item \textbf{Tie:} This means the model-based tuners and the model-free counterpart have no statistically significant difference regardless of the average deviation in performance value.
\end{itemize}


\begin{figure}[t!]
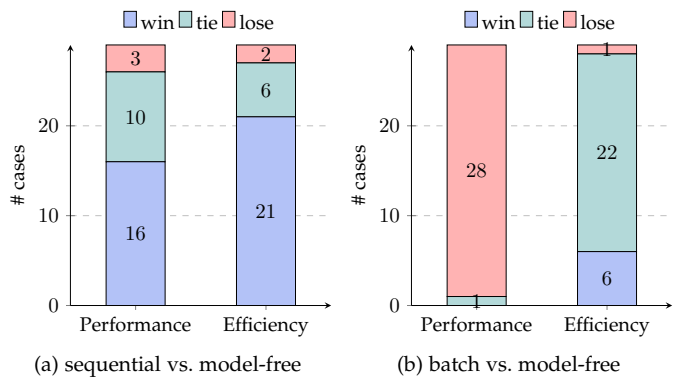

\centering
\subfloat[sequential vs. model-free]{\includestandalone[width=0.49\columnwidth]{Figures/rq1.1-summary-1}}
~\hfill
\subfloat[batch vs. model-free]{\includestandalone[width=0.49\columnwidth]{Figures/rq1.1-summary-2}}

\caption{Comparing the best tuner-model pair against the best model-free tuner under each system for all 29 cases (30 runs). We count how many cases the best tuner-model pair wins, loses, or ties.}
\label{fig:rq1.1-sum}
\end{figure}

\subsubsection{Results}

From Figure~\ref{fig:rq1.1-sum}, for sequential model-based tuner, it is clear that the models are overwhelmingly helpful against the counterpart when the model is absent. Even with the statistical test, out of the 29 cases, the sequential model-based tuner still wins over the model-free counterpart on 16 cases; draws on 10, and loses only on 3 in terms of performance. For efficiency, the best model-tuner pair has a remarkable win on 21 cases with 6 and 2 cases of tie and loss, respectively. Looking at the detailed results in Table~\ref{tb:rq1.1}\footnote{If we were to match the table and the Figure~\ref{fig:rq1.1-sum}, a win/loss in the figure corresponds to a colored cell in the table with bold texts; a tie in the figure reflects any remaining cells in the table.}, the model can help the sequential model-based tuner to achieve a maximum of 160\% performance improvement and up to $50\times$ better for efficiency.

\input{Tables/RQ4/MFO}

We also found that the best tuner-model pair for the sequential model-based tuners is system-dependent for both performance and efficiency; the same applies to the model-free tuners. However, \texttt{Flash} and \texttt{SMAC}/\texttt{ROBOTune} appear to be often part of the best tuner-model pair for performance and efficiency, respectively. For the model-free counterparts, both the \texttt{BestConfig} and \texttt{ParamILS} tend to be more competitive than the others.

The divergence of tuning quality between the two types of tuner appears to be larger on small systems (e.g., \textsc{Brotli}) than the others. This is possible since being small in terms of the search space does not necessarily mean that they are simple, because the landscape can still be radically complex to be modeled and contain many local optima traps.

\begin{tcbitemize}[%
    raster columns=1, 
    raster rows=1
    ]
  \tcbitem[myhbox={}{Finding \thefindingcount}]   \textit{Models can considerably improve the sequential model-based tuner over its model-free counterpart in 16 (up to 160\% improvement) and 21 (up to $5\times$ better) out of 29 cases for performance and efficiency, respectively, demonstrating its usefulness in guiding the tuning when they can be continuously updated.}
\end{tcbitemize}
\addtocounter{findingcount}{1}

In contrast, we observe rather different results when comparing batch model-based tuners with the model-free ones. From Figure~\ref{fig:rq1.1-sum}, we see that the best tuner-model pair from the batch type is commonly worse than the best model-free counterpart for performance. While it appears to be better than model-free tuner in efficiency, the improvements are often not statistically significant. In particular, the batch model-based tuner loses 28 cases with 1 tie for performance; it wins 6 cases and loses 1 case, together with 22 ties, for efficiency. The detailed results in Table~\ref{tb:rq1.2} show that the degradation of batch model-based tuner over the model-free one can be up to 200\%. Such a discrepancy, compared with the results of sequential model-based tuners, is due to the fact that the models are not updated progressively during tuning when used for batch model-based tuners.

As for the generally best tuner-model pair under batch model-based tuners, again, we found no generally best one across the systems. For both types of tuners, the \texttt{BestConfig} and \texttt{ParamILS} remain robust on both tuning quality metrics. Again, the quality divergence tends to be larger on the small systems compared with the others due to the same reason mentioned before.

\begin{tcbitemize}[%
    raster columns=1, 
    raster rows=1
    ]
  \tcbitem[myhbox={}{Finding \thefindingcount}]  \textit{Conversely, models generally have marginal or even harmful impact on the tuning quality of batch model-based tuner, as out of the 29 cases, they lose on 28 cases on performance while drawing or losing on 23 cases for efficiency against the model-free one.}
\end{tcbitemize}
\addtocounter{findingcount}{1}



\begin{figure}[t!]
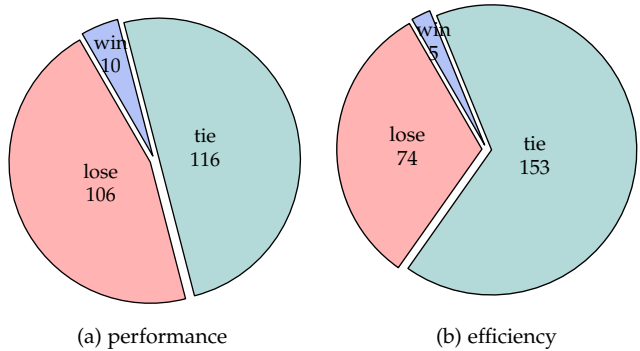

\centering
\subfloat[performance]{\includestandalone[width=0.49\columnwidth]{Figures/rq1.2-summary-1}}
~\hfill
\subfloat[efficiency]{\includestandalone[width=0.48\columnwidth]{Figures/rq1.2-summary-2}}

\caption{Comparing the original model in the sequential model-based tuner against the best alternative model when tuning under each combination of system and tuner for all 232 cases (30 runs). We count how many cases the origin wins, loses, or ties.}
\label{fig:rq1.2-sum}
\end{figure}

\subsection{RQ2: Do the Chosen Models Work the Best on Tuning Quality?}

\subsubsection{Method}

Another interesting investigation related to the model usefulness is to explore whether the fixed model, originally chosen for sequential model-based tuners (so-called origin), is indeed the optimal setting in terms of tuning quality, e.g., whether the \textsc{CART} used in \textsc{FLASH} is the best option. To that end, under each combination of system and tuner, we compare the best representative from all tuner-model pairs (selected in the same way as \textbf{RQ1} according to a quality metric) against the original tuner-model pair. In total, we have 29 systems $\times$ 8 sequential model-based tuners $=232$ cases to discuss for each metric. Other settings are the same as \textbf{RQ1}.

\subsubsection{Results}

Surprisingly, from Figure~\ref{fig:rq1.2-sum}, we see that the origins in sequential model-based tuners are often not as effective as we expected, as compared with some arbitrarily new tuner-model pairs, they lose on 46\% of the cases (106/232) with 50\% tie (116/232) on performance. Similar results can also be observed for efficiency: the origin has 32\% (74/232 cases) loss and 66\% (153/232 cases) tie.

Specifically, Tables~\ref{tb:rq2.1} and~\ref{tb:rq2.2} show the detailed results on different systems. Clearly, the performance can be improved by up to 286.62\% when we simply replace \texttt{OtterTune}'s \texttt{GP} with \texttt{DECART} under \textsc{kanzi}. Likewise, the efficiency improvement can reach up to 699.63\% by modifying \texttt{OtterTune}'s \texttt{GP} to \texttt{DaL} under \textsc{ExaStencils}. All of the above suggests that the originally chosen tuner-model pairs for sequential model-based tuners are far away from being optimal on any single metric of the tuning quality.

We also see that, relatively, the tuning quality of certain tuners can be more significantly improved by replacing the original model. This means that the impact of model replacement depends on the tuner's nature, e.g., \texttt{OtterTune} relies on gradient descent that works on the landscape predicted by the model to explore the configuration space, thereby any subtle change of the landscape caused by the model replacement can be much more influential to its behaviors. We also observe that there has been no consistent pattern regarding the systems, as the models can be sensitive to the characteristics of configuration data, e.g., sparsity.


\begin{tcbitemize}[%
    raster columns=1, 
    raster rows=1
    ]
  \tcbitem[myhbox={}{Finding \thefindingcount}]  \textit{The original model used in sequential model-based tuners has marginal difference or can be significantly improved (for over 95\% of the cases) compared with some other, ``newly created'' tuner-model pairs.}
\end{tcbitemize}
\addtocounter{findingcount}{1}

\input{Tables/RQ5/SMBO1}
\input{Tables/RQ5/SMBO2}

\subsection{RQ3: Does the Goodness of Model Consistent with the Resulted Tuning Quality?}


\subsubsection{Method}

To understand \textbf{RQ3}, for each combination of the system and model-based tuner, we select representatives of the best and worst according to a metric. Those are chosen using Scott-Knott ESD test and the average metric value as before. Our goal is to analyze whether the best/worst model, selected by using an accuracy metric, can still lead to the best/worst tuning outcome after the model is used to guide the tuning, assessed by one of the tuning quality metrics. As a result, for sequential model-based tuners, we have 29 systems $\times$ 8 tuners $=232$ cases under an accuracy-quality metric set. Similarly, for the batch model-based tuners, there are 21 large systems $\times$ 8 tuners + 8 small systems $=176$ cases\footnote{Note that for small systems, only \texttt{Brute-force} is needed as the space can be reasonably covered.} per accuracy-quality metric set. Other settings are the same as before.

\begin{figure}[t!]
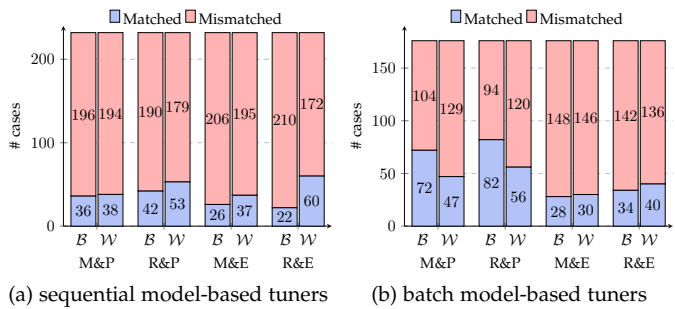

\centering
\subfloat[sequential model-based tuners]{\includestandalone[width=0.49\columnwidth]{Figures/rq3-summary-smbt}}
~\hfill
\subfloat[batch model-based tuners]{\includestandalone[width=0.49\columnwidth]{Figures/rq3-summary-bmbt}}

\caption{Counting the number of cases where the best (worst) model chosen based on accuracy matches with the model that leads to the best (worst) tuning quality with 30 repeats. $\mathcal{B}$ and $\mathcal{W}$ refer to the scenario of best and worst representative, respectively. M\&P, R\&P, M\&E, and R\&E respectively denote the combined metric set of MAPE-performance, $\mu$RD-performance, MAPE-efficiency, and $\mu$RD-efficiency.}
\label{fig:rq3-sum}
\end{figure}

\subsubsection{Results}

For sequential model-based tuners, as from Figure~\ref{fig:rq3-sum}, we see a rather low consistency between the model accuracy and the model's tuning performance: there are 16\% and 18\% matched cases (36 and 42 out of 232 cases) on the best representative for MAPE and $\mu$RD, respectively. This is similarly low for the worst representative, exhibiting 16\% and 23\% for MAPE and $\mu$RD (38 and 53 out of 232 cases), respectively. An even lower percentage of matching can be found between model accuracy and tuning efficiency, leading to 26 and 22 out of 232 cases for MAPE and $\mu$RD, respectively, under best representative; 37 and 60 out of 232 cases for MAPE and $\mu$RD, respectively, under worst representative.

The results tend to be more positive for the batch model-based tuners (Figure~\ref{fig:rq3-sum}) in terms of performance, but there are still less than 50\% of the cases show a match on both MAPE and $\mu$RD for the best representative. Similar results can be seen for the scenario of the worst representative: the percentage becomes rather low again when considering efficiency ---16\% and 19\% for MAPE and $\mu$RD (28 and 34 out of 176 cases), respectively, for the best representative. This becomes 17\% and 23\% for the worst representative on MAPE and $\mu$RD (30 and 40 out of 176 cases), respectively.

\input{Tables/RQ2/rq2_smbt}
\input{Tables/RQ2/rq2_bmbt}

The deviated matching between sequential and batch model-based tuners is not hard to understand, as the model continues to evolve in the former while it remains fixed for the latter. Therefore, the most/least accurate model could be more indicative and stable of their best/worst performance for batch model-based tuners than that of their sequential counterparts. Yet, for efficiency, the consistency remains low due to the sparse nature of configuration tuning.

\begin{figure*}[t!]
\centering
\subfloat{\includegraphics[width=0.6\textwidth]{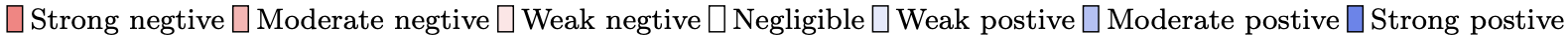}}
\vspace{-0.8cm}
\end{figure*}
\begin{figure*}[t!]
\subfloat[sequential model-based tuners]{\includestandalone[width=0.49\textwidth]{Figures/rq4-summary-1}}
~\hfill
\subfloat[batch model-based tuners]{\includestandalone[width=0.49\textwidth]{Figures/rq4-summary-2}}

\caption{Counting the number of cases on the nature/strength for the correlation between accuracy and tuning quality with 30 repeats. The abbreviations are the same as Figure~\ref{fig:rq3-sum}.}
\label{fig:rq4-sum}
\end{figure*}

Tables~\ref{tb:rq3.1} and~\ref{tb:rq3.2} and illustrate the detailed results. We see that, when using certain tuners, it is more likely to have a good consistency between model accuracy and tuning quality of the model than others, e.g., \texttt{OtterTune}, \texttt{ParamILS} and \texttt{ConEx}. The reason behind this can be dependent on how well the tuner balances exploitation and exploration by leveraging the model (see Section~\ref{sec:pair}). Interestingly, we observed that even for \texttt{Brute-force}, there is still a considerable number of mismatched cases. This is because while the overall accuracy might be high, the best-predicted configuration might still be rather erroneous, which prevents it from being effective in guiding the tuning. We will further examine how accurate is enough in \textbf{RQ5}. It is also clear that certain systems lead to more matched cases depending on the tuner type, e.g., \textsc{MongoDB} for sequential and \textsc{MariaDB} for batch model-based tuners. This is due to the tuning landscape of the systems: if the landscape emulated by the model happens to be well-described by the accuracy, then its better or worse is more likely to be consistent with the best/worst tuning quality\footnote{Detailed data can be found at:  \textcolor{blue}{\texttt{\url{https://github.com/ideas-labo/model-impact/blob/main/RQ_supplementary/RQ3/supplement.pdf}}}.}.


\begin{tcbitemize}[%
    raster columns=1, 
    raster rows=1
    ]
  \tcbitem[myhbox={}{Finding \thefindingcount}]  \textit{The sequential model-based tuners lead to generally inconsistent best/worst match between the accuracy and tuning quality of the model (matching in less than 20\% of the cases). The batch model-based tuners, in contrast, have similar results for efficiency but are likely to be more consistent in performance. However, the matching cases are merely up to 45\%.}
\end{tcbitemize}
\addtocounter{findingcount}{1}

\input{Tables/RQ3/rq3_smbt_perf}
\input{Tables/RQ3/rq3_smbt_eff}
\input{Tables/RQ3/rq3_bmbt_perf}
\input{Tables/RQ3/rq3_bmbt_eff}

\subsection{RQ4: How Do the Model Accuracy and Tuning Quality Correlated?}

\subsubsection{Methods}

For \textbf{RQ4}, our goal is to study the detailed correlation between the changes in accuracy and that of the tuning quality affected by the models. Therefore, we leverage Spearman's rank correlation and the interpretation mentioned in Section~\ref{sec:sta} to analyze the trend therein. We also use statistical tests to verify whether the correlation is statistically significant. Again, there are 232 and 176 cases to investigate for sequential and batch model-based tuners, respectively. Other settings are identical to those of \textbf{RQ3}. Note that to maintain consistency in the correlation analysis, we convert the systems wherein the performance should be maximized by using the additive inverse of the performance value instead. Therefore, for all systems, a negative correlation means the tuning quality becomes worse when the accuracy improves, and vice versa.

\subsubsection{Results}

As from the Figure~\ref{fig:rq4-sum}a, we observe that a negligible correlation between any accuracy and tuning quality has been prevalent for sequential model-based tuners, ranging from 35\% for $\mu$RD-performance to 58\% for $\mu$RD-efficiency. There are also a considerable amount of cases with weak correlations (from 31\% to 47\% cases). This means that, together for between 82\% and 97\% of the cases, the accuracy is unlikely to significantly influence the tuning quality. It is interesting to see that there is a small but noticeable proportion of negative correlation (weak or moderate)---the worse accuracy actually leads to better configuration tuning by a model. This, albeit counter-intuitive, is indeed possible due to the complex change in the structure of the configuration landscape, as we will discuss in Section~\ref{sec:discussion}.

We see similar results between accuracy and tuning efficiency for batch model-based tuners but negative correlations are less likely, as shown in Figure~\ref{fig:rq4-sum}b. When examining the accuracy and performance, batch model-based tuners are more commonly to exhibit positive correlations than their sequential counterparts, i.e., 32\% and 40\% cases respectively for MAPE and $\mu$RD wherein the correlation is positively moderate or above. This matches with the observations of \textbf{RQ3}, where the models provide more inductive and stable guidance for batch model-based tuners. Yet, only up to 9\% cases of the correlations are strong and positive as implied in the general belief.

Tables~\ref{tb:rq4_smbt_perf}---\ref{tb:rq4_bmbt_eff} show the detailed results. Again, we see that certain tuners show generally stronger correlations than others, e.g., \texttt{OtterTune}, \texttt{Flash}, \texttt{SMAC}, and \texttt{GA}, due to the same reasons mentioned in \textbf{RQ3}. Interestingly, \texttt{Random} also exhibits considerable strength of correlations due to its nature, since it has no sophisticated heuristics and hence the tuning quality is more relevant to the accuracy of the mode that guides it. Some tuners also easier to have negative correlations because they tend to leverage the model more loosely, e.g., in \texttt{ROBOTune}, the model predictions are guided by an ensemble of three acquisition functions, hence weakening the positive effect of the model.

Despite some systems being more likely to result in negative correlation due to their complex tuning landscapes (e.g., \textsc{Apache}), we see no consistent pattern in the strength of correlation with respect to the scale of systems. This is because the scale might not be the only factor that influences the correlation between model accuracy and its tuning quality, but more importantly the characteristic of the corresponding configuration landscape (as we will discuss in Section~\ref{sec:discussion}).


\begin{tcbitemize}[%
    raster columns=1, 
    raster rows=1
    ]
  \tcbitem[myhbox={}{Finding \thefindingcount}]  \textit{The accuracy of the model generally has a negligible or weak correlation to the tuning quality it guides for sequential model-based tuners (from 82\% to 97\% of the cases). The batch counterparts exhibit mostly similar results but with more obvious correlations between model accuracy and the resulting performance. The strength is often not strong though.}
\end{tcbitemize}
\addtocounter{findingcount}{1}

\subsection{RQ5: How Much Accuracy Change Do We Need?}

\subsubsection{Methods}

Finally, we are particularly interested in understanding to what extent the model accuracy needs to differ in order to create nontrivial improvement in the tuning quality. To that end, in \textbf{RQ5}, we perform the following steps:

\begin{enumerate}
    \item Pick a model accuracy metric and a tuning quality metric.
    \item For MAPE, we divide the values into 11 ranges, i.e., $\{0-10,10-20,...,>100\}$. Similarly, for $\mu$RD, we use four ranges, i.e., $\{0-0.1,0.1-0.2,0.2-0.3,>0.3\}$. Those ranges are determined based on the distribution of their accuracy values we observed. 
    \item Pick a tuner and a system studied.
    \item Within each range, we find the set of tuner-model pairs (under the chosen tuner and system), denoted as $\mathbfcal{A}$, whose average accuracy values across the runs fit within the range. For each pair $\vect{a} \in \mathbfcal{A}$ (an anchor), we calculate its difference on the accuracy metric against any other pairs for the same tuner and system, says $\vect{b}$, such that $\vect{b} \in \mathbfcal{B}_a$ achieves significantly better results than that of $\vect{a}$ on the tuning quality metric according to Mann-Whitney U-test ($\vect{b}$ could fall in any range). 
        \item We find the $\vect{b_s} \in \mathbfcal{B}_a$ such that the accuracy difference $\Delta$ between $\vect{a}$ and $\vect{b_s}$ is the smallest across any other $\vect{b} \in \mathbfcal{B}_a$ (we distinguish the cases where $\vect{b}$ has better accuracy from those where $\vect{b}$ has worse accuracy than $\vect{a}$, since the correlation between accuracy and tuning quality can be negative). This $\Delta$ serves as a sample. 
    \item Repeat from 3) until all combinations of tuners and systems are covered.
    
    \item Repeat from 1) until all combinations of accuracy-quality metrics have been considered.
\end{enumerate}

\begin{figure*}[t!]
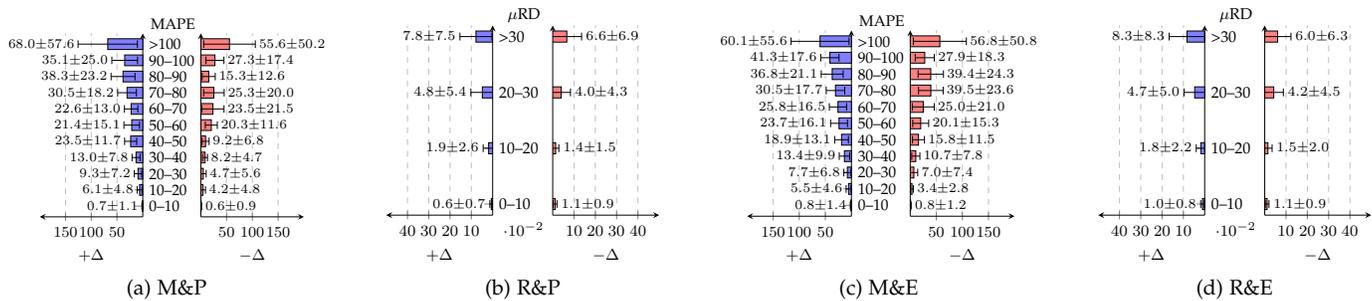

\centering
\subfloat[M\&P]{\includestandalone[width=0.24\textwidth]{Figures/RQ5/smbt-mp}}
~\hfill
\subfloat[R\&P]{\includestandalone[width=0.2\textwidth]{Figures/RQ5/smbt-rp}}
~\hfill
\subfloat[M\&E]{\includestandalone[width=0.24\textwidth]{Figures/RQ5/smbt-me}}
~\hfill
\subfloat[R\&E]{\includestandalone[width=0.2\textwidth]{Figures/RQ5/smbt-re}}

\caption{The $\Delta$ of model accuracy for creating significant improvements on tuning quality under sequential model-based tuners. $+\Delta$ and $-\Delta$ denote the improvements in tuning quality are the results of enhanced and worsened accuracy, respectively. The abbreviations are the same as Figure~\ref{fig:rq3-sum}. Detailed figures can be found at here: \textcolor{blue}{\texttt{\url{https://github.com/ideas-labo/model-impact/blob/main/RQ_supplementary/RQ5/supplement.pdf}}}.}
\label{fig:rq5-1}
\end{figure*}

\begin{figure*}[t!]
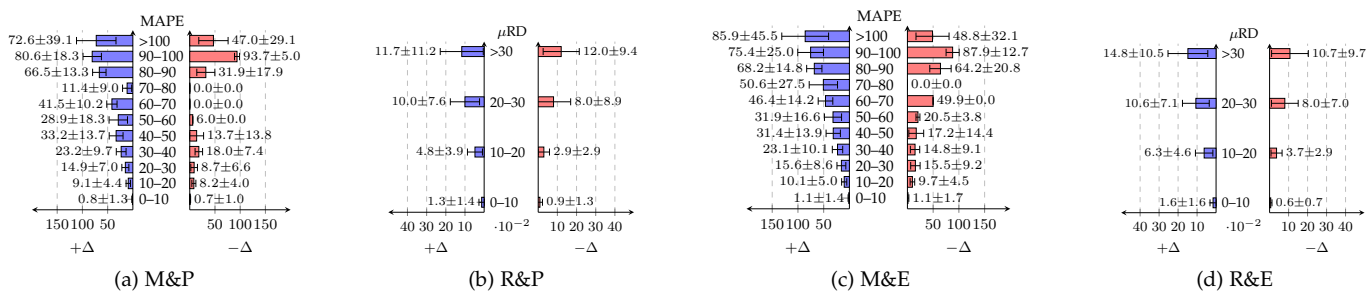

\centering
\subfloat[M\&P]{\includestandalone[width=0.222\textwidth]{Figures/RQ5/bmbt-mp}}
~\hfill
\subfloat[R\&P]{\includestandalone[width=0.2\textwidth]{Figures/RQ5/bmbt-rp}}
~\hfill
\subfloat[M\&E]{\includestandalone[width=0.24\textwidth]{Figures/RQ5/bmbt-me}}
~\hfill
\subfloat[R\&E]{\includestandalone[width=0.2\textwidth]{Figures/RQ5/bmbt-re}}

\caption{The $\Delta$ of model accuracy for creating significant improvements on tuning quality under batch model-based tuners. 0.0$\pm$0.0 implies no samples can significantly improve the tuning for a range. The formats are the same as Figure~\ref{fig:rq5-1}. Detailed figures can be found at here: \textcolor{blue}{\texttt{\url{https://github.com/ideas-labo/model-impact/blob/main/RQ_supplementary/RQ5/supplement.pdf}}}.}
\label{fig:rq5-2}
\end{figure*}

For example, suppose that we aim to study how much MAPE changes can significantly influence the tuning performance while the ranges of interest have been defined and there are four models (\texttt{DaL}, \texttt{DeepPerf}, \texttt{GP}, and \texttt{DECART}) with an average MAPE of 5\%, 15\%, 40\%, and 45\%, respectively (Steps 1-3), we at first pick a tuner and system, e.g., \texttt{Flash} and \textsc{MariaDB}. At Step 4, assuming that the best performance of the tuner-model pair is \texttt{Flash}-\texttt{DaL} followed by \texttt{Flash}-\texttt{DeepPerf}, \texttt{Flash}-\texttt{GP} and then \texttt{Flash}-\texttt{DECART}, such that all of them except that between \texttt{Flash}-\texttt{GP} and \texttt{Flash}-\texttt{DECART} are statistically different according to the U-test, if we treat \texttt{Flash}-\texttt{DECART} as $\vect{a}$ (an anchor), then we knows its MAPE is within 40\% and 50\% while it corresponds to two $\vect{b}$: \texttt{Flash}-\texttt{DaL} and \texttt{Flash}-\texttt{DeepPerf}, which can significantly improve the tuning results of \texttt{Flash}-\texttt{DECART} (Step 4). Now, in Step 5, we know that \texttt{DaL} and \texttt{DeepPerf} improves the MAPE from \texttt{DECART} by 40\% and 30\%, respectively. Therefore, to answer \textbf{RQ5}, a sample (for the $+\Delta$ case) we found for MAPE between 40\% and 50\% is that it needs at least 30\% improvement to significantly boost the tuning performance. We can then treat the other pair as  $\vect{a}$ to collect more samples. The same can be performed on other tuners, systems, and metrics (Steps 6-7).

For each accuracy-quality combination, we plot the mean and standard deviation of the accuracy changes from the samples for both sequential and batch model-based tuners.

\subsubsection{Results}


From the results plotted in Figures~\ref{fig:rq5-1} and~\ref{fig:rq5-2}, we can suggest that for a model that has an accuracy range of $[x,y]$, it needs to improve (or worsen due to the findings of \textbf{RQ4}) its accuracy by at least an average $\Delta$ in general to significantly enhance the tuning quality. Clearly, we see that for both sequential and batch model-based tuners, certain ranges of the model accuracy only require a small accuracy change to improve the tuning quality, e.g., when the model has between 0\% - 10\% MAPE under sequential model-based tuners, an increase of as small as 0.7\% in average would have already led to a considerable performance improvement on tuning. Other ranges, in contrast, exhibit the need for a relatively larger accuracy change to realize nontrivial tuning improvement. For example, when the MAPE is worse than 100\%, it needs around 85.9\% MAPE improvement to achieve significantly better efficiency of batch model-based tuners.

Notably, there is a clear trend where models with better accuracy tend to require smaller $\Delta$ for significantly improving the tuning quality over all cases. This is a consistent pattern regardless of whether the model accuracy enhances (blue) or degrades (red). 


\begin{tcbitemize}[%
    raster columns=1, 
    raster rows=1
    ]
  \tcbitem[myhbox={}{Finding \thefindingcount}]  \textit{Depending on the ranges of model accuracy, it requires distinct changes in the accuracy (improve or worsen) to significantly ameliorate any tuning quality metric for both sequential and batch model-based tuners ($[0.6,85.9]$ for MAPE and $[0.006,0.148]$ for $\mu$RD). In particular, the better the accuracy, the smaller the accuracy change is needed to considerably enhance the tuning.}
\end{tcbitemize}
\addtocounter{findingcount}{1}

%% file: Tables/RQ4/SMBO.tex
\begin{table*}[t!]
\caption{Comparing the best tuner-model pairs of sequential model-based tuners and their best model-free counterparts over 30 runs. $\Delta$\% refers to the ratio of ${{f-b} \over f} \times 100$ (or ${{b-f} \over f} \times 100$ if the quality metric is to be maximized) such that $b$ and $f$ are the values of the tuning quality metric for a model-based tuner and corresponding model-free tuner, respectively. $^\dagger$ and $^\star$ denote $p<0.001$ and $0.001 \leq p<0.05$, respectively. Regardless of statistical significance, \colorbox{steel!30}{blue cells} indicate that the pairs under model-based tuners lead to better average performance than their model-free counterparts; \colorbox{red!20}{red cells} mean the tuner-model pairs have worse average performance. Statistically significant comparisons are highlighted in \textbf{bold}. There is a \textsc{Storm} case where the performance value is too small as a decimal value, and hence showing as 0; but the \% of difference can still be very large.} 
\label{tb:rq1.1}
\centering
\footnotesize
\begin{adjustbox}{width=\textwidth,center}
\begin{tabular}{l|lll|ll|r|lll|ll|r}


\toprule

\multirow{2}{*}{\textbf{System}}&\multicolumn{6}{c|}{\textbf{Performance}}&\multicolumn{6}{c}{\textbf{Efficiency}}\\

\cmidrule{2-13}
 &\textbf{Sequential}&\textbf{Model}&\textbf{Performance}&\textbf{Model-free}&\textbf{Performance}&\textbf{$\Delta$\%}&\textbf{Sequential}&\textbf{Model}&\textbf{Efficiency}&\textbf{Model-free}&\textbf{Efficiency}&\textbf{$\Delta$\%}\\

\midrule

\textsc{Brotli}&\cellcolor{steel!30}\textbf{\texttt{FLASH}}&\cellcolor{steel!30}\textbf{\texttt{DeP}}&\cellcolor{steel!30}\textbf{0.54$\pm$0.00}&\cellcolor{steel!30}\textbf{\texttt{R\_search}}&\cellcolor{steel!30}\textbf{1.46$\pm$0.01}&\cellcolor{steel!30}\textbf{{169.00\%$^\dagger$}}&\cellcolor{steel!30}\textbf{\texttt{SMAC}}&\cellcolor{steel!30}\textbf{\texttt{GP}}&\cellcolor{steel!30}\textbf{0.01$\pm$0.01}&\cellcolor{steel!30}\textbf{\texttt{SWAY}}&\cellcolor{steel!30}\textbf{0.19$\pm$0.05}&\cellcolor{steel!30}\textbf{{1421.69\%$^\dagger$}}\\
\textsc{LLVM}&\cellcolor{steel!30}\textbf{\texttt{ROBOTune}}&\cellcolor{steel!30}\textbf{\texttt{DaL}}&\cellcolor{steel!30}\textbf{199.68$\pm$0.00}&\cellcolor{steel!30}\textbf{\texttt{ParamILS}}&\cellcolor{steel!30}\textbf{199.91$\pm$0.24}&\cellcolor{steel!30}\textbf{{0.11\%$^\dagger$}}&\cellcolor{steel!30}\textbf{\texttt{SMAC}}&\cellcolor{steel!30}\textbf{\texttt{GP}}&\cellcolor{steel!30}\textbf{0.01$\pm$0.01}&\cellcolor{steel!30}\textbf{\texttt{GGA}}&\cellcolor{steel!30}\textbf{0.43$\pm$0.26}&\cellcolor{steel!30}\textbf{{2949.30\%$^\dagger$}}\\
\textsc{Lrzip}&\cellcolor{steel!30}\texttt{ResTune}&\cellcolor{steel!30}\texttt{RF}&\cellcolor{steel!30}3.12$\pm$0.00&\cellcolor{steel!30}\texttt{ConEx}&\cellcolor{steel!30}3.12$\pm$0.00&\cellcolor{steel!30}{0.02\%$^{\,\,\,}$}&\cellcolor{steel!30}\textbf{\texttt{SMAC}}&\cellcolor{steel!30}\textbf{\texttt{GP}}&\cellcolor{steel!30}\textbf{0.01$\pm$0.01}&\cellcolor{steel!30}\textbf{\texttt{ConEx}}&\cellcolor{steel!30}\textbf{0.36$\pm$0.23}&\cellcolor{steel!30}\textbf{{3493.17\%$^\dagger$}}\\
\textsc{xgboost}&\cellcolor{steel!30}\textbf{\texttt{FLASH}}&\cellcolor{steel!30}\textbf{\texttt{DeP}}&\cellcolor{steel!30}\textbf{10.32$\pm$0.00}&\cellcolor{steel!30}\textbf{\texttt{ParamILS}}&\cellcolor{steel!30}\textbf{11.39$\pm$2.15}&\cellcolor{steel!30}\textbf{{10.43\%$^\star$}}&\cellcolor{steel!30}\textbf{\texttt{OtterTune}}&\cellcolor{steel!30}\textbf{\texttt{DeP}}&\cellcolor{steel!30}\textbf{0.13$\pm$0.08}&\cellcolor{steel!30}\textbf{\texttt{ParamILS}}&\cellcolor{steel!30}\textbf{0.31$\pm$0.24}&\cellcolor{steel!30}\textbf{{131.47\%$^\star$}}\\
\textsc{noc-CM-log}&\texttt{ATConf}&\texttt{RF}&4.31$\pm$0.00&\texttt{BestConfig}&4.31$\pm$0.00&{0.00\%$^{\,\,\,}$}&\cellcolor{steel!30}\textbf{\texttt{SMAC}}&\cellcolor{steel!30}\textbf{\texttt{GP}}&\cellcolor{steel!30}\textbf{0.01$\pm$0.00}&\cellcolor{steel!30}\textbf{\texttt{ParamILS}}&\cellcolor{steel!30}\textbf{0.13$\pm$0.05}&\cellcolor{steel!30}\textbf{{1219.58\%$^\dagger$}}\\
\textsc{DeepArch}&\cellcolor{steel!30}\textbf{\texttt{FLASH}}&\cellcolor{steel!30}\textbf{\texttt{DeP}}&\cellcolor{steel!30}\textbf{1.05$\pm$0.00}&\cellcolor{steel!30}\textbf{\texttt{GGA}}&\cellcolor{steel!30}\textbf{1.05$\pm$0.00}&\cellcolor{steel!30}\textbf{{0.10\%$^\dagger$}}&\cellcolor{steel!30}\textbf{\texttt{FLASH}}&\cellcolor{steel!30}\textbf{\texttt{RF}}&\cellcolor{steel!30}\textbf{0.11$\pm$0.15}&\cellcolor{steel!30}\textbf{\texttt{GGA}}&\cellcolor{steel!30}\textbf{0.43$\pm$0.24}&\cellcolor{steel!30}\textbf{{297.07\%$^\dagger$}}\\
\textsc{BDB\_C}&\cellcolor{steel!30}\textbf{\texttt{SMAC}}&\cellcolor{steel!30}\textbf{\texttt{DT}}&\cellcolor{steel!30}\textbf{0.35$\pm$0.00}&\cellcolor{steel!30}\textbf{\texttt{ConEx}}&\cellcolor{steel!30}\textbf{0.35$\pm$0.00}&\cellcolor{steel!30}\textbf{{0.19\%$^\dagger$}}&\cellcolor{steel!30}\textbf{\texttt{SMAC}}&\cellcolor{steel!30}\textbf{\texttt{GP}}&\cellcolor{steel!30}\textbf{0.08$\pm$0.06}&\cellcolor{steel!30}\textbf{\texttt{ParamILS}}&\cellcolor{steel!30}\textbf{0.32$\pm$0.19}&\cellcolor{steel!30}\textbf{{331.56\%$^\dagger$}}\\
\textsc{HSQLDB}&\cellcolor{steel!30}\textbf{\texttt{BOCA}}&\cellcolor{steel!30}\textbf{\texttt{SPL}}&\cellcolor{steel!30}\textbf{248.20$\pm$0.00}&\cellcolor{steel!30}\textbf{\texttt{BestConfig}}&\cellcolor{steel!30}\textbf{248.23$\pm$0.07}&\cellcolor{steel!30}\textbf{{0.01\%$^\star$}}&\cellcolor{steel!30}\textbf{\texttt{SMAC}}&\cellcolor{steel!30}\textbf{\texttt{GP}}&\cellcolor{steel!30}\textbf{0.01$\pm$0.00}&\cellcolor{steel!30}\textbf{\texttt{BestConfig}}&\cellcolor{steel!30}\textbf{0.34$\pm$0.31}&\cellcolor{steel!30}\textbf{{5016.25\%$^\dagger$}}\\
\textsc{DConvert}&\cellcolor{steel!30}\textbf{\texttt{FLASH}}&\cellcolor{steel!30}\textbf{\texttt{DaL}}&\cellcolor{steel!30}\textbf{1.76$\pm$0.00}&\cellcolor{steel!30}\textbf{\texttt{GGA}}&\cellcolor{steel!30}\textbf{1.80$\pm$0.02}&\cellcolor{steel!30}\textbf{{2.38\%$^\dagger$}}&\cellcolor{steel!30}\textbf{\texttt{ROBOTune}}&\cellcolor{steel!30}\textbf{\texttt{SVR}}&\cellcolor{steel!30}\textbf{0.03$\pm$0.02}&\cellcolor{steel!30}\textbf{\texttt{ParamILS}}&\cellcolor{steel!30}\textbf{0.34$\pm$0.27}&\cellcolor{steel!30}\textbf{{1132.27\%$^\dagger$}}\\
\textsc{7z}&\cellcolor{steel!30}\textbf{\texttt{SMAC}}&\cellcolor{steel!30}\textbf{\texttt{DCT}}&\cellcolor{steel!30}\textbf{4238.93$\pm$35.88}&\cellcolor{steel!30}\textbf{\texttt{GGA}}&\cellcolor{steel!30}\textbf{4272.85$\pm$38.87}&\cellcolor{steel!30}\textbf{{0.80\%$^\star$}}&\cellcolor{steel!30}\texttt{OtterTune}&\cellcolor{steel!30}\texttt{DaL}&\cellcolor{steel!30}0.28$\pm$0.25&\cellcolor{steel!30}\texttt{ParamILS}&\cellcolor{steel!30}0.37$\pm$0.31&\cellcolor{steel!30}{33.94\%$^{\,\,\,}$}\\
\textsc{Apache}&\cellcolor{steel!30}\texttt{SMAC}&\cellcolor{steel!30}\texttt{RF}&\cellcolor{steel!30}170.82$\pm$0.03&\cellcolor{steel!30}\texttt{GA}&\cellcolor{steel!30}170.80$\pm$0.11&\cellcolor{steel!30}{0.01\%$^{\,\,\,}$}&\cellcolor{steel!30}\textbf{\texttt{SMAC}}&\cellcolor{steel!30}\textbf{\texttt{GP}}&\cellcolor{steel!30}\textbf{0.02$\pm$0.01}&\cellcolor{steel!30}\textbf{\texttt{GGA}}&\cellcolor{steel!30}\textbf{0.43$\pm$0.25}&\cellcolor{steel!30}\textbf{{2056.01\%$^\dagger$}}\\
\textsc{HSMGP}&\cellcolor{steel!30}\textbf{\texttt{OtterTune}}&\cellcolor{steel!30}\textbf{\texttt{HIP}}&\cellcolor{steel!30}\textbf{100.31$\pm$0.00}&\cellcolor{steel!30}\textbf{\texttt{GGA}}&\cellcolor{steel!30}\textbf{100.51$\pm$0.38}&\cellcolor{steel!30}\textbf{{0.19\%$^\star$}}&\cellcolor{steel!30}\textbf{\texttt{ROBOTune}}&\cellcolor{steel!30}\textbf{\texttt{SPL}}&\cellcolor{steel!30}\textbf{0.02$\pm$0.02}&\cellcolor{steel!30}\textbf{\texttt{ParamILS}}&\cellcolor{steel!30}\textbf{0.21$\pm$0.13}&\cellcolor{steel!30}\textbf{{814.74\%$^\dagger$}}\\
\textsc{MongoDB}&\cellcolor{steel!30}\textbf{\texttt{SMAC}}&\cellcolor{steel!30}\textbf{\texttt{RF}}&\cellcolor{steel!30}\textbf{206356.00$\pm$0.00}&\cellcolor{steel!30}\textbf{\texttt{BestConfig}}&\cellcolor{steel!30}\textbf{206723.43$\pm$548.92}&\cellcolor{steel!30}\textbf{{0.18\%$^\dagger$}}&\cellcolor{steel!30}\textbf{\texttt{ROBOTune}}&\cellcolor{steel!30}\textbf{\texttt{SVR}}&\cellcolor{steel!30}\textbf{0.02$\pm$0.01}&\cellcolor{steel!30}\textbf{\texttt{BestConfig}}&\cellcolor{steel!30}\textbf{0.47$\pm$0.37}&\cellcolor{steel!30}\textbf{{2726.02\%$^\dagger$}}\\
\textsc{PostgreSQL}&\cellcolor{steel!30}\textbf{\texttt{FLASH}}&\cellcolor{steel!30}\textbf{\texttt{DeP}}&\cellcolor{steel!30}\textbf{45999.31$\pm$49.26}&\cellcolor{steel!30}\textbf{\texttt{GA}}&\cellcolor{steel!30}\textbf{50680.19$\pm$12.07}&\cellcolor{steel!30}\textbf{{10.18\%$^\dagger$}}&\cellcolor{steel!30}\textbf{\texttt{FLASH}}&\cellcolor{steel!30}\textbf{\texttt{GP}}&\cellcolor{steel!30}\textbf{0.01$\pm$0.00}&\cellcolor{steel!30}\textbf{\texttt{SWAY}}&\cellcolor{steel!30}\textbf{0.26$\pm$0.13}&\cellcolor{steel!30}\textbf{{4743.40\%$^\dagger$}}\\
\textsc{ExaStencils}&\cellcolor{steel!30}\texttt{FLASH}&\cellcolor{steel!30}\texttt{DeP}&\cellcolor{steel!30}4646.99$\pm$4.55&\cellcolor{steel!30}\texttt{ConEx}&\cellcolor{steel!30}4658.07$\pm$28.37&\cellcolor{steel!30}{0.24\%$^{\,\,\,}$}&\cellcolor{red!20}\texttt{ROBOTune}&\cellcolor{red!20}\texttt{GP}&\cellcolor{red!20}0.46$\pm$0.25&\cellcolor{red!20}\texttt{ParamILS}&\cellcolor{red!20}0.41$\pm$0.32&\cellcolor{red!20}{$-$10.65\%$^{\,\,\,}$}\\
\textsc{kanzi}&\cellcolor{steel!30}\textbf{\texttt{FLASH}}&\cellcolor{steel!30}\textbf{\texttt{DeP}}&\cellcolor{steel!30}\textbf{3.56$\pm$0.00}&\cellcolor{steel!30}\textbf{\texttt{ConEx}}&\cellcolor{steel!30}\textbf{3.76$\pm$0.20}&\cellcolor{steel!30}\textbf{{5.55\%$^\dagger$}}&\cellcolor{steel!30}\textbf{\texttt{ROBOTune}}&\cellcolor{steel!30}\textbf{\texttt{SVR}}&\cellcolor{steel!30}\textbf{0.06$\pm$0.02}&\cellcolor{steel!30}\textbf{\texttt{ParamILS}}&\cellcolor{steel!30}\textbf{0.24$\pm$0.14}&\cellcolor{steel!30}\textbf{{343.38\%$^\dagger$}}\\
\textsc{jump3r}&\cellcolor{steel!30}\textbf{\texttt{OtterTune}}&\cellcolor{steel!30}\textbf{\texttt{DT}}&\cellcolor{steel!30}\textbf{0.60$\pm$0.00}&\cellcolor{steel!30}\textbf{\texttt{GGA}}&\cellcolor{steel!30}\textbf{0.63$\pm$0.03}&\cellcolor{steel!30}\textbf{{5.17\%$^\dagger$}}&\cellcolor{steel!30}\textbf{\texttt{ROBOTune}}&\cellcolor{steel!30}\textbf{\texttt{SVR}}&\cellcolor{steel!30}\textbf{0.04$\pm$0.05}&\cellcolor{steel!30}\textbf{\texttt{ParamILS}}&\cellcolor{steel!30}\textbf{0.21$\pm$0.14}&\cellcolor{steel!30}\textbf{{411.69\%$^\dagger$}}\\
\textsc{MariaDB}&\cellcolor{steel!30}\textbf{\texttt{ResTune}}&\cellcolor{steel!30}\textbf{\texttt{LR}}&\cellcolor{steel!30}\textbf{55.97$\pm$0.00}&\cellcolor{steel!30}\textbf{\texttt{BestConfig}}&\cellcolor{steel!30}\textbf{56.09$\pm$0.30}&\cellcolor{steel!30}\textbf{{0.21\%$^\star$}}&\cellcolor{steel!30}\textbf{\texttt{SMAC}}&\cellcolor{steel!30}\textbf{\texttt{GP}}&\cellcolor{steel!30}\textbf{0.01$\pm$0.01}&\cellcolor{steel!30}\textbf{\texttt{BestConfig}}&\cellcolor{steel!30}\textbf{0.32$\pm$0.29}&\cellcolor{steel!30}\textbf{{4022.32\%$^\dagger$}}\\
\textsc{polly}&\cellcolor{steel!30}\texttt{SMAC}&\cellcolor{steel!30}\texttt{RF}&\cellcolor{steel!30}4.25$\pm$0.02&\cellcolor{steel!30}\texttt{BestConfig}&\cellcolor{steel!30}4.26$\pm$0.01&\cellcolor{steel!30}{0.10\%$^{\,\,\,}$}&\cellcolor{steel!30}\texttt{ROBOTune}&\cellcolor{steel!30}\texttt{SVR}&\cellcolor{steel!30}0.07$\pm$0.06&\cellcolor{steel!30}\texttt{BestConfig}&\cellcolor{steel!30}0.19$\pm$0.23&\cellcolor{steel!30}{192.60\%$^{\,\,\,}$}\\
\textsc{SQL}&\cellcolor{red!20}\textbf{\texttt{ResTune}}&\cellcolor{red!20}\textbf{\texttt{RF}}&\cellcolor{red!20}\textbf{12.57$\pm$0.09}&\cellcolor{red!20}\textbf{\texttt{BestConfig}}&\cellcolor{red!20}\textbf{12.51$\pm$0.00}&\cellcolor{red!20}\textbf{{$-$0.42\%$^\star$}}&\cellcolor{red!20}\textbf{\texttt{ROBOTune}}&\cellcolor{red!20}\textbf{\texttt{SVR}}&\cellcolor{red!20}\textbf{0.06$\pm$0.05}&\cellcolor{red!20}\textbf{\texttt{BestConfig}}&\cellcolor{red!20}\textbf{0.05$\pm$0.07}&\cellcolor{red!20}\textbf{{$-$10.91\%$^\star$}}\\
\textsc{vp9}&\cellcolor{steel!30}\texttt{ATConf}&\cellcolor{steel!30}\texttt{SPL}&\cellcolor{steel!30}41.26$\pm$0.00&\cellcolor{steel!30}\texttt{BestConfig}&\cellcolor{steel!30}41.28$\pm$0.13&\cellcolor{steel!30}{0.06\%$^{\,\,\,}$}&\cellcolor{red!20}\texttt{SMAC}&\cellcolor{red!20}\texttt{SPL}&\cellcolor{red!20}0.17$\pm$0.09&\cellcolor{red!20}\texttt{ParamILS}&\cellcolor{red!20}0.16$\pm$0.06&\cellcolor{red!20}{$-$4.11\%$^{\,\,\,}$}\\
\textsc{Spark}&\cellcolor{steel!30}\textbf{\texttt{FLASH}}&\cellcolor{steel!30}\textbf{\texttt{RF}}&\cellcolor{steel!30}\textbf{3427.07$\pm$59.91}&\cellcolor{steel!30}\textbf{\texttt{BestConfig}}&\cellcolor{steel!30}\textbf{3370.70$\pm$60.49}&\cellcolor{steel!30}\textbf{{1.64\%$^\dagger$}}&\cellcolor{steel!30}\textbf{\texttt{SMAC}}&\cellcolor{steel!30}\textbf{\texttt{GP}}&\cellcolor{steel!30}\textbf{0.05$\pm$0.02}&\cellcolor{steel!30}\textbf{\texttt{ParamILS}}&\cellcolor{steel!30}\textbf{0.13$\pm$0.08}&\cellcolor{steel!30}\textbf{{162.00\%$^\dagger$}}\\
\textsc{HIPAcc}&\cellcolor{red!20}\textbf{\texttt{FLASH}}&\cellcolor{red!20}\textbf{\texttt{SVR}}&\cellcolor{red!20}\textbf{21.18$\pm$0.00}&\cellcolor{red!20}\textbf{\texttt{BestConfig}}&\cellcolor{red!20}\textbf{21.17$\pm$0.00}&\cellcolor{red!20}\textbf{{$-$0.03\%$^\dagger$}}&\cellcolor{steel!30}\textbf{\texttt{ROBOTune}}&\cellcolor{steel!30}\textbf{\texttt{SVR}}&\cellcolor{steel!30}\textbf{0.05$\pm$0.02}&\cellcolor{steel!30}\textbf{\texttt{ParamILS}}&\cellcolor{steel!30}\textbf{0.12$\pm$0.06}&\cellcolor{steel!30}\textbf{{156.33\%$^\dagger$}}\\
\textsc{Redis}&\cellcolor{steel!30}\texttt{FLASH}&\cellcolor{steel!30}\texttt{SPL}&\cellcolor{steel!30}86005.16$\pm$1585.10&\cellcolor{steel!30}\texttt{R\_search}&\cellcolor{steel!30}85960.53$\pm$1148.89&\cellcolor{steel!30}{0.05\%$^{\,\,\,}$}&\cellcolor{steel!30}\textbf{\texttt{SMAC}}&\cellcolor{steel!30}\textbf{\texttt{GP}}&\cellcolor{steel!30}\textbf{0.06$\pm$0.03}&\cellcolor{steel!30}\textbf{\texttt{ConEx}}&\cellcolor{steel!30}\textbf{0.25$\pm$0.22}&\cellcolor{steel!30}\textbf{{356.11\%$^\dagger$}}\\
\textsc{storm}&\cellcolor{steel!30}\texttt{OtterTune}&\cellcolor{steel!30}\texttt{GP}&\cellcolor{steel!30}$\approx$ 0&\cellcolor{steel!30}\texttt{ConEx}&\cellcolor{steel!30}$\approx$ 0&\cellcolor{steel!30}{200.00\%$^{\,\,\,}$}&\cellcolor{steel!30}\textbf{\texttt{SMAC}}&\cellcolor{steel!30}\textbf{\texttt{DeP}}&\cellcolor{steel!30}\textbf{0.04$\pm$0.02}&\cellcolor{steel!30}\textbf{\texttt{ConEx}}&\cellcolor{steel!30}\textbf{0.36$\pm$0.30}&\cellcolor{steel!30}\textbf{{914.43\%$^\dagger$}}\\
\textsc{SaC}&\cellcolor{red!20}\textbf{\texttt{SMAC}}&\cellcolor{red!20}\textbf{\texttt{DCT}}&\cellcolor{red!20}\textbf{0.39$\pm$0.00}&\cellcolor{red!20}\textbf{\texttt{BestConfig}}&\cellcolor{red!20}\textbf{0.39$\pm$0.00}&\cellcolor{red!20}\textbf{{$-$0.68\%$^\star$}}&\cellcolor{red!20}\textbf{\texttt{ROBOTune}}&\cellcolor{red!20}\textbf{\texttt{SVR}}&\cellcolor{red!20}\textbf{0.03$\pm$0.02}&\cellcolor{red!20}\textbf{\texttt{BestConfig}}&\cellcolor{red!20}\textbf{0.02$\pm$0.06}&\cellcolor{red!20}\textbf{{$-$6.36\%$^\star$}}\\
\textsc{Hadoop}&\cellcolor{steel!30}\texttt{FLASH}&\cellcolor{steel!30}\texttt{GP}&\cellcolor{steel!30}1416.67$\pm$13.50&\cellcolor{steel!30}\texttt{SWAY}&\cellcolor{steel!30}1410.57$\pm$16.97&\cellcolor{steel!30}{0.43\%$^{\,\,\,}$}&\cellcolor{steel!30}\texttt{ATConf}&\cellcolor{steel!30}\texttt{SPL}&\cellcolor{steel!30}0.30$\pm$0.15&\cellcolor{steel!30}\texttt{ParamILS}&\cellcolor{steel!30}0.30$\pm$0.13&\cellcolor{steel!30}{2.19\%$^{\,\,\,}$}\\
\textsc{Tomcat}&\cellcolor{steel!30}\texttt{FLASH}&\cellcolor{steel!30}\texttt{DT}&\cellcolor{steel!30}1338.49$\pm$26.87&\cellcolor{steel!30}\texttt{R\_search}&\cellcolor{steel!30}1332.90$\pm$29.22&\cellcolor{steel!30}{0.42\%$^{\,\,\,}$}&\cellcolor{steel!30}\texttt{OtterTune}&\cellcolor{steel!30}\texttt{RF}&\cellcolor{steel!30}0.16$\pm$0.12&\cellcolor{steel!30}\texttt{GA}&\cellcolor{steel!30}0.35$\pm$0.37&\cellcolor{steel!30}{122.32\%$^{\,\,\,}$}\\
\textsc{JavaGC}&\cellcolor{steel!30}\textbf{\texttt{FLASH}}&\cellcolor{steel!30}\textbf{\texttt{DCT}}&\cellcolor{steel!30}\textbf{395.57$\pm$0.25}&\cellcolor{steel!30}\textbf{\texttt{GGA}}&\cellcolor{steel!30}\textbf{402.27$\pm$3.28}&\cellcolor{steel!30}\textbf{{1.69\%$^\dagger$}}&\cellcolor{steel!30}\textbf{\texttt{FLASH}}&\cellcolor{steel!30}\textbf{\texttt{RF}}&\cellcolor{steel!30}\textbf{0.10$\pm$0.06}&\cellcolor{steel!30}\textbf{\texttt{ParamILS}}&\cellcolor{steel!30}\textbf{0.19$\pm$0.15}&\cellcolor{steel!30}\textbf{{88.31\%$^\star$}}\\

\bottomrule
\end{tabular}

\end{adjustbox}
\end{table*}

%% file: Figures/rq1.1-summary-1.tex
\begin{tikzpicture}
\begin{axis}[
 ybar stacked,
  ymajorgrids=true,
   grid style = {dashed},
    width=6cm,
    height=6cm,
    ymax=29,
    ymin=0,
    xmin=1,
    xmax=5,
    bar width=1cm,
    	nodes near coords, 
 enlarge x limits=-0.05,
 axis x line  = bottom,
    axis y line  = left  ,
    legend style={at={(0.5,1.15)},draw=none,
      anchor=north,legend columns=3},
    ylabel={\# cases},
    legend cell align={left},
    symbolic x coords={1,2,3,4,5},
     xticklabels={Performance, Efficiency},
         xtick=data
    ]

\addplot[ybar, fill=steel!30, draw=black] coordinates {(2,16) (4,21)};

\addplot[ybar, fill=teal!30, draw=black] coordinates {(2,10) (4,6)};

\addplot[ybar, fill=red!30, draw=black] coordinates {(2,3) (4,2)};


\legend{win, tie, lose}
\end{axis}
\end{tikzpicture}

%% file: Figures/rq1.1-summary-2.tex
\begin{tikzpicture}
\begin{axis}[
 ybar stacked,
   ymajorgrids=true,
   grid style = {dashed},
    width=6cm,
    height=6cm,
    ymax=29,
    ymin=0,
    xmin=1,
    xmax=5,
    bar width=1cm,
    	nodes near coords, 
 axis x line  = bottom,
    axis y line  = left  ,
    legend style={at={(0.5,1.15)},draw=none,
      anchor=north,legend columns=3},
    ylabel={\# cases},
    legend cell align={left},
    symbolic x coords={1,2,3,4,5},
     xticklabels={Performance, Efficiency},
         xtick=data
    ]

\addplot[ybar, fill=steel!30, draw=black] coordinates {(2,0) (4,6)};

\addplot[ybar, fill=teal!30, draw=black] coordinates {(2,1) (4,22)};

\addplot[ybar, fill=red!30, draw=black] coordinates {(2,28) (4,1)};

\legend{win, tie, lose}
\end{axis}
\end{tikzpicture}

%% file: Tables/RQ4/MFO.tex
\begin{table*}[t!]
\caption{Comparing the best tuner-model pairs of batch model-based tuners and their best model-free counterparts over 30 runs. The format is the same as Table~\ref{tb:rq1.1}.}
\label{tb:rq1.2}
\centering
\footnotesize
\begin{adjustbox}{width=\textwidth,center}
\begin{tabular}{l|lll|ll|r|lll|ll|r}


\toprule

\multirow{2}{*}{\textbf{System}}&\multicolumn{6}{c|}{\textbf{Performance}}&\multicolumn{6}{c}{\textbf{Efficiency}}\\

\cmidrule{2-13}

 &\textbf{Batch}&\textbf{Model}&\textbf{Performance}&\textbf{Model-free}&\textbf{Performance}&\textbf{$\Delta$\%}&\textbf{Batch}&\textbf{Model}&\textbf{Efficiency}&\textbf{Model-free}&\textbf{Efficiency}&\textbf{$\Delta$\%}\\

\midrule

\textsc{Brotli}&\cellcolor{red!20}\textbf{\texttt{Brute-force}}&\cellcolor{red!20}\textbf{\texttt{SVR}}&\cellcolor{red!20}\textbf{1.55$\pm$0.05}&\cellcolor{red!20}\textbf{\texttt{BestConfig}}&\cellcolor{red!20}\textbf{1.47$\pm$0.01}&\cellcolor{red!20}\textbf{{$-$4.91\%$^\dagger$}}&\cellcolor{steel!30}\texttt{Brute-force}&\cellcolor{steel!30}\texttt{SVR}&\cellcolor{steel!30}0.19$\pm$0.00&\cellcolor{steel!30}\texttt{SWAY}&\cellcolor{steel!30}0.25$\pm$0.08&\cellcolor{steel!30}{35.16\%$^{\,\,\,}$}\\
\textsc{LLVM}&\cellcolor{red!20}\textbf{\texttt{Brute-force}}&\cellcolor{red!20}\textbf{\texttt{RF}}&\cellcolor{red!20}\textbf{203.54$\pm$2.88}&\cellcolor{red!20}\textbf{\texttt{ParamILS}}&\cellcolor{red!20}\textbf{199.92$\pm$0.25}&\cellcolor{red!20}\textbf{{$-$1.78\%$^\dagger$}}&\cellcolor{steel!30}\texttt{Brute-force}&\cellcolor{steel!30}\texttt{HIP}&\cellcolor{steel!30}0.04$\pm$0.00&\cellcolor{steel!30}\texttt{ParamILS}&\cellcolor{steel!30}0.43$\pm$0.33&\cellcolor{steel!30}{985.48\%$^{\,\,\,}$}\\
\textsc{Lrzip}&\cellcolor{red!20}\textbf{\texttt{Brute-force}}&\cellcolor{red!20}\textbf{\texttt{SVR}}&\cellcolor{red!20}\textbf{3.68$\pm$0.74}&\cellcolor{red!20}\textbf{\texttt{Random}}&\cellcolor{red!20}\textbf{3.14$\pm$0.02}&\cellcolor{red!20}\textbf{{$-$14.46\%$^\dagger$}}&\cellcolor{steel!30}\texttt{Brute-force}&\cellcolor{steel!30}\texttt{SVR}&\cellcolor{steel!30}0.05$\pm$0.00&\cellcolor{steel!30}\texttt{ConEx}&\cellcolor{steel!30}0.30$\pm$0.24&\cellcolor{steel!30}{460.09\%$^{\,\,\,}$}\\
\textsc{xgboost}&\cellcolor{red!20}\textbf{\texttt{Brute-force}}&\cellcolor{red!20}\textbf{\texttt{RF}}&\cellcolor{red!20}\textbf{33.61$\pm$15.08}&\cellcolor{red!20}\textbf{\texttt{ParamILS}}&\cellcolor{red!20}\textbf{14.87$\pm$6.84}&\cellcolor{red!20}\textbf{{$-$55.75\%$^\dagger$}}&\cellcolor{steel!30}\texttt{Brute-force}&\cellcolor{steel!30}\texttt{RF}&\cellcolor{steel!30}0.02$\pm$0.00&\cellcolor{steel!30}\texttt{SWAY}&\cellcolor{steel!30}0.41$\pm$0.32&\cellcolor{steel!30}{2167.55\%$^{\,\,\,}$}\\
\textsc{noc-CM-log}&\cellcolor{red!20}\textbf{\texttt{Brute-force}}&\cellcolor{red!20}\textbf{\texttt{SVR}}&\cellcolor{red!20}\textbf{4.36$\pm$0.04}&\cellcolor{red!20}\textbf{\texttt{ParamILS}}&\cellcolor{red!20}\textbf{4.31$\pm$0.00}&\cellcolor{red!20}\textbf{{$-$1.17\%$^\dagger$}}&\cellcolor{red!20}\texttt{Brute-force}&\cellcolor{red!20}\texttt{SVR}&\cellcolor{red!20}1.00$\pm$0.00&\cellcolor{red!20}\texttt{ParamILS}&\cellcolor{red!20}0.22$\pm$0.13&\cellcolor{red!20}{$-$77.74\%$^{\,\,\,}$}\\
\textsc{DeepArch}&\cellcolor{red!20}\textbf{\texttt{Brute-force}}&\cellcolor{red!20}\textbf{\texttt{RF}}&\cellcolor{red!20}\textbf{1.29$\pm$0.45}&\cellcolor{red!20}\textbf{\texttt{SWAY}}&\cellcolor{red!20}\textbf{1.09$\pm$0.12}&\cellcolor{red!20}\textbf{{$-$15.20\%$^\dagger$}}&\cellcolor{steel!30}\texttt{Brute-force}&\cellcolor{steel!30}\texttt{RF}&\cellcolor{steel!30}0.33$\pm$0.00&\cellcolor{steel!30}\texttt{SWAY}&\cellcolor{steel!30}0.41$\pm$0.31&\cellcolor{steel!30}{24.37\%$^{\,\,\,}$}\\
\textsc{BDB\_C}&\cellcolor{red!20}\textbf{\texttt{Brute-force}}&\cellcolor{red!20}\textbf{\texttt{RF}}&\cellcolor{red!20}\textbf{0.38$\pm$0.03}&\cellcolor{red!20}\textbf{\texttt{BestConfig}}&\cellcolor{red!20}\textbf{0.36$\pm$0.00}&\cellcolor{red!20}\textbf{{$-$5.75\%$^\dagger$}}&\cellcolor{steel!30}\texttt{Brute-force}&\cellcolor{steel!30}\texttt{HIP}&\cellcolor{steel!30}0.16$\pm$0.00&\cellcolor{steel!30}\texttt{BestConfig}&\cellcolor{steel!30}0.40$\pm$0.35&\cellcolor{steel!30}{150.33\%$^{\,\,\,}$}\\
\textsc{HSQLDB}&\cellcolor{red!20}\textbf{\texttt{Brute-force}}&\cellcolor{red!20}\textbf{\texttt{RF}}&\cellcolor{red!20}\textbf{248.80$\pm$0.43}&\cellcolor{red!20}\textbf{\texttt{ParamILS}}&\cellcolor{red!20}\textbf{248.29$\pm$0.10}&\cellcolor{red!20}\textbf{{$-$0.21\%$^\dagger$}}&\cellcolor{steel!30}\texttt{Brute-force}&\cellcolor{steel!30}\texttt{RF}&\cellcolor{steel!30}0.12$\pm$0.00&\cellcolor{steel!30}\texttt{BestConfig}&\cellcolor{steel!30}0.35$\pm$0.28&\cellcolor{steel!30}{182.97\%$^{\,\,\,}$}\\
\textsc{DConvert}&\cellcolor{red!20}\textbf{\texttt{GGA}}&\cellcolor{red!20}\textbf{\texttt{DT}}&\cellcolor{red!20}\textbf{1.89$\pm$0.06}&\cellcolor{red!20}\textbf{\texttt{GGA}}&\cellcolor{red!20}\textbf{1.80$\pm$0.02}&\cellcolor{red!20}\textbf{{$-$4.69\%$^\dagger$}}&\cellcolor{steel!30}\textbf{\texttt{BestConfig}}&\cellcolor{steel!30}\textbf{\texttt{DeP}}&\cellcolor{steel!30}\textbf{0.08$\pm$0.03}&\cellcolor{steel!30}\textbf{\texttt{ParamILS}}&\cellcolor{steel!30}\textbf{0.33$\pm$0.25}&\cellcolor{steel!30}\textbf{{314.72\%$^\dagger$}}\\
\textsc{7z}&\cellcolor{red!20}\textbf{\texttt{ConEx}}&\cellcolor{red!20}\textbf{\texttt{DaL}}&\cellcolor{red!20}\textbf{4502.71$\pm$174.88}&\cellcolor{red!20}\textbf{\texttt{GGA}}&\cellcolor{red!20}\textbf{4270.68$\pm$36.02}&\cellcolor{red!20}\textbf{{$-$5.15\%$^\dagger$}}&\cellcolor{red!20}\texttt{BestConfig}&\cellcolor{red!20}\texttt{RF}&\cellcolor{red!20}0.48$\pm$0.33&\cellcolor{red!20}\texttt{ParamILS}&\cellcolor{red!20}0.34$\pm$0.31&\cellcolor{red!20}{$-$28.05\%$^{\,\,\,}$}\\
\textsc{Apache}&\cellcolor{red!20}\textbf{\texttt{ParamILS}}&\cellcolor{red!20}\textbf{\texttt{GP}}&\cellcolor{red!20}\textbf{168.91$\pm$0.68}&\cellcolor{red!20}\textbf{\texttt{GA}}&\cellcolor{red!20}\textbf{170.80$\pm$0.11}&\cellcolor{red!20}\textbf{{$-$1.12\%$^\dagger$}}&\cellcolor{steel!30}\textbf{\texttt{ParamILS}}&\cellcolor{steel!30}\textbf{\texttt{LR}}&\cellcolor{steel!30}\textbf{0.20$\pm$0.10}&\cellcolor{steel!30}\textbf{\texttt{GGA}}&\cellcolor{steel!30}\textbf{0.36$\pm$0.22}&\cellcolor{steel!30}\textbf{{78.95\%$^\star$}}\\
\textsc{HSMGP}&\cellcolor{red!20}\textbf{\texttt{ConEx}}&\cellcolor{red!20}\textbf{\texttt{HIP}}&\cellcolor{red!20}\textbf{102.43$\pm$1.64}&\cellcolor{red!20}\textbf{\texttt{ConEx}}&\cellcolor{red!20}\textbf{100.47$\pm$0.20}&\cellcolor{red!20}\textbf{{$-$1.92\%$^\dagger$}}&\cellcolor{steel!30}\textbf{\texttt{ParamILS}}&\cellcolor{steel!30}\textbf{\texttt{HIP}}&\cellcolor{steel!30}\textbf{0.09$\pm$0.04}&\cellcolor{steel!30}\textbf{\texttt{ParamILS}}&\cellcolor{steel!30}\textbf{0.17$\pm$0.09}&\cellcolor{steel!30}\textbf{{95.64\%$^\star$}}\\
\textsc{MongoDB}&\cellcolor{red!20}\textbf{\texttt{BestConfig}}&\cellcolor{red!20}\textbf{\texttt{HIP}}&\cellcolor{red!20}\textbf{208786.93$\pm$1540.66}&\cellcolor{red!20}\textbf{\texttt{BestConfig}}&\cellcolor{red!20}\textbf{206561.38$\pm$386.20}&\cellcolor{red!20}\textbf{{$-$1.07\%$^\dagger$}}&\cellcolor{steel!30}\textbf{\texttt{ParamILS}}&\cellcolor{steel!30}\textbf{\texttt{LR}}&\cellcolor{steel!30}\textbf{0.20$\pm$0.15}&\cellcolor{steel!30}\textbf{\texttt{ConEx}}&\cellcolor{steel!30}\textbf{0.40$\pm$0.19}&\cellcolor{steel!30}\textbf{{99.51\%$^\dagger$}}\\
\textsc{PostgreSQL}&\cellcolor{red!20}\textbf{\texttt{GA}}&\cellcolor{red!20}\textbf{\texttt{RF}}&\cellcolor{red!20}\textbf{51025.81$\pm$233.26}&\cellcolor{red!20}\textbf{\texttt{ParamILS}}&\cellcolor{red!20}\textbf{50690.25$\pm$19.25}&\cellcolor{red!20}\textbf{{$-$0.66\%$^\dagger$}}&\cellcolor{red!20}\texttt{Random}&\cellcolor{red!20}\texttt{RF}&\cellcolor{red!20}0.27$\pm$0.26&\cellcolor{red!20}\texttt{ConEx}&\cellcolor{red!20}0.19$\pm$0.12&\cellcolor{red!20}{$-$29.34\%$^{\,\,\,}$}\\
\textsc{ExaStencils}&\cellcolor{red!20}\textbf{\texttt{ParamILS}}&\cellcolor{red!20}\textbf{\texttt{RF}}&\cellcolor{red!20}\textbf{4841.37$\pm$113.92}&\cellcolor{red!20}\textbf{\texttt{ParamILS}}&\cellcolor{red!20}\textbf{4671.02$\pm$99.33}&\cellcolor{red!20}\textbf{{$-$3.52\%$^\dagger$}}&\cellcolor{red!20}\textbf{\texttt{ParamILS}}&\cellcolor{red!20}\textbf{\texttt{DeP}}&\cellcolor{red!20}\textbf{0.65$\pm$0.33}&\cellcolor{red!20}\textbf{\texttt{ParamILS}}&\cellcolor{red!20}\textbf{0.30$\pm$0.22}&\cellcolor{red!20}\textbf{{$-$54.37\%$^\dagger$}}\\
\textsc{kanzi}&\cellcolor{red!20}\textbf{\texttt{ParamILS}}&\cellcolor{red!20}\textbf{\texttt{RF}}&\cellcolor{red!20}\textbf{4.79$\pm$1.26}&\cellcolor{red!20}\textbf{\texttt{ConEx}}&\cellcolor{red!20}\textbf{3.80$\pm$0.20}&\cellcolor{red!20}\textbf{{$-$20.61\%$^\dagger$}}&\cellcolor{steel!30}\texttt{ParamILS}&\cellcolor{steel!30}\texttt{RF}&\cellcolor{steel!30}0.24$\pm$0.21&\cellcolor{steel!30}\texttt{ParamILS}&\cellcolor{steel!30}0.27$\pm$0.18&\cellcolor{steel!30}{12.91\%$^{\,\,\,}$}\\
\textsc{jump3r}&\cellcolor{red!20}\texttt{ParamILS}&\cellcolor{red!20}\texttt{SVR}&\cellcolor{red!20}0.64$\pm$0.06&\cellcolor{red!20}\texttt{GGA}&\cellcolor{red!20}0.63$\pm$0.04&\cellcolor{red!20}{$-$2.12\%$^{\,\,\,}$}&\cellcolor{steel!30}\textbf{\texttt{ParamILS}}&\cellcolor{steel!30}\textbf{\texttt{SVR}}&\cellcolor{steel!30}\textbf{0.13$\pm$0.10}&\cellcolor{steel!30}\textbf{\texttt{ParamILS}}&\cellcolor{steel!30}\textbf{0.22$\pm$0.16}&\cellcolor{steel!30}\textbf{{71.75\%$^\star$}}\\
\textsc{MariaDB}&\cellcolor{red!20}\textbf{\texttt{GA}}&\cellcolor{red!20}\textbf{\texttt{RF}}&\cellcolor{red!20}\textbf{57.51$\pm$1.22}&\cellcolor{red!20}\textbf{\texttt{BestConfig}}&\cellcolor{red!20}\textbf{56.11$\pm$0.38}&\cellcolor{red!20}\textbf{{$-$2.43\%$^\dagger$}}&\cellcolor{steel!30}\texttt{ParamILS}&\cellcolor{steel!30}\texttt{LR}&\cellcolor{steel!30}0.21$\pm$0.16&\cellcolor{steel!30}\texttt{BestConfig}&\cellcolor{steel!30}0.24$\pm$0.25&\cellcolor{steel!30}{13.20\%$^{\,\,\,}$}\\
\textsc{polly}&\cellcolor{red!20}\textbf{\texttt{BestConfig}}&\cellcolor{red!20}\textbf{\texttt{DaL}}&\cellcolor{red!20}\textbf{4.76$\pm$0.60}&\cellcolor{red!20}\textbf{\texttt{BestConfig}}&\cellcolor{red!20}\textbf{4.26$\pm$0.02}&\cellcolor{red!20}\textbf{{$-$10.51\%$^\dagger$}}&\cellcolor{red!20}\texttt{BestConfig}&\cellcolor{red!20}\texttt{GP}&\cellcolor{red!20}0.19$\pm$0.26&\cellcolor{red!20}\texttt{BestConfig}&\cellcolor{red!20}0.17$\pm$0.20&\cellcolor{red!20}{$-$11.27\%$^{\,\,\,}$}\\
\textsc{SQL}&\cellcolor{red!20}\textbf{\texttt{GA}}&\cellcolor{red!20}\textbf{\texttt{DaL}}&\cellcolor{red!20}\textbf{13.11$\pm$0.30}&\cellcolor{red!20}\textbf{\texttt{BestConfig}}&\cellcolor{red!20}\textbf{12.51$\pm$0.00}&\cellcolor{red!20}\textbf{{$-$4.58\%$^\dagger$}}&\cellcolor{steel!30}\texttt{BestConfig}&\cellcolor{steel!30}\texttt{DeP}&\cellcolor{steel!30}0.04$\pm$0.04&\cellcolor{steel!30}\texttt{BestConfig}&\cellcolor{steel!30}0.05$\pm$0.07&\cellcolor{steel!30}{33.12\%$^{\,\,\,}$}\\
\textsc{vp9}&\cellcolor{red!20}\textbf{\texttt{BestConfig}}&\cellcolor{red!20}\textbf{\texttt{RF}}&\cellcolor{red!20}\textbf{42.39$\pm$1.18}&\cellcolor{red!20}\textbf{\texttt{BestConfig}}&\cellcolor{red!20}\textbf{41.28$\pm$0.13}&\cellcolor{red!20}\textbf{{$-$2.61\%$^\dagger$}}&\cellcolor{steel!30}\texttt{BestConfig}&\cellcolor{steel!30}\texttt{DeP}&\cellcolor{steel!30}0.05$\pm$0.02&\cellcolor{steel!30}\texttt{BestConfig}&\cellcolor{steel!30}0.14$\pm$0.19&\cellcolor{steel!30}{205.60\%$^{\,\,\,}$}\\
\textsc{Spark}&\cellcolor{red!20}\textbf{\texttt{GA}}&\cellcolor{red!20}\textbf{\texttt{SPL}}&\cellcolor{red!20}\textbf{3095.87$\pm$99.26}&\cellcolor{red!20}\textbf{\texttt{BestConfig}}&\cellcolor{red!20}\textbf{3355.47$\pm$52.27}&\cellcolor{red!20}\textbf{{$-$8.39\%$^\dagger$}}&\cellcolor{steel!30}\texttt{ParamILS}&\cellcolor{steel!30}\texttt{LR}&\cellcolor{steel!30}0.12$\pm$0.05&\cellcolor{steel!30}\texttt{ParamILS}&\cellcolor{steel!30}0.13$\pm$0.07&\cellcolor{steel!30}{11.30\%$^{\,\,\,}$}\\
\textsc{HIPAcc}&\cellcolor{red!20}\textbf{\texttt{ParamILS}}&\cellcolor{red!20}\textbf{\texttt{RF}}&\cellcolor{red!20}\textbf{21.24$\pm$0.08}&\cellcolor{red!20}\textbf{\texttt{BestConfig}}&\cellcolor{red!20}\textbf{21.17$\pm$0.00}&\cellcolor{red!20}\textbf{{$-$0.31\%$^\dagger$}}&\cellcolor{red!20}\texttt{BestConfig}&\cellcolor{red!20}\texttt{DeP}&\cellcolor{red!20}0.10$\pm$0.12&\cellcolor{red!20}\texttt{ParamILS}&\cellcolor{red!20}0.10$\pm$0.05&\cellcolor{red!20}{$-$2.49\%$^{\,\,\,}$}\\
\textsc{Redis}&\cellcolor{red!20}\textbf{\texttt{Random}}&\cellcolor{red!20}\textbf{\texttt{LR}}&\cellcolor{red!20}\textbf{73750.87$\pm$4036.21}&\cellcolor{red!20}\textbf{\texttt{Random}}&\cellcolor{red!20}\textbf{85716.19$\pm$1162.73}&\cellcolor{red!20}\textbf{{$-$16.22\%$^\dagger$}}&\cellcolor{steel!30}\texttt{ParamILS}&\cellcolor{steel!30}\texttt{HIP}&\cellcolor{steel!30}0.26$\pm$0.10&\cellcolor{steel!30}\texttt{ConEx}&\cellcolor{steel!30}0.28$\pm$0.24&\cellcolor{steel!30}{8.22\%$^{\,\,\,}$}\\
\textsc{storm}&\cellcolor{red!20}\textbf{\texttt{ConEx}}&\cellcolor{red!20}\textbf{\texttt{RF}}&\cellcolor{red!20}\textbf{$\approx$ 0}&\cellcolor{red!20}\textbf{\texttt{BestConfig}}&\cellcolor{red!20}\textbf{$\approx$ 0}&\cellcolor{red!20}\textbf{{$-$200.00\%$^\dagger$}}&\cellcolor{steel!30}\textbf{\texttt{BestConfig}}&\cellcolor{steel!30}\textbf{\texttt{DeP}}&\cellcolor{steel!30}\textbf{0.11$\pm$0.03}&\cellcolor{steel!30}\textbf{\texttt{BestConfig}}&\cellcolor{steel!30}\textbf{0.29$\pm$0.23}&\cellcolor{steel!30}\textbf{{155.03\%$^\dagger$}}\\
\textsc{SaC}&\cellcolor{red!20}\textbf{\texttt{SWAY}}&\cellcolor{red!20}\textbf{\texttt{DT}}&\cellcolor{red!20}\textbf{0.41$\pm$0.02}&\cellcolor{red!20}\textbf{\texttt{BestConfig}}&\cellcolor{red!20}\textbf{0.39$\pm$0.00}&\cellcolor{red!20}\textbf{{$-$3.94\%$^\dagger$}}&\cellcolor{steel!30}\texttt{BestConfig}&\cellcolor{steel!30}\texttt{SPL}&\cellcolor{steel!30}0.01$\pm$0.03&\cellcolor{steel!30}\texttt{BestConfig}&\cellcolor{steel!30}0.02$\pm$0.05&\cellcolor{steel!30}{32.64\%$^{\,\,\,}$}\\
\textsc{Hadoop}&\cellcolor{red!20}\textbf{\texttt{irace}}&\cellcolor{red!20}\textbf{\texttt{HIP}}&\cellcolor{red!20}\textbf{1313.47$\pm$39.33}&\cellcolor{red!20}\textbf{\texttt{SWAY}}&\cellcolor{red!20}\textbf{1405.13$\pm$18.03}&\cellcolor{red!20}\textbf{{$-$6.98\%$^\dagger$}}&\cellcolor{steel!30}\texttt{ParamILS}&\cellcolor{steel!30}\texttt{DaL}&\cellcolor{steel!30}0.26$\pm$0.16&\cellcolor{steel!30}\texttt{ParamILS}&\cellcolor{steel!30}0.34$\pm$0.18&\cellcolor{steel!30}{33.20\%$^{\,\,\,}$}\\
\textsc{Tomcat}&\cellcolor{red!20}\textbf{\texttt{BestConfig}}&\cellcolor{red!20}\textbf{\texttt{SVR}}&\cellcolor{red!20}\textbf{1164.20$\pm$56.09}&\cellcolor{red!20}\textbf{\texttt{Random}}&\cellcolor{red!20}\textbf{1332.96$\pm$29.15}&\cellcolor{red!20}\textbf{{$-$14.50\%$^\dagger$}}&\cellcolor{steel!30}\texttt{ParamILS}&\cellcolor{steel!30}\texttt{GP}&\cellcolor{steel!30}0.22$\pm$0.07&\cellcolor{steel!30}\texttt{GA}&\cellcolor{steel!30}0.34$\pm$0.36&\cellcolor{steel!30}{58.10\%$^{\,\,\,}$}\\
\textsc{JavaGC}&\cellcolor{red!20}\textbf{\texttt{ParamILS}}&\cellcolor{red!20}\textbf{\texttt{RF}}&\cellcolor{red!20}\textbf{407.33$\pm$5.65}&\cellcolor{red!20}\textbf{\texttt{GGA}}&\cellcolor{red!20}\textbf{401.15$\pm$2.18}&\cellcolor{red!20}\textbf{{$-$1.52\%$^\dagger$}}&\cellcolor{red!20}\texttt{ParamILS}&\cellcolor{red!20}\texttt{HIP}&\cellcolor{red!20}0.12$\pm$0.09&\cellcolor{red!20}\texttt{ParamILS}&\cellcolor{red!20}0.11$\pm$0.06&\cellcolor{red!20}{$-$14.26\%$^{\,\,\,}$}\\

\bottomrule
\end{tabular}

\end{adjustbox}
\end{table*}

%% file: Figures/rq1.2-summary-1.tex
    \begin{tikzpicture}
      \pie[rotate=120, font=\large,text=inside,color={red!30, teal!30, steel!30},explode = 0.1, sum = auto, scale=1]
      {106/lose, 116/tie, 10/win}
    \end{tikzpicture}

%% file: Figures/rq1.2-summary-2.tex
    \begin{tikzpicture}
      \pie[rotate=120, font=\large,text=inside,color={red!30, teal!30, steel!30},explode = 0.1, sum = auto, scale=1]
      {74/lose, 153/tie, 5/win}
    \end{tikzpicture}

%% file: Tables/RQ5/SMBO1.tex
\begin{table*}[t!]
\caption{Comparing the original models and the best new models in terms of performance for sequential model-based tuners over 30 runs. $\Delta$\% refers to the ratio of ${{f-b} \over f} \times 100$ (or ${{b-f} \over f} \times 100$ if the quality metric is to be maximized) such that $b$ and $f$ are the values of the tuning quality metric when using the best new model and the original model, respectively. In $\mathcal{N}/\mathcal{O}$, $\mathcal{N}$ and $\mathcal{O}$ denote the best new model and the original model, respectively. \colorbox{steel!30}{Blue cells} indicate that the original model wins while \colorbox{red!20}{red cells} mean the original model loses. Statistically significant comparisons are highlighted in \textbf{bold} and the best model for a system is \underline{underlined}. Other formats are the same as Table~\ref{tb:rq1.1}.} 

\label{tb:rq2.1}
\setlength{\tabcolsep}{1mm}
\centering 
\footnotesize
\begin{adjustbox}{width=\textwidth,center}
\begin{tabular}{llr|lr|lr|lr|lr|lr|lr|lr}


\toprule

\multirow{2}{*}{\textbf{System}}&\multicolumn{2}{c|}{\textbf{\texttt{BOCA}}}&\multicolumn{2}{c|}{\textbf{\texttt{ATConf}}}&\multicolumn{2}{c|}{\textbf{\texttt{FLASH}}}&\multicolumn{2}{c|}{\textbf{\texttt{OtterTune}}}&\multicolumn{2}{c|}{\textbf{\texttt{ResTune}}}&\multicolumn{2}{c|}{\textbf{\texttt{ROBOTune}}}&\multicolumn{2}{c|}{\textbf{\texttt{SMAC}}}&\multicolumn{2}{c}{\textbf{\texttt{Tuneful}}}\\

&\textbf{Model} ($\mathcal{N}/\mathcal{O}$)&\textbf{$\Delta$\%}&\textbf{Model} ($\mathcal{N}/\mathcal{O}$)&\textbf{$\Delta$\%}&\textbf{Model} ($\mathcal{N}/\mathcal{O}$)&\textbf{$\Delta$\%}&\textbf{Model} ($\mathcal{N}/\mathcal{O}$)&\textbf{$\Delta$\%}&\textbf{Model} ($\mathcal{N}/\mathcal{O}$)&\textbf{$\Delta$\%}&\textbf{Model} ($\mathcal{N}/\mathcal{O}$)&\textbf{$\Delta$\%}&\textbf{Model} ($\mathcal{N}/\mathcal{O}$)&\textbf{$\Delta$\%}&\textbf{Model} ($\mathcal{N}/\mathcal{O}$)&\textbf{$\Delta$\%}\\

\midrule


\textsc{Brotli}&\cellcolor{red!20}\texttt{DT/RF}&\cellcolor{red!20}{$-$0.42\%$^{\,\,\,}$}&\cellcolor{steel!30}\textbf{\texttt{HIP/GP}}&\cellcolor{steel!30}\textbf{{1.54\%$^\dagger$}}&\cellcolor{red!20}\underline{\textbf{\texttt{DeP}}}\textbf{\texttt{/DT}}&\cellcolor{red!20}\textbf{{$-$6.43\%$^\dagger$}}&\cellcolor{red!20}\textbf{\texttt{DCT/GP}}&\cellcolor{red!20}\textbf{{$-$0.49\%$^\dagger$}}&\cellcolor{red!20}\textbf{\texttt{DaL/GP}}&\cellcolor{red!20}\textbf{{$-$0.69\%$^\dagger$}}&\cellcolor{red!20}\textbf{\texttt{RF/GP}}&\cellcolor{red!20}\textbf{{$-$14.34\%$^\dagger$}}&\cellcolor{steel!30}\texttt{HIP/RF}&\cellcolor{steel!30}{0.32\%$^{\,\,\,}$}&\cellcolor{red!20}\textbf{\texttt{DaL/GP}}&\cellcolor{red!20}\textbf{{$-$0.66\%$^\star$}}\\
\bottomrule
\end{tabular}
\end{adjustbox}

\end{table*}

%% file: Tables/RQ5/SMBO2.tex
\begin{table*}[t!]
\caption{Comparing the original models and the best other models in terms of efficiency for sequential model-based tuners over 30 runs. The format and abbreviations are the same as Table~\ref{tb:rq2.1}.} 
\label{tb:rq2.2}
\setlength{\tabcolsep}{1mm}
\centering 
\footnotesize

\begin{adjustbox}{width=\textwidth,center}
\begin{tabular}{llr|lr|lr|lr|lr|lr|lr|lr}


\toprule

\multirow{2}{*}{\textbf{System}}&\multicolumn{2}{c|}{\textbf{\texttt{BOCA}}}&\multicolumn{2}{c|}{\textbf{\texttt{ATConf}}}&\multicolumn{2}{c|}{\textbf{\texttt{FLASH}}}&\multicolumn{2}{c|}{\textbf{\texttt{OtterTune}}}&\multicolumn{2}{c|}{\textbf{\texttt{ResTune}}}&\multicolumn{2}{c|}{\textbf{\texttt{ROBOTune}}}&\multicolumn{2}{c|}{\textbf{\texttt{SMAC}}}&\multicolumn{2}{c}{\textbf{\texttt{Tuneful}}}\\

&\textbf{Model} ($\mathcal{N}/\mathcal{O}$)&\textbf{$\Delta$\%}&\textbf{Model} ($\mathcal{N}/\mathcal{O}$)&\textbf{$\Delta$\%}&\textbf{Model} ($\mathcal{N}/\mathcal{O}$)&\textbf{$\Delta$\%}&\textbf{Model} ($\mathcal{N}/\mathcal{O}$)&\textbf{$\Delta$\%}&\textbf{Model} ($\mathcal{N}/\mathcal{O}$)&\textbf{$\Delta$\%}&\textbf{Model} ($\mathcal{N}/\mathcal{O}$)&\textbf{$\Delta$\%}&\textbf{Model} ($\mathcal{N}/\mathcal{O}$)&\textbf{$\Delta$\%}&\textbf{Model} ($\mathcal{N}/\mathcal{O}$)&\textbf{$\Delta$\%}\\

\midrule


\textsc{Brotli}&\cellcolor{red!20}\textbf{\texttt{GP/RF}}&\cellcolor{red!20}\textbf{{$-$212.84\%$^\dagger$}}&\cellcolor{red!20}\textbf{\texttt{LR/GP}}&\cellcolor{red!20}\textbf{{$-$54.20\%$^\star$}}&\cellcolor{red!20}\textbf{\texttt{LR/DT}}&\cellcolor{red!20}\textbf{{$-$119.01\%$^\dagger$}}&\cellcolor{red!20}\texttt{SVR/GP}&\cellcolor{red!20}{$-$10.68\%$^{\,\,\,}$}&\cellcolor{red!20}\textbf{\texttt{LR/GP}}&\cellcolor{red!20}\textbf{{$-$85.16\%$^\star$}}&\cellcolor{red!20}\textbf{\texttt{DaL/GP}}&\cellcolor{red!20}\textbf{{$-$285.96\%$^\dagger$}}&\cellcolor{red!20}\underline{\textbf{\texttt{GP}}}\textbf{\texttt{/RF}}&\cellcolor{red!20}\textbf{{$-$103.42\%$^\dagger$}}&\cellcolor{red!20}\texttt{SPL/GP}&\cellcolor{red!20}{$-$61.48\%$^{\,\,\,}$}\\
\bottomrule
\end{tabular}
\end{adjustbox}

\end{table*}

%% file: Figures/rq3-summary-smbt.tex
\begin{tikzpicture}
\begin{axis}[
 ybar stacked,
 ymajorgrids=true,
   grid style = {dashed},
    width=7.5cm,
    height=6cm,
    ymax=240,
    ymin=0,
    xmin=1,
    xmax=9,
    bar width=0.55cm,
    	nodes near coords, 
 enlarge x limits=-0.05,
 axis x line  = bottom,
    axis y line  = left  ,
    legend style={draw=none,at={(0.5,1.1)},
      anchor=north,legend columns=2},
    ylabel={\# cases},
    legend cell align={left},
    symbolic x coords={1,2,3,4,5,6,7,8,9,10},
      xticklabels={M\&P, R\&P, M\&E, R\&E},
     xticklabel style={yshift=-0.5cm},
     bar shift=-8.5pt,
         xtick=data
    ]

\addplot[ybar, fill=steel!30, draw=black] coordinates {(2,36) (4,42) (6,26) (8,22)};


\addplot[ybar, fill=red!30, draw=black] coordinates {(2,196) (4,190) (6,206) (8,210)};

\addlegendentry{Matched}
\addlegendentry{Mismatched}

\end{axis}

\node at (0.4,-0.3) {$\mathcal{B}$};
\node at (1,-0.3) {$\mathcal{W}$};

\node at (1.9,-0.3) {$\mathcal{B}$};
\node at (2.5,-0.3) {$\mathcal{W}$};

\node at (3.4,-0.3) {$\mathcal{B}$};
\node at (4,-0.3) {$\mathcal{W}$};

\node at (4.9,-0.3) {$\mathcal{B}$};
\node at (5.4,-0.3) {$\mathcal{W}$};

\begin{axis}[ybar stacked,
    width=7.5cm,
    height=6cm,
    ymax=240,
    ymin=0,
    xmin=1,
    xmax=9,
    bar width=0.55cm,
    	nodes near coords, 
 enlarge x limits=-0.05,
 axis x line  = bottom,
    axis y line  = left  ,
    legend style={at={(1.3,-0.07)},
      anchor=north,legend columns=1},
    ylabel={\# cases},
    legend style={draw=none,at={(1.35,1)}},
    legend cell align={left},
    symbolic x coords={1,2,3,4,5,6,7,8,9,10},
      xticklabels={M\&P, R\&P, M\&E, R\&E},
         xtick=data,
         bar shift=8.5pt,
         hide axis]
\addplot[ybar, fill=steel!30, draw=black] coordinates {(2,38) (4,53) (6,37) (8,60)};

\addplot[ybar, fill=red!30, draw=black] coordinates {(2,194) (4,179) (6,195) (8,172)};
\end{axis}
\end{tikzpicture}

%% file: Figures/rq3-summary-bmbt.tex
\begin{tikzpicture}
\begin{axis}[
 ybar stacked,
 ymajorgrids=true,
   grid style = {dashed},
    width=7.5cm,
    height=6cm,
    ymax=190,
    ymin=0,
    xmin=1,
    xmax=9,
    bar width=0.55cm,
    	nodes near coords, 
 enlarge x limits=-0.05,
 axis x line  = bottom,
    axis y line  = left  ,
    legend style={draw=none,at={(0.5,1.1)},
      anchor=north,legend columns=2},
    ylabel={\# cases},
    legend cell align={left},
    symbolic x coords={1,2,3,4,5,6,7,8,9,10},
     xticklabels={M\&P, R\&P, M\&E, R\&E},
       xticklabel style={yshift=-0.5cm},
     bar shift=-8.5pt,
         xtick=data
    ]

\addplot[ybar, fill=steel!30, draw=black] coordinates {(2,72) (4,82) (6,28) (8,34)};
\addplot[ybar, fill=red!30, draw=black] coordinates {(2,104) (4,94) (6,148) (8,142)};

\addlegendentry{Matched}
\addlegendentry{Mismatched}

\end{axis}

\node at (0.4,-0.3) {$\mathcal{B}$};
\node at (1,-0.3) {$\mathcal{W}$};

\node at (1.9,-0.3) {$\mathcal{B}$};
\node at (2.5,-0.3) {$\mathcal{W}$};

\node at (3.4,-0.3) {$\mathcal{B}$};
\node at (4,-0.3) {$\mathcal{W}$};

\node at (4.9,-0.3) {$\mathcal{B}$};
\node at (5.4,-0.3) {$\mathcal{W}$};

\begin{axis}[ybar stacked,
    width=7.5cm,
    height=6cm,
    ymax=190,
    ymin=0,
    xmin=1,
    xmax=9,
    bar width=0.55cm,
    	nodes near coords, 
 enlarge x limits=-0.05,
 axis x line  = bottom,
    axis y line  = left  ,
    legend style={at={(1.3,-0.07)},
      anchor=north,legend columns=1},
    ylabel={\# cases},
    legend style={draw=none,at={(1.35,1)}},
    legend cell align={left},
    symbolic x coords={1,2,3,4,5,6,7,8,9,10},
      xticklabels={M\&P, R\&P, M\&E, R\&E},
         xtick=data,
         bar shift=8.5pt,
         hide axis]
\addplot[ybar, fill=steel!30, draw=black] coordinates {(2,47) (4,56) (6,30) (8,40)};

\addplot[ybar, fill=red!30, draw=black] coordinates {(2,129) (4,120) (6,146) (8,136)};
\end{axis}
\end{tikzpicture}

%% file: Tables/RQ2/rq2_smbt.tex
\begin{table*}[t!]
\caption{The consistency between the best (worst) model, chosen according to the accuracy, and the model that leads to the best (worst) tuning quality under sequential model-based tuners with 30 repeats. \cmark~and \xmark~denote a match and mismatch, respectively. The formats and abbreviations are the same as Figure~\ref{fig:rq3-sum}. }

\label{tb:rq3.1}
\setlength{\tabcolsep}{0.5mm}
\centering
\footnotesize
\begin{adjustbox}{width=\textwidth,center}
\begin{tabular}{lcccccccc|cccccccc|cccccccc|cccccccc|cccccccc|cccccccc|cccccccc|cccccccc}


\toprule

\multirow{3}{*}{\textbf{System}}&\multicolumn{8}{c|}{\textbf{\texttt{BOCA}}}&\multicolumn{8}{c|}{\textbf{\texttt{ATConf}}}&\multicolumn{8}{c|}{\textbf{\texttt{FLASH}}}&\multicolumn{8}{c|}{\textbf{\texttt{OtterTune}}}&\multicolumn{8}{c|}{\textbf{\texttt{ResTune}}}&\multicolumn{8}{c|}{\textbf{\texttt{ROBOTune}}}&\multicolumn{8}{c|}{\textbf{\texttt{SMAC}}}&\multicolumn{8}{c}{\textbf{\texttt{Tuneful}}}\\

\cmidrule(l){2-65}

&\multicolumn{2}{c}{\textbf{M\&P}}&\multicolumn{2}{c}{\textbf{R\&P}}&\multicolumn{2}{c}{\textbf{M\&E}}&\multicolumn{2}{c|}{\textbf{R\&E}}&\multicolumn{2}{c}{\textbf{M\&P}}&\multicolumn{2}{c}{\textbf{R\&P}}&\multicolumn{2}{c}{\textbf{M\&E}}&\multicolumn{2}{c|}{\textbf{R\&E}}&\multicolumn{2}{c}{\textbf{M\&P}}&\multicolumn{2}{c}{\textbf{R\&P}}&\multicolumn{2}{c}{\textbf{M\&E}}&\multicolumn{2}{c|}{\textbf{R\&E}}&\multicolumn{2}{c}{\textbf{M\&P}}&\multicolumn{2}{c}{\textbf{R\&P}}&\multicolumn{2}{c}{\textbf{M\&E}}&\multicolumn{2}{c|}{\textbf{R\&E}}&\multicolumn{2}{c}{\textbf{M\&P}}&\multicolumn{2}{c}{\textbf{R\&P}}&\multicolumn{2}{c}{\textbf{M\&E}}&\multicolumn{2}{c|}{\textbf{R\&E}}&\multicolumn{2}{c}{\textbf{M\&P}}&\multicolumn{2}{c}{\textbf{R\&P}}&\multicolumn{2}{c}{\textbf{M\&E}}&\multicolumn{2}{c|}{\textbf{R\&E}}&\multicolumn{2}{c}{\textbf{M\&P}}&\multicolumn{2}{c}{\textbf{R\&P}}&\multicolumn{2}{c}{\textbf{M\&E}}&\multicolumn{2}{c|}{\textbf{R\&E}}&\multicolumn{2}{c}{\textbf{M\&P}}&\multicolumn{2}{c}{\textbf{R\&P}}&\multicolumn{2}{c}{\textbf{M\&E}}&\multicolumn{2}{c}{\textbf{R\&E}}\\

&$\mathcal{B}$&$\mathcal{W}$&$\mathcal{B}$&$\mathcal{W}$&$\mathcal{B}$&$\mathcal{W}$&$\mathcal{B}$&$\mathcal{W}$&$\mathcal{B}$&$\mathcal{W}$&$\mathcal{B}$&$\mathcal{W}$&$\mathcal{B}$&$\mathcal{W}$&$\mathcal{B}$&$\mathcal{W}$&$\mathcal{B}$&$\mathcal{W}$&$\mathcal{B}$&$\mathcal{W}$&$\mathcal{B}$&$\mathcal{W}$&$\mathcal{B}$&$\mathcal{W}$&$\mathcal{B}$&$\mathcal{W}$&$\mathcal{B}$&$\mathcal{W}$&$\mathcal{B}$&$\mathcal{W}$&$\mathcal{B}$&$\mathcal{W}$&$\mathcal{B}$&$\mathcal{W}$&$\mathcal{B}$&$\mathcal{W}$&$\mathcal{B}$&$\mathcal{W}$&$\mathcal{B}$&$\mathcal{W}$&$\mathcal{B}$&$\mathcal{W}$&$\mathcal{B}$&$\mathcal{W}$&$\mathcal{B}$&$\mathcal{W}$&$\mathcal{B}$&$\mathcal{W}$&$\mathcal{B}$&$\mathcal{W}$&$\mathcal{B}$&$\mathcal{W}$&$\mathcal{B}$&$\mathcal{W}$&$\mathcal{B}$&$\mathcal{W}$&$\mathcal{B}$&$\mathcal{W}$&$\mathcal{B}$&$\mathcal{W}$&$\mathcal{B}$&$\mathcal{W}$&$\mathcal{B}$&$\mathcal{W}$\\

\midrule

\textsc{Brotli}&\xmark&\xmark&\xmark&\xmark&\xmark&\xmark&\xmark&\cmark&\xmark&\xmark&\xmark&\xmark&\xmark&\xmark&\xmark&\xmark&\xmark&\xmark&\xmark&\xmark&\xmark&\xmark&\xmark&\xmark&\xmark&\xmark&\xmark&\xmark&\xmark&\xmark&\xmark&\xmark&\xmark&\xmark&\xmark&\xmark&\xmark&\xmark&\xmark&\xmark&\cmark&\xmark&\cmark&\xmark&\xmark&\xmark&\xmark&\xmark&\xmark&\xmark&\cmark&\xmark&\xmark&\xmark&\xmark&\cmark&\xmark&\xmark&\cmark&\xmark&\xmark&\xmark&\xmark&\xmark\\
\textsc{LLVM}&\xmark&\xmark&\xmark&\cmark&\xmark&\xmark&\xmark&\xmark&\xmark&\xmark&\xmark&\xmark&\xmark&\xmark&\xmark&\xmark&\xmark&\xmark&\xmark&\xmark&\cmark&\xmark&\xmark&\xmark&\xmark&\xmark&\xmark&\cmark&\cmark&\xmark&\cmark&\cmark&\xmark&\xmark&\xmark&\xmark&\xmark&\xmark&\xmark&\xmark&\xmark&\xmark&\xmark&\xmark&\xmark&\xmark&\xmark&\xmark&\xmark&\cmark&\xmark&\xmark&\xmark&\xmark&\xmark&\xmark&\xmark&\xmark&\xmark&\xmark&\xmark&\xmark&\xmark&\xmark\\
\textsc{Lrzip}&\xmark&\xmark&\xmark&\cmark&\xmark&\cmark&\xmark&\xmark&\cmark&\xmark&\xmark&\cmark&\cmark&\xmark&\xmark&\xmark&\cmark&\xmark&\xmark&\xmark&\xmark&\xmark&\xmark&\xmark&\cmark&\xmark&\cmark&\cmark&\xmark&\xmark&\xmark&\xmark&\xmark&\xmark&\xmark&\xmark&\xmark&\xmark&\xmark&\xmark&\xmark&\xmark&\xmark&\xmark&\xmark&\xmark&\xmark&\xmark&\xmark&\xmark&\xmark&\cmark&\xmark&\xmark&\xmark&\xmark&\xmark&\xmark&\xmark&\xmark&\xmark&\xmark&\xmark&\xmark\\
\textsc{xgboost}&\xmark&\xmark&\cmark&\xmark&\xmark&\xmark&\cmark&\xmark&\xmark&\xmark&\xmark&\xmark&\xmark&\xmark&\xmark&\xmark&\xmark&\xmark&\xmark&\xmark&\xmark&\xmark&\xmark&\xmark&\xmark&\cmark&\xmark&\xmark&\xmark&\xmark&\xmark&\cmark&\xmark&\xmark&\xmark&\cmark&\xmark&\xmark&\xmark&\xmark&\cmark&\xmark&\xmark&\xmark&\xmark&\xmark&\xmark&\xmark&\cmark&\xmark&\xmark&\xmark&\xmark&\cmark&\xmark&\cmark&\xmark&\xmark&\xmark&\xmark&\xmark&\cmark&\xmark&\xmark\\
\textsc{noc-CM-log}&\xmark&\xmark&\cmark&\xmark&\xmark&\xmark&\xmark&\xmark&\cmark&\xmark&\cmark&\cmark&\xmark&\xmark&\xmark&\cmark&\xmark&\xmark&\xmark&\xmark&\xmark&\xmark&\xmark&\xmark&\xmark&\xmark&\cmark&\xmark&\xmark&\xmark&\xmark&\xmark&\xmark&\xmark&\xmark&\xmark&\xmark&\xmark&\xmark&\xmark&\xmark&\xmark&\xmark&\xmark&\xmark&\xmark&\xmark&\xmark&\xmark&\xmark&\xmark&\xmark&\xmark&\xmark&\xmark&\xmark&\xmark&\xmark&\xmark&\xmark&\xmark&\xmark&\xmark&\xmark\\
\textsc{DeepArch}&\xmark&\cmark&\xmark&\cmark&\xmark&\xmark&\xmark&\xmark&\xmark&\xmark&\xmark&\xmark&\xmark&\xmark&\xmark&\xmark&\xmark&\xmark&\xmark&\xmark&\xmark&\xmark&\xmark&\xmark&\xmark&\xmark&\xmark&\xmark&\xmark&\xmark&\xmark&\xmark&\cmark&\xmark&\xmark&\xmark&\xmark&\xmark&\xmark&\xmark&\xmark&\xmark&\xmark&\xmark&\xmark&\xmark&\xmark&\xmark&\cmark&\xmark&\cmark&\xmark&\xmark&\cmark&\xmark&\cmark&\xmark&\xmark&\xmark&\xmark&\xmark&\xmark&\xmark&\xmark\\
\textsc{BDB\_C}&\xmark&\xmark&\xmark&\xmark&\xmark&\xmark&\xmark&\xmark&\xmark&\xmark&\xmark&\xmark&\xmark&\xmark&\xmark&\xmark&\xmark&\xmark&\xmark&\cmark&\xmark&\xmark&\xmark&\cmark&\xmark&\xmark&\xmark&\xmark&\xmark&\xmark&\xmark&\xmark&\xmark&\xmark&\xmark&\cmark&\xmark&\xmark&\xmark&\xmark&\xmark&\xmark&\cmark&\xmark&\xmark&\xmark&\xmark&\xmark&\xmark&\xmark&\xmark&\xmark&\xmark&\xmark&\xmark&\xmark&\xmark&\xmark&\xmark&\xmark&\xmark&\xmark&\xmark&\xmark\\
\textsc{HSQLDB}&\xmark&\xmark&\xmark&\xmark&\xmark&\xmark&\xmark&\xmark&\xmark&\xmark&\xmark&\cmark&\xmark&\xmark&\xmark&\cmark&\xmark&\cmark&\xmark&\cmark&\xmark&\cmark&\xmark&\cmark&\xmark&\cmark&\xmark&\cmark&\xmark&\cmark&\xmark&\cmark&\xmark&\cmark&\xmark&\cmark&\xmark&\xmark&\xmark&\xmark&\cmark&\xmark&\cmark&\xmark&\xmark&\xmark&\xmark&\xmark&\cmark&\cmark&\xmark&\cmark&\xmark&\xmark&\xmark&\xmark&\xmark&\cmark&\xmark&\cmark&\xmark&\xmark&\xmark&\xmark\\
\textsc{DConvert}&\xmark&\xmark&\xmark&\xmark&\xmark&\cmark&\xmark&\cmark&\xmark&\cmark&\xmark&\xmark&\xmark&\xmark&\xmark&\xmark&\xmark&\cmark&\xmark&\cmark&\cmark&\cmark&\cmark&\cmark&\xmark&\cmark&\xmark&\cmark&\xmark&\cmark&\xmark&\cmark&\xmark&\xmark&\xmark&\xmark&\xmark&\xmark&\xmark&\xmark&\xmark&\xmark&\xmark&\xmark&\xmark&\xmark&\xmark&\xmark&\xmark&\cmark&\cmark&\cmark&\xmark&\cmark&\xmark&\cmark&\xmark&\cmark&\xmark&\cmark&\xmark&\xmark&\xmark&\xmark\\
\textsc{7z}&\xmark&\xmark&\xmark&\xmark&\xmark&\xmark&\xmark&\xmark&\xmark&\xmark&\cmark&\xmark&\xmark&\xmark&\xmark&\xmark&\xmark&\xmark&\xmark&\xmark&\xmark&\xmark&\xmark&\xmark&\xmark&\xmark&\xmark&\xmark&\xmark&\xmark&\xmark&\xmark&\xmark&\xmark&\xmark&\xmark&\xmark&\xmark&\xmark&\xmark&\cmark&\xmark&\cmark&\xmark&\cmark&\xmark&\cmark&\cmark&\xmark&\xmark&\xmark&\xmark&\xmark&\xmark&\cmark&\xmark&\xmark&\xmark&\xmark&\xmark&\xmark&\xmark&\xmark&\xmark\\
\textsc{Apache}&\xmark&\xmark&\xmark&\xmark&\xmark&\cmark&\xmark&\cmark&\xmark&\cmark&\xmark&\xmark&\xmark&\xmark&\xmark&\xmark&\cmark&\xmark&\cmark&\xmark&\xmark&\xmark&\xmark&\xmark&\xmark&\xmark&\cmark&\xmark&\xmark&\cmark&\xmark&\cmark&\xmark&\xmark&\xmark&\xmark&\xmark&\xmark&\xmark&\cmark&\xmark&\xmark&\xmark&\xmark&\xmark&\xmark&\xmark&\xmark&\xmark&\xmark&\xmark&\xmark&\xmark&\xmark&\xmark&\xmark&\xmark&\xmark&\xmark&\xmark&\xmark&\xmark&\xmark&\cmark\\
\textsc{HSMGP}&\xmark&\xmark&\xmark&\xmark&\cmark&\xmark&\cmark&\xmark&\xmark&\xmark&\xmark&\xmark&\xmark&\xmark&\xmark&\xmark&\cmark&\xmark&\cmark&\xmark&\cmark&\xmark&\cmark&\xmark&\xmark&\cmark&\xmark&\cmark&\cmark&\cmark&\cmark&\cmark&\xmark&\xmark&\xmark&\xmark&\xmark&\xmark&\xmark&\xmark&\xmark&\xmark&\cmark&\xmark&\xmark&\xmark&\xmark&\cmark&\xmark&\xmark&\cmark&\xmark&\xmark&\xmark&\xmark&\xmark&\xmark&\xmark&\xmark&\xmark&\xmark&\xmark&\xmark&\xmark\\
\textsc{MongoDB}&\xmark&\cmark&\cmark&\cmark&\xmark&\cmark&\xmark&\cmark&\cmark&\cmark&\xmark&\xmark&\xmark&\xmark&\xmark&\cmark&\xmark&\cmark&\xmark&\cmark&\cmark&\cmark&\cmark&\cmark&\xmark&\cmark&\xmark&\cmark&\cmark&\cmark&\xmark&\cmark&\xmark&\cmark&\xmark&\cmark&\xmark&\cmark&\xmark&\cmark&\xmark&\xmark&\xmark&\xmark&\xmark&\xmark&\xmark&\xmark&\xmark&\cmark&\xmark&\cmark&\xmark&\xmark&\xmark&\xmark&\cmark&\cmark&\cmark&\cmark&\cmark&\xmark&\cmark&\xmark\\
\textsc{PostgreSQL}&\cmark&\xmark&\xmark&\xmark&\xmark&\xmark&\xmark&\xmark&\xmark&\xmark&\xmark&\xmark&\xmark&\xmark&\xmark&\xmark&\xmark&\xmark&\xmark&\xmark&\xmark&\xmark&\xmark&\xmark&\xmark&\xmark&\xmark&\cmark&\cmark&\xmark&\cmark&\xmark&\xmark&\xmark&\xmark&\xmark&\xmark&\xmark&\xmark&\cmark&\xmark&\xmark&\xmark&\xmark&\xmark&\xmark&\xmark&\cmark&\xmark&\xmark&\xmark&\xmark&\xmark&\xmark&\xmark&\xmark&\xmark&\xmark&\xmark&\xmark&\xmark&\xmark&\xmark&\cmark\\
\textsc{ExaStencils}&\xmark&\xmark&\xmark&\xmark&\xmark&\xmark&\xmark&\xmark&\cmark&\xmark&\cmark&\xmark&\cmark&\xmark&\cmark&\xmark&\xmark&\xmark&\xmark&\xmark&\cmark&\xmark&\cmark&\xmark&\xmark&\xmark&\xmark&\xmark&\cmark&\xmark&\cmark&\xmark&\xmark&\xmark&\xmark&\xmark&\xmark&\xmark&\xmark&\xmark&\xmark&\xmark&\xmark&\xmark&\xmark&\xmark&\xmark&\cmark&\cmark&\xmark&\cmark&\xmark&\cmark&\xmark&\cmark&\xmark&\cmark&\xmark&\cmark&\xmark&\cmark&\xmark&\cmark&\xmark\\
\textsc{kanzi}&\xmark&\xmark&\xmark&\xmark&\xmark&\xmark&\xmark&\xmark&\xmark&\xmark&\xmark&\xmark&\xmark&\xmark&\xmark&\xmark&\xmark&\xmark&\xmark&\cmark&\xmark&\xmark&\xmark&\cmark&\xmark&\xmark&\xmark&\cmark&\xmark&\xmark&\xmark&\cmark&\xmark&\xmark&\xmark&\xmark&\xmark&\xmark&\xmark&\xmark&\xmark&\xmark&\cmark&\xmark&\xmark&\xmark&\xmark&\cmark&\xmark&\xmark&\xmark&\cmark&\xmark&\xmark&\xmark&\xmark&\xmark&\xmark&\xmark&\xmark&\xmark&\xmark&\xmark&\xmark\\
\textsc{jump3r}&\xmark&\xmark&\xmark&\xmark&\xmark&\xmark&\xmark&\xmark&\xmark&\xmark&\xmark&\xmark&\xmark&\xmark&\xmark&\xmark&\xmark&\xmark&\xmark&\xmark&\xmark&\xmark&\xmark&\xmark&\cmark&\xmark&\xmark&\xmark&\xmark&\xmark&\xmark&\xmark&\xmark&\xmark&\xmark&\xmark&\xmark&\xmark&\xmark&\xmark&\cmark&\xmark&\cmark&\xmark&\xmark&\xmark&\xmark&\xmark&\xmark&\xmark&\cmark&\xmark&\xmark&\xmark&\xmark&\xmark&\xmark&\xmark&\xmark&\xmark&\xmark&\xmark&\xmark&\xmark\\
\textsc{MariaDB}&\xmark&\xmark&\xmark&\xmark&\xmark&\xmark&\xmark&\xmark&\xmark&\xmark&\xmark&\xmark&\xmark&\xmark&\xmark&\xmark&\xmark&\xmark&\xmark&\cmark&\cmark&\xmark&\xmark&\cmark&\xmark&\cmark&\xmark&\xmark&\xmark&\xmark&\xmark&\cmark&\xmark&\cmark&\xmark&\xmark&\xmark&\xmark&\xmark&\xmark&\cmark&\xmark&\xmark&\xmark&\xmark&\xmark&\xmark&\xmark&\xmark&\xmark&\xmark&\xmark&\xmark&\xmark&\xmark&\cmark&\xmark&\xmark&\xmark&\xmark&\xmark&\xmark&\xmark&\xmark\\
\textsc{polly}&\xmark&\xmark&\xmark&\xmark&\xmark&\xmark&\xmark&\xmark&\cmark&\xmark&\cmark&\xmark&\cmark&\cmark&\cmark&\xmark&\cmark&\xmark&\xmark&\xmark&\cmark&\cmark&\xmark&\xmark&\xmark&\xmark&\cmark&\xmark&\xmark&\cmark&\xmark&\cmark&\xmark&\cmark&\xmark&\cmark&\xmark&\xmark&\xmark&\xmark&\xmark&\xmark&\xmark&\xmark&\cmark&\xmark&\xmark&\xmark&\xmark&\xmark&\cmark&\cmark&\cmark&\cmark&\xmark&\xmark&\xmark&\cmark&\xmark&\cmark&\xmark&\xmark&\xmark&\xmark\\
\textsc{SQL}&\cmark&\xmark&\xmark&\cmark&\xmark&\xmark&\xmark&\cmark&\xmark&\xmark&\xmark&\xmark&\xmark&\cmark&\xmark&\xmark&\xmark&\xmark&\xmark&\xmark&\cmark&\xmark&\xmark&\cmark&\xmark&\xmark&\xmark&\cmark&\cmark&\xmark&\xmark&\xmark&\xmark&\xmark&\xmark&\xmark&\xmark&\xmark&\xmark&\xmark&\cmark&\xmark&\cmark&\xmark&\xmark&\xmark&\xmark&\cmark&\xmark&\xmark&\xmark&\xmark&\xmark&\cmark&\xmark&\xmark&\xmark&\xmark&\xmark&\xmark&\xmark&\xmark&\xmark&\cmark\\
\textsc{vp9}&\xmark&\xmark&\xmark&\cmark&\xmark&\xmark&\xmark&\cmark&\xmark&\xmark&\xmark&\xmark&\xmark&\xmark&\xmark&\xmark&\xmark&\xmark&\xmark&\xmark&\xmark&\xmark&\xmark&\xmark&\xmark&\xmark&\cmark&\cmark&\xmark&\xmark&\xmark&\cmark&\xmark&\cmark&\xmark&\cmark&\xmark&\xmark&\xmark&\xmark&\xmark&\xmark&\xmark&\xmark&\xmark&\xmark&\xmark&\xmark&\xmark&\xmark&\xmark&\cmark&\xmark&\xmark&\xmark&\cmark&\xmark&\cmark&\xmark&\cmark&\xmark&\xmark&\xmark&\xmark\\
\textsc{Spark}&\xmark&\cmark&\xmark&\xmark&\xmark&\xmark&\xmark&\xmark&\xmark&\xmark&\xmark&\xmark&\xmark&\xmark&\xmark&\cmark&\xmark&\xmark&\xmark&\xmark&\xmark&\cmark&\xmark&\xmark&\xmark&\xmark&\cmark&\xmark&\xmark&\xmark&\xmark&\xmark&\cmark&\cmark&\cmark&\xmark&\xmark&\xmark&\xmark&\cmark&\xmark&\xmark&\xmark&\xmark&\xmark&\xmark&\xmark&\cmark&\xmark&\xmark&\cmark&\xmark&\xmark&\xmark&\xmark&\xmark&\xmark&\xmark&\xmark&\xmark&\xmark&\xmark&\xmark&\cmark\\
\textsc{HIPAcc}&\xmark&\xmark&\xmark&\xmark&\xmark&\xmark&\xmark&\xmark&\xmark&\xmark&\xmark&\xmark&\xmark&\xmark&\xmark&\xmark&\xmark&\xmark&\xmark&\xmark&\xmark&\xmark&\xmark&\xmark&\xmark&\cmark&\cmark&\cmark&\xmark&\xmark&\xmark&\xmark&\xmark&\cmark&\xmark&\cmark&\xmark&\cmark&\xmark&\cmark&\cmark&\xmark&\xmark&\xmark&\xmark&\xmark&\xmark&\xmark&\cmark&\xmark&\cmark&\xmark&\xmark&\xmark&\xmark&\xmark&\xmark&\xmark&\xmark&\cmark&\xmark&\xmark&\xmark&\cmark\\
\textsc{Redis}&\cmark&\xmark&\xmark&\xmark&\xmark&\xmark&\xmark&\xmark&\xmark&\xmark&\xmark&\cmark&\xmark&\xmark&\xmark&\cmark&\xmark&\xmark&\xmark&\xmark&\xmark&\xmark&\xmark&\xmark&\xmark&\xmark&\xmark&\xmark&\xmark&\xmark&\xmark&\xmark&\xmark&\xmark&\xmark&\xmark&\xmark&\xmark&\xmark&\xmark&\xmark&\xmark&\xmark&\xmark&\cmark&\xmark&\cmark&\xmark&\xmark&\xmark&\cmark&\xmark&\xmark&\xmark&\xmark&\xmark&\xmark&\xmark&\xmark&\xmark&\xmark&\xmark&\xmark&\xmark\\
\textsc{storm}&\xmark&\xmark&\xmark&\xmark&\xmark&\xmark&\xmark&\xmark&\xmark&\xmark&\xmark&\xmark&\xmark&\xmark&\xmark&\xmark&\xmark&\cmark&\xmark&\xmark&\xmark&\xmark&\xmark&\xmark&\xmark&\xmark&\xmark&\xmark&\xmark&\xmark&\xmark&\xmark&\xmark&\xmark&\xmark&\cmark&\xmark&\xmark&\xmark&\xmark&\cmark&\xmark&\xmark&\xmark&\xmark&\xmark&\xmark&\xmark&\xmark&\xmark&\xmark&\xmark&\xmark&\xmark&\xmark&\xmark&\xmark&\xmark&\xmark&\xmark&\xmark&\xmark&\xmark&\cmark\\
\textsc{SaC}&\xmark&\xmark&\xmark&\xmark&\xmark&\xmark&\xmark&\xmark&\xmark&\xmark&\xmark&\xmark&\xmark&\xmark&\xmark&\xmark&\xmark&\cmark&\xmark&\xmark&\xmark&\cmark&\xmark&\xmark&\xmark&\xmark&\xmark&\xmark&\xmark&\cmark&\xmark&\xmark&\xmark&\cmark&\xmark&\cmark&\xmark&\cmark&\xmark&\cmark&\xmark&\xmark&\xmark&\xmark&\xmark&\cmark&\xmark&\xmark&\cmark&\cmark&\xmark&\xmark&\xmark&\cmark&\xmark&\xmark&\xmark&\xmark&\xmark&\xmark&\xmark&\xmark&\xmark&\xmark\\
\textsc{Hadoop}&\cmark&\xmark&\xmark&\xmark&\xmark&\xmark&\cmark&\xmark&\xmark&\xmark&\xmark&\xmark&\xmark&\xmark&\xmark&\xmark&\xmark&\cmark&\xmark&\cmark&\xmark&\cmark&\xmark&\cmark&\xmark&\xmark&\xmark&\xmark&\xmark&\cmark&\xmark&\cmark&\xmark&\xmark&\xmark&\cmark&\xmark&\xmark&\xmark&\xmark&\xmark&\xmark&\xmark&\xmark&\xmark&\xmark&\cmark&\xmark&\xmark&\xmark&\xmark&\xmark&\xmark&\cmark&\cmark&\cmark&\xmark&\xmark&\xmark&\xmark&\cmark&\xmark&\xmark&\xmark\\
\textsc{Tomcat}&\xmark&\xmark&\xmark&\xmark&\xmark&\xmark&\xmark&\xmark&\xmark&\xmark&\xmark&\xmark&\xmark&\xmark&\xmark&\xmark&\xmark&\cmark&\xmark&\cmark&\xmark&\cmark&\xmark&\cmark&\xmark&\xmark&\xmark&\xmark&\xmark&\cmark&\xmark&\cmark&\xmark&\xmark&\xmark&\xmark&\xmark&\xmark&\xmark&\xmark&\xmark&\xmark&\xmark&\xmark&\xmark&\xmark&\cmark&\xmark&\xmark&\xmark&\xmark&\xmark&\xmark&\xmark&\xmark&\xmark&\xmark&\xmark&\xmark&\xmark&\xmark&\xmark&\xmark&\xmark\\
\textsc{JavaGC}&\xmark&\xmark&\cmark&\xmark&\xmark&\xmark&\xmark&\xmark&\xmark&\xmark&\xmark&\xmark&\xmark&\xmark&\xmark&\xmark&\xmark&\xmark&\xmark&\cmark&\xmark&\cmark&\xmark&\xmark&\cmark&\xmark&\cmark&\xmark&\xmark&\xmark&\xmark&\xmark&\xmark&\xmark&\xmark&\xmark&\xmark&\cmark&\xmark&\cmark&\xmark&\xmark&\cmark&\xmark&\xmark&\xmark&\xmark&\xmark&\cmark&\xmark&\cmark&\xmark&\xmark&\xmark&\xmark&\xmark&\xmark&\xmark&\xmark&\xmark&\xmark&\xmark&\xmark&\xmark\\


\bottomrule
\end{tabular}

\end{adjustbox}

\end{table*}

%% file: Tables/RQ2/rq2_bmbt.tex
\begin{table*}[t!]
\caption{The consistency between the best (worst) model, chosen according to the accuracy, and the model that leads to the best (worst) tuning quality under batch model-based tuners with 30 repeats. The format is the same as Table~\ref{tb:rq3.1}.}
\label{tb:rq3.2}
\setlength{\tabcolsep}{0.5mm}
\centering
\footnotesize
\begin{adjustbox}{width=\textwidth,center}
\begin{tabular}{lcccccccc|cccccccc|cccccccc|cccccccc|cccccccc|cccccccc|cccccccc|cccccccc|cccccccc}


\toprule
\multirow{3}{*}{\textbf{System}}&\multicolumn{8}{c|}{\textbf{\texttt{BestConfig}}}&\multicolumn{8}{c|}{\textbf{\texttt{Irace}}}&\multicolumn{8}{c|}{\textbf{\texttt{GGA}}}&\multicolumn{8}{c|}{\textbf{\texttt{ParamILS}}}&\multicolumn{8}{c|}{\textbf{\texttt{Random}}}&\multicolumn{8}{c|}{\textbf{\texttt{GA}}}&\multicolumn{8}{c|}{\textbf{\texttt{SWAY}}}&\multicolumn{8}{c|}{\textbf{\texttt{ConEx}}}&\multicolumn{8}{c}{\textbf{\texttt{Brute-force}}}\\

\cmidrule(l){2-73}

&\multicolumn{2}{c}{\textbf{M\&P}}&\multicolumn{2}{c}{\textbf{R\&P}}&\multicolumn{2}{c}{\textbf{M\&E}}&\multicolumn{2}{c|}{\textbf{R\&E}}&\multicolumn{2}{c}{\textbf{M\&P}}&\multicolumn{2}{c}{\textbf{R\&P}}&\multicolumn{2}{c}{\textbf{M\&E}}&\multicolumn{2}{c|}{\textbf{R\&E}}&\multicolumn{2}{c}{\textbf{M\&P}}&\multicolumn{2}{c}{\textbf{R\&P}}&\multicolumn{2}{c}{\textbf{M\&E}}&\multicolumn{2}{c|}{\textbf{R\&E}}&\multicolumn{2}{c}{\textbf{M\&P}}&\multicolumn{2}{c}{\textbf{R\&P}}&\multicolumn{2}{c}{\textbf{M\&E}}&\multicolumn{2}{c|}{\textbf{R\&E}}&\multicolumn{2}{c}{\textbf{M\&P}}&\multicolumn{2}{c}{\textbf{R\&P}}&\multicolumn{2}{c}{\textbf{M\&E}}&\multicolumn{2}{c|}{\textbf{R\&E}}&\multicolumn{2}{c}{\textbf{M\&P}}&\multicolumn{2}{c}{\textbf{R\&P}}&\multicolumn{2}{c}{\textbf{M\&E}}&\multicolumn{2}{c|}{\textbf{R\&E}}&\multicolumn{2}{c}{\textbf{M\&P}}&\multicolumn{2}{c}{\textbf{R\&P}}&\multicolumn{2}{c}{\textbf{M\&E}}&\multicolumn{2}{c|}{\textbf{R\&E}}&\multicolumn{2}{c}{\textbf{M\&P}}&\multicolumn{2}{c}{\textbf{R\&P}}&\multicolumn{2}{c}{\textbf{M\&E}}&\multicolumn{2}{c}{\textbf{R\&E}}\\

&$\mathcal{B}$&$\mathcal{W}$&$\mathcal{B}$&$\mathcal{W}$&$\mathcal{B}$&$\mathcal{W}$&$\mathcal{B}$&$\mathcal{W}$&$\mathcal{B}$&$\mathcal{W}$&$\mathcal{B}$&$\mathcal{W}$&$\mathcal{B}$&$\mathcal{W}$&$\mathcal{B}$&$\mathcal{W}$&$\mathcal{B}$&$\mathcal{W}$&$\mathcal{B}$&$\mathcal{W}$&$\mathcal{B}$&$\mathcal{W}$&$\mathcal{B}$&$\mathcal{W}$&$\mathcal{B}$&$\mathcal{W}$&$\mathcal{B}$&$\mathcal{W}$&$\mathcal{B}$&$\mathcal{W}$&$\mathcal{B}$&$\mathcal{W}$&$\mathcal{B}$&$\mathcal{W}$&$\mathcal{B}$&$\mathcal{W}$&$\mathcal{B}$&$\mathcal{W}$&$\mathcal{B}$&$\mathcal{W}$&$\mathcal{B}$&$\mathcal{W}$&$\mathcal{B}$&$\mathcal{W}$&$\mathcal{B}$&$\mathcal{W}$&$\mathcal{B}$&$\mathcal{W}$&$\mathcal{B}$&$\mathcal{W}$&$\mathcal{B}$&$\mathcal{W}$&$\mathcal{B}$&$\mathcal{W}$&$\mathcal{B}$&$\mathcal{W}$&$\mathcal{B}$&$\mathcal{W}$&$\mathcal{B}$&$\mathcal{W}$&$\mathcal{B}$&$\mathcal{W}$&$\mathcal{B}$&$\mathcal{W}$&$\mathcal{B}$&$\mathcal{W}$&$\mathcal{B}$&$\mathcal{W}$&$\mathcal{B}$&$\mathcal{W}$&$\mathcal{B}$&$\mathcal{W}$\\

\midrule
\textsc{Brotli}&-&-&-&-&-&-&-&-&-&-&-&-&-&-&-&-&-&-&-&-&-&-&-&-&-&-&-&-&-&-&-&-&-&-&-&-&-&-&-&-&-&-&-&-&-&-&-&-&-&-&-&-&-&-&-&-&-&-&-&-&-&-&-&-&\xmark&\xmark&\xmark&\xmark&\xmark&\xmark&\xmark&\cmark\\
\textsc{LLVM}&-&-&-&-&-&-&-&-&-&-&-&-&-&-&-&-&-&-&-&-&-&-&-&-&-&-&-&-&-&-&-&-&-&-&-&-&-&-&-&-&-&-&-&-&-&-&-&-&-&-&-&-&-&-&-&-&-&-&-&-&-&-&-&-&\cmark&\xmark&\cmark&\xmark&\xmark&\xmark&\xmark&\cmark\\
\textsc{Lrzip}&-&-&-&-&-&-&-&-&-&-&-&-&-&-&-&-&-&-&-&-&-&-&-&-&-&-&-&-&-&-&-&-&-&-&-&-&-&-&-&-&-&-&-&-&-&-&-&-&-&-&-&-&-&-&-&-&-&-&-&-&-&-&-&-&\xmark&\cmark&\cmark&\cmark&\xmark&\xmark&\cmark&\xmark\\
\textsc{xgboost}&-&-&-&-&-&-&-&-&-&-&-&-&-&-&-&-&-&-&-&-&-&-&-&-&-&-&-&-&-&-&-&-&-&-&-&-&-&-&-&-&-&-&-&-&-&-&-&-&-&-&-&-&-&-&-&-&-&-&-&-&-&-&-&-&\cmark&\xmark&\cmark&\xmark&\cmark&\xmark&\cmark&\cmark\\
\textsc{noc-CM-log}&-&-&-&-&-&-&-&-&-&-&-&-&-&-&-&-&-&-&-&-&-&-&-&-&-&-&-&-&-&-&-&-&-&-&-&-&-&-&-&-&-&-&-&-&-&-&-&-&-&-&-&-&-&-&-&-&-&-&-&-&-&-&-&-&\xmark&\xmark&\xmark&\xmark&\xmark&\xmark&\xmark&\xmark\\
\textsc{DeepArch}&-&-&-&-&-&-&-&-&-&-&-&-&-&-&-&-&-&-&-&-&-&-&-&-&-&-&-&-&-&-&-&-&-&-&-&-&-&-&-&-&-&-&-&-&-&-&-&-&-&-&-&-&-&-&-&-&-&-&-&-&-&-&-&-&\cmark&\xmark&\cmark&\cmark&\cmark&\xmark&\cmark&\xmark\\
\textsc{BDB\_C}&-&-&-&-&-&-&-&-&-&-&-&-&-&-&-&-&-&-&-&-&-&-&-&-&-&-&-&-&-&-&-&-&-&-&-&-&-&-&-&-&-&-&-&-&-&-&-&-&-&-&-&-&-&-&-&-&-&-&-&-&-&-&-&-&\xmark&\xmark&\xmark&\xmark&\xmark&\xmark&\xmark&\xmark\\
\textsc{HSQLDB}&-&-&-&-&-&-&-&-&-&-&-&-&-&-&-&-&-&-&-&-&-&-&-&-&-&-&-&-&-&-&-&-&-&-&-&-&-&-&-&-&-&-&-&-&-&-&-&-&-&-&-&-&-&-&-&-&-&-&-&-&-&-&-&-&\xmark&\xmark&\cmark&\xmark&\xmark&\xmark&\cmark&\xmark\\
\textsc{DConvert}&\xmark&\xmark&\xmark&\xmark&\xmark&\xmark&\xmark&\xmark&\xmark&\xmark&\xmark&\xmark&\xmark&\xmark&\xmark&\xmark&\cmark&\xmark&\cmark&\xmark&\xmark&\xmark&\xmark&\xmark&\xmark&\xmark&\xmark&\xmark&\xmark&\xmark&\xmark&\xmark&\xmark&\cmark&\xmark&\cmark&\xmark&\cmark&\xmark&\cmark&\xmark&\xmark&\xmark&\xmark&\xmark&\xmark&\xmark&\xmark&\xmark&\xmark&\xmark&\xmark&\xmark&\xmark&\xmark&\xmark&\xmark&\xmark&\xmark&\xmark&\xmark&\xmark&\xmark&\xmark&-&-&-&-&-&-&-&-\\
\textsc{7z}&\xmark&\xmark&\xmark&\xmark&\xmark&\xmark&\xmark&\xmark&\xmark&\xmark&\xmark&\xmark&\cmark&\xmark&\cmark&\xmark&\xmark&\xmark&\xmark&\xmark&\xmark&\xmark&\xmark&\xmark&\xmark&\xmark&\xmark&\xmark&\xmark&\cmark&\xmark&\xmark&\xmark&\xmark&\xmark&\cmark&\cmark&\xmark&\cmark&\cmark&\xmark&\xmark&\xmark&\xmark&\xmark&\xmark&\xmark&\xmark&\cmark&\xmark&\cmark&\xmark&\xmark&\xmark&\xmark&\xmark&\xmark&\xmark&\xmark&\xmark&\xmark&\xmark&\xmark&\xmark&-&-&-&-&-&-&-&-\\
\textsc{Apache}&\xmark&\xmark&\xmark&\xmark&\xmark&\xmark&\xmark&\xmark&\xmark&\cmark&\xmark&\cmark&\xmark&\xmark&\xmark&\xmark&\xmark&\cmark&\xmark&\cmark&\xmark&\cmark&\xmark&\cmark&\xmark&\cmark&\xmark&\cmark&\cmark&\xmark&\cmark&\xmark&\xmark&\cmark&\xmark&\cmark&\xmark&\cmark&\xmark&\cmark&\xmark&\cmark&\xmark&\cmark&\xmark&\xmark&\xmark&\xmark&\xmark&\cmark&\xmark&\cmark&\xmark&\cmark&\xmark&\cmark&\xmark&\cmark&\xmark&\cmark&\cmark&\xmark&\cmark&\xmark&-&-&-&-&-&-&-&-\\
\textsc{HSMGP}&\xmark&\xmark&\xmark&\xmark&\xmark&\xmark&\xmark&\cmark&\xmark&\xmark&\xmark&\xmark&\xmark&\xmark&\xmark&\xmark&\cmark&\xmark&\cmark&\xmark&\xmark&\xmark&\xmark&\xmark&\cmark&\xmark&\cmark&\xmark&\cmark&\xmark&\cmark&\xmark&\cmark&\xmark&\cmark&\cmark&\cmark&\xmark&\cmark&\cmark&\xmark&\xmark&\xmark&\xmark&\xmark&\xmark&\xmark&\xmark&\cmark&\xmark&\cmark&\xmark&\xmark&\xmark&\xmark&\xmark&\cmark&\xmark&\cmark&\xmark&\cmark&\xmark&\cmark&\xmark&-&-&-&-&-&-&-&-\\
\textsc{MongoDB}&\xmark&\xmark&\xmark&\xmark&\xmark&\cmark&\xmark&\cmark&\xmark&\xmark&\xmark&\xmark&\xmark&\xmark&\xmark&\xmark&\xmark&\xmark&\xmark&\xmark&\cmark&\xmark&\cmark&\xmark&\xmark&\xmark&\xmark&\xmark&\xmark&\xmark&\xmark&\xmark&\xmark&\cmark&\xmark&\cmark&\xmark&\cmark&\xmark&\cmark&\xmark&\xmark&\xmark&\xmark&\xmark&\xmark&\xmark&\xmark&\xmark&\xmark&\xmark&\xmark&\xmark&\xmark&\xmark&\xmark&\xmark&\xmark&\xmark&\xmark&\xmark&\xmark&\xmark&\xmark&-&-&-&-&-&-&-&-\\
\textsc{PostgreSQL}&\xmark&\xmark&\xmark&\xmark&\xmark&\xmark&\xmark&\xmark&\cmark&\xmark&\xmark&\xmark&\xmark&\xmark&\xmark&\xmark&\cmark&\xmark&\xmark&\xmark&\xmark&\xmark&\xmark&\xmark&\xmark&\xmark&\xmark&\xmark&\xmark&\xmark&\xmark&\xmark&\xmark&\cmark&\xmark&\cmark&\xmark&\cmark&\xmark&\cmark&\xmark&\xmark&\xmark&\xmark&\xmark&\xmark&\xmark&\xmark&\xmark&\xmark&\xmark&\xmark&\xmark&\xmark&\xmark&\xmark&\xmark&\xmark&\xmark&\xmark&\xmark&\xmark&\cmark&\xmark&-&-&-&-&-&-&-&-\\
\textsc{ExaStencils}&\xmark&\xmark&\xmark&\xmark&\xmark&\xmark&\xmark&\xmark&\cmark&\xmark&\cmark&\xmark&\xmark&\xmark&\xmark&\xmark&\cmark&\xmark&\cmark&\xmark&\cmark&\xmark&\cmark&\xmark&\cmark&\xmark&\cmark&\xmark&\xmark&\xmark&\xmark&\xmark&\cmark&\cmark&\cmark&\cmark&\xmark&\xmark&\xmark&\xmark&\cmark&\xmark&\cmark&\xmark&\xmark&\xmark&\xmark&\xmark&\cmark&\xmark&\cmark&\xmark&\xmark&\xmark&\xmark&\xmark&\cmark&\xmark&\cmark&\xmark&\xmark&\xmark&\xmark&\xmark&-&-&-&-&-&-&-&-\\
\textsc{kanzi}&\xmark&\cmark&\xmark&\xmark&\xmark&\xmark&\xmark&\xmark&\cmark&\xmark&\cmark&\cmark&\xmark&\xmark&\xmark&\xmark&\cmark&\xmark&\cmark&\cmark&\cmark&\xmark&\cmark&\xmark&\cmark&\xmark&\cmark&\cmark&\cmark&\xmark&\cmark&\xmark&\cmark&\xmark&\cmark&\cmark&\cmark&\xmark&\cmark&\cmark&\cmark&\xmark&\cmark&\cmark&\cmark&\xmark&\cmark&\xmark&\cmark&\xmark&\cmark&\cmark&\xmark&\xmark&\xmark&\xmark&\cmark&\xmark&\cmark&\cmark&\cmark&\xmark&\cmark&\xmark&-&-&-&-&-&-&-&-\\
\textsc{jump3r}&\xmark&\cmark&\xmark&\xmark&\cmark&\xmark&\xmark&\xmark&\xmark&\xmark&\cmark&\xmark&\xmark&\xmark&\xmark&\xmark&\xmark&\xmark&\cmark&\xmark&\xmark&\xmark&\cmark&\xmark&\xmark&\cmark&\cmark&\xmark&\xmark&\xmark&\cmark&\xmark&\xmark&\xmark&\cmark&\xmark&\xmark&\cmark&\cmark&\xmark&\xmark&\xmark&\cmark&\cmark&\xmark&\xmark&\xmark&\xmark&\xmark&\cmark&\cmark&\xmark&\xmark&\xmark&\cmark&\xmark&\xmark&\xmark&\cmark&\cmark&\xmark&\xmark&\cmark&\cmark&-&-&-&-&-&-&-&-\\
\textsc{MariaDB}&\cmark&\cmark&\cmark&\cmark&\xmark&\cmark&\xmark&\cmark&\cmark&\cmark&\cmark&\cmark&\xmark&\xmark&\xmark&\xmark&\cmark&\cmark&\cmark&\cmark&\xmark&\cmark&\xmark&\cmark&\cmark&\cmark&\cmark&\cmark&\xmark&\cmark&\xmark&\cmark&\cmark&\xmark&\cmark&\xmark&\xmark&\xmark&\xmark&\xmark&\cmark&\cmark&\cmark&\cmark&\xmark&\cmark&\xmark&\cmark&\cmark&\cmark&\cmark&\cmark&\cmark&\cmark&\cmark&\cmark&\cmark&\cmark&\cmark&\cmark&\cmark&\cmark&\cmark&\cmark&-&-&-&-&-&-&-&-\\
\textsc{polly}&\xmark&\cmark&\xmark&\cmark&\xmark&\xmark&\xmark&\xmark&\cmark&\xmark&\cmark&\xmark&\cmark&\xmark&\cmark&\xmark&\cmark&\xmark&\cmark&\xmark&\xmark&\cmark&\xmark&\cmark&\cmark&\cmark&\cmark&\cmark&\xmark&\cmark&\xmark&\cmark&\cmark&\xmark&\cmark&\xmark&\xmark&\cmark&\xmark&\cmark&\xmark&\xmark&\xmark&\xmark&\xmark&\xmark&\xmark&\xmark&\cmark&\xmark&\cmark&\xmark&\xmark&\xmark&\xmark&\xmark&\cmark&\xmark&\cmark&\xmark&\xmark&\cmark&\xmark&\cmark&-&-&-&-&-&-&-&-\\
\textsc{SQL}&\xmark&\xmark&\xmark&\xmark&\xmark&\xmark&\xmark&\xmark&\cmark&\xmark&\cmark&\xmark&\xmark&\xmark&\xmark&\xmark&\cmark&\xmark&\cmark&\xmark&\xmark&\xmark&\xmark&\xmark&\cmark&\xmark&\cmark&\xmark&\xmark&\xmark&\xmark&\xmark&\xmark&\xmark&\xmark&\xmark&\xmark&\xmark&\xmark&\xmark&\xmark&\xmark&\xmark&\xmark&\xmark&\xmark&\xmark&\xmark&\xmark&\xmark&\xmark&\xmark&\xmark&\xmark&\xmark&\xmark&\cmark&\xmark&\cmark&\xmark&\xmark&\xmark&\xmark&\xmark&-&-&-&-&-&-&-&-\\
\textsc{vp9}&\xmark&\xmark&\xmark&\xmark&\xmark&\xmark&\xmark&\xmark&\xmark&\cmark&\xmark&\cmark&\xmark&\xmark&\xmark&\xmark&\xmark&\cmark&\xmark&\cmark&\xmark&\cmark&\xmark&\cmark&\xmark&\cmark&\xmark&\cmark&\xmark&\cmark&\xmark&\cmark&\xmark&\cmark&\xmark&\cmark&\xmark&\cmark&\xmark&\cmark&\xmark&\cmark&\xmark&\cmark&\xmark&\xmark&\xmark&\xmark&\xmark&\cmark&\xmark&\cmark&\cmark&\xmark&\cmark&\xmark&\xmark&\cmark&\xmark&\cmark&\xmark&\cmark&\xmark&\cmark&-&-&-&-&-&-&-&-\\
\textsc{Spark}&\xmark&\xmark&\xmark&\xmark&\xmark&\xmark&\xmark&\xmark&\cmark&\xmark&\cmark&\xmark&\xmark&\xmark&\xmark&\xmark&\cmark&\xmark&\cmark&\xmark&\xmark&\xmark&\xmark&\xmark&\cmark&\cmark&\cmark&\cmark&\xmark&\xmark&\xmark&\xmark&\cmark&\xmark&\cmark&\xmark&\xmark&\cmark&\xmark&\cmark&\xmark&\xmark&\xmark&\xmark&\xmark&\xmark&\xmark&\xmark&\cmark&\xmark&\cmark&\xmark&\xmark&\xmark&\xmark&\xmark&\cmark&\cmark&\cmark&\cmark&\xmark&\xmark&\xmark&\xmark&-&-&-&-&-&-&-&-\\
\textsc{HIPAcc}&\xmark&\xmark&\xmark&\xmark&\xmark&\xmark&\xmark&\xmark&\xmark&\xmark&\xmark&\xmark&\xmark&\xmark&\xmark&\xmark&\xmark&\xmark&\xmark&\xmark&\xmark&\xmark&\xmark&\xmark&\xmark&\xmark&\xmark&\xmark&\xmark&\xmark&\xmark&\xmark&\xmark&\cmark&\xmark&\cmark&\xmark&\cmark&\xmark&\cmark&\xmark&\xmark&\xmark&\xmark&\xmark&\xmark&\xmark&\xmark&\xmark&\xmark&\xmark&\xmark&\xmark&\xmark&\xmark&\xmark&\xmark&\xmark&\xmark&\xmark&\xmark&\xmark&\xmark&\xmark&-&-&-&-&-&-&-&-\\
\textsc{Redis}&\xmark&\xmark&\xmark&\xmark&\xmark&\cmark&\xmark&\cmark&\xmark&\xmark&\cmark&\xmark&\xmark&\xmark&\xmark&\xmark&\xmark&\xmark&\cmark&\xmark&\xmark&\xmark&\xmark&\xmark&\xmark&\cmark&\cmark&\cmark&\xmark&\xmark&\xmark&\xmark&\xmark&\xmark&\cmark&\xmark&\xmark&\xmark&\xmark&\xmark&\xmark&\cmark&\cmark&\cmark&\xmark&\xmark&\xmark&\xmark&\xmark&\xmark&\cmark&\xmark&\xmark&\xmark&\xmark&\xmark&\xmark&\xmark&\cmark&\xmark&\xmark&\xmark&\xmark&\xmark&-&-&-&-&-&-&-&-\\
\textsc{storm}&\xmark&\xmark&\xmark&\xmark&\xmark&\cmark&\xmark&\xmark&\xmark&\xmark&\xmark&\xmark&\xmark&\xmark&\xmark&\xmark&\cmark&\xmark&\xmark&\xmark&\xmark&\xmark&\xmark&\xmark&\xmark&\xmark&\xmark&\xmark&\cmark&\xmark&\xmark&\xmark&\cmark&\xmark&\xmark&\cmark&\xmark&\xmark&\xmark&\cmark&\cmark&\xmark&\xmark&\xmark&\xmark&\xmark&\xmark&\xmark&\xmark&\xmark&\xmark&\xmark&\xmark&\xmark&\xmark&\xmark&\cmark&\xmark&\xmark&\xmark&\xmark&\xmark&\xmark&\xmark&-&-&-&-&-&-&-&-\\
\textsc{SaC}&\xmark&\xmark&\xmark&\cmark&\xmark&\xmark&\xmark&\cmark&\xmark&\xmark&\xmark&\xmark&\xmark&\xmark&\xmark&\xmark&\xmark&\xmark&\xmark&\cmark&\xmark&\xmark&\xmark&\xmark&\xmark&\xmark&\xmark&\cmark&\xmark&\xmark&\xmark&\cmark&\xmark&\cmark&\xmark&\xmark&\cmark&\xmark&\cmark&\xmark&\cmark&\xmark&\cmark&\cmark&\cmark&\xmark&\cmark&\cmark&\cmark&\xmark&\cmark&\xmark&\cmark&\xmark&\cmark&\xmark&\xmark&\xmark&\xmark&\cmark&\xmark&\xmark&\xmark&\xmark&-&-&-&-&-&-&-&-\\
\textsc{Hadoop}&\cmark&\xmark&\cmark&\xmark&\xmark&\xmark&\xmark&\xmark&\cmark&\xmark&\cmark&\xmark&\xmark&\xmark&\xmark&\xmark&\cmark&\xmark&\cmark&\xmark&\xmark&\xmark&\xmark&\xmark&\cmark&\xmark&\cmark&\xmark&\xmark&\xmark&\xmark&\xmark&\cmark&\cmark&\cmark&\cmark&\xmark&\cmark&\xmark&\cmark&\cmark&\cmark&\cmark&\cmark&\xmark&\cmark&\xmark&\cmark&\cmark&\cmark&\cmark&\cmark&\xmark&\xmark&\xmark&\xmark&\cmark&\xmark&\cmark&\xmark&\xmark&\xmark&\xmark&\xmark&-&-&-&-&-&-&-&-\\
\textsc{Tomcat}&\cmark&\xmark&\cmark&\xmark&\xmark&\xmark&\xmark&\cmark&\cmark&\cmark&\cmark&\xmark&\xmark&\cmark&\xmark&\xmark&\xmark&\cmark&\xmark&\xmark&\cmark&\xmark&\cmark&\xmark&\cmark&\cmark&\cmark&\xmark&\xmark&\xmark&\xmark&\xmark&\cmark&\cmark&\cmark&\xmark&\cmark&\xmark&\cmark&\cmark&\cmark&\cmark&\cmark&\xmark&\xmark&\xmark&\xmark&\xmark&\cmark&\cmark&\cmark&\xmark&\xmark&\xmark&\xmark&\xmark&\cmark&\xmark&\cmark&\cmark&\xmark&\xmark&\xmark&\xmark&-&-&-&-&-&-&-&-\\
\textsc{JavaGC}&\xmark&\xmark&\xmark&\xmark&\xmark&\xmark&\xmark&\xmark&\xmark&\xmark&\xmark&\xmark&\xmark&\xmark&\xmark&\xmark&\xmark&\xmark&\xmark&\xmark&\xmark&\xmark&\xmark&\xmark&\xmark&\xmark&\xmark&\xmark&\xmark&\xmark&\xmark&\xmark&\xmark&\xmark&\xmark&\cmark&\cmark&\xmark&\cmark&\xmark&\xmark&\xmark&\xmark&\xmark&\xmark&\xmark&\xmark&\xmark&\cmark&\xmark&\cmark&\xmark&\xmark&\xmark&\xmark&\xmark&\xmark&\xmark&\xmark&\xmark&\xmark&\xmark&\xmark&\xmark&-&-&-&-&-&-&-&-\\

\bottomrule
\end{tabular}

\end{adjustbox}

\end{table*}

%% file: Figures/rq4-summary-1.tex
  
  \begin{tikzpicture}
      \begin{axis}[
          ymajorgrids=true,
          ybar,
          bar width=.2cm,
          width=10cm,
          height=4cm,
           axis x line  = bottom,
    axis y line  = left  ,
    legend cell align={left},
          legend image code/.code={%
                    \draw[#1] (0cm,-0.1cm) rectangle (0.15cm,0.15cm);
                }, 
          legend style={at={(-0.05,1.3)},
              anchor=west,legend columns=4,font=\footnotesize,nodes={scale=1, transform shape},draw=none,fill=none,},
          xtick=data,
          xticklabels={M\&P, R\&P, M\&E, R\&E},
          nodes near coords = {\tiny\pgfmathprintnumber[fixed, precision=1]{\pgfplotspointmeta}},
          nodes near coords align={vertical},
          ymin=-5, ymax=140,
          font=\small,
          ylabel={\# cases},
          enlarge x limits={abs=1cm},  
          grid style = {dashed},
          ]
          
          \addplot[fill=red!50] coordinates {(1, 0) (2, 0) (3, 0) (4, 0)};
          \addplot[fill=red!30] coordinates {(1, 7) (2, 1) (3, 3) (4, 5)};
          \addplot[fill=red!10] coordinates {(1, 26) (2, 21) (3, 52) (4, 29)};
          \addplot[fill=white!10] coordinates {(1, 108) (2, 82) (3, 116) (4, 135)};
          \addplot[fill=steel!10]  coordinates {(1, 63) (2, 87) (3, 58) (4, 55)};
          \addplot[fill=steel!30] coordinates {(1, 27) (2, 39) (3, 3) (4, 8)};
          \addplot[fill=steel!60] coordinates {(1, 1) (2, 2) (3, 0) (4, 0)};

      \end{axis}
  \end{tikzpicture}%

 

%% file: Figures/rq4-summary-2.tex
  \begin{tikzpicture}
      \begin{axis}[
          ymajorgrids=true,
          ybar,
          bar width=.2cm,
          width=10cm,
          height=4cm,
           axis x line  = bottom,
    axis y line  = left  ,
    legend cell align={left},
          legend image code/.code={%
                    \draw[#1] (0cm,-0.1cm) rectangle (0.15cm,0.15cm);
                }, 
          legend style={at={(-0.05,1.3)},
              anchor=west,legend columns=4,font=\footnotesize,nodes={scale=1, transform shape},draw=none,fill=none,},
          xtick=data,
          xticklabels={M\&P, R\&P, M\&E, R\&E},
          nodes near coords = {\tiny\pgfmathprintnumber[fixed, precision=1]{\pgfplotspointmeta}},
          nodes near coords align={vertical},
          ymin=-5, ymax=100,
          font=\small,
          ylabel={\# cases},
          enlarge x limits={abs=1cm},  
          grid style = {dashed},
          ]
          \addplot[fill=red!50] coordinates {(1, 0) (2, 0) (3, 0) (4, 0)};
          \addplot[fill=red!30] coordinates {(1, 0) (2, 0) (3, 0) (4, 0)};
          \addplot[fill=red!10] coordinates {(1, 2) (2, 0) (3, 7) (4, 9)};
          \addplot[fill=white!10] coordinates {(1, 35) (2, 20) (3, 97) (4, 85)};
          \addplot[fill=steel!10]  coordinates {(1, 73) (2, 81) (3, 55) (4, 63)};
          \addplot[fill=steel!30] coordinates {(1, 53) (2, 53) (3, 14) (4, 15)};
          \addplot[fill=steel!60] coordinates {(1, 13) (2, 22) (3, 3) (4, 4)};
          
      \end{axis}
  \end{tikzpicture}%

%% file: Tables/RQ3/rq3_smbt_perf.tex
\begin{table*}[t!]
\caption{Spearman correlations between model accuracy and its resulting tuning performance for sequential model-based tuners. $^\dagger$ and $^\star$ denote $p<0.001$ and $0.001 \leq p<0.05$, respectively. \colorbox{steel!30}{Blue bars} and \colorbox{red!30}{red bars} indicate positive and negative correlation, respectively. The number and bar length show strength.}
\label{tb:rq4_smbt_perf}

\centering

\setlength{\tabcolsep}{0.7mm}
\centering
\footnotesize
\begin{adjustbox}{width=\textwidth,center} 

\begin{tabular}{lc|c|c|c|c|c|c|c||c|c|c|c|c|c|c|c}


\toprule

\multirow{2}{*}{\textbf{System}}&\multicolumn{8}{c||}{\textbf{MAPE and Performance}}&\multicolumn{8}{c}{\textbf{$\mu$RD and Performance}}\\

\cmidrule{2-17}

&\textbf{\texttt{BOCA}}&\textbf{\texttt{ATConf}}&\textbf{\texttt{FLASH}}&\textbf{\texttt{OtterTune}}&\textbf{\texttt{ResTune}}&\textbf{\texttt{ROBOTune}}&\textbf{\texttt{SMAC}}&\textbf{\texttt{Tuneful}}&\textbf{\texttt{BOCA}}&\textbf{\texttt{ATConf}}&\textbf{\texttt{FLASH}}&\textbf{\texttt{OtterTune}}&\textbf{\texttt{ResTune}}&\textbf{\texttt{ROBOTune}}&\textbf{\texttt{SMAC}}&\textbf{\texttt{Tuneful}}\\

\midrule




\textsc{Brotli}&\phantom{$-$}\percent[p]{0.36}{1}{0.36$^\dagger$}{40}&$-$\percent[n]{0.53}{1}{0.53$^\dagger$}{60}&$-$\percent[n]{0.02}{1}{0.02$^{\,\,\,}$}{20}&\phantom{$-$}\percent[p]{0.26}{1}{0.26$^\dagger$}{40}&$-$\percent[n]{0.03}{1}{0.03$^{\,\,\,}$}{20}&\phantom{$-$}\percent[p]{0.30}{1}{0.30$^\dagger$}{40}&\phantom{$-$}\percent[p]{0.44}{1}{0.44$^\dagger$}{60}&$-$\percent[n]{0.05}{1}{0.05$^{\,\,\,}$}{20}&\phantom{$-$}\percent[p]{0.39}{1}{0.39$^\dagger$}{60}&$-$\percent[n]{0.09}{1}{0.09$^{\,\,\,}$}{40}&\phantom{$-$}\percent[p]{0.32}{1}{0.32$^\dagger$}{40}&\phantom{$-$}\percent[p]{0.17}{1}{0.17$^\star$}{40}&$-$\percent[n]{0.03}{1}{0.03$^{\,\,\,}$}{20}&\phantom{$-$}\percent[p]{0.21}{1}{0.21$^\dagger$}{40}&\phantom{$-$}\percent[p]{0.45}{1}{0.45$^\dagger$}{60}&$-$\percent[n]{0.03}{1}{0.03$^{\,\,\,}$}{20}\\
\bottomrule
\end{tabular}

\end{adjustbox}

\end{table*}

%% file: Tables/RQ3/rq3_smbt_eff.tex
\begin{table*}[t!]
\caption{Spearman correlations between model accuracy and its resulting tuning efficiency for sequential model-based tuners. The formats are the same as Table~\ref{tb:rq4_smbt_perf}.}
\label{tb:rq4_smbt_eff}

\centering

\setlength{\tabcolsep}{0.7mm}
\centering
\footnotesize
\begin{adjustbox}{width=\textwidth,center} 

\begin{tabular}{lc|c|c|c|c|c|c|c||c|c|c|c|c|c|c|c}


\toprule

\multirow{2}{*}{\textbf{System}}&\multicolumn{8}{c||}{\textbf{MAPE and Efficiency}}&\multicolumn{8}{c}{\textbf{$\mu$RD and Efficiency}}\\

\cmidrule{2-17}

&\textbf{\texttt{BOCA}}&\textbf{\texttt{ATConf}}&\textbf{\texttt{FLASH}}&\textbf{\texttt{OtterTune}}&\textbf{\texttt{ResTune}}&\textbf{\texttt{ROBOTune}}&\textbf{\texttt{SMAC}}&\textbf{\texttt{Tuneful}}&\textbf{\texttt{BOCA}}&\textbf{\texttt{ATConf}}&\textbf{\texttt{FLASH}}&\textbf{\texttt{OtterTune}}&\textbf{\texttt{ResTune}}&\textbf{\texttt{ROBOTune}}&\textbf{\texttt{SMAC}}&\textbf{\texttt{Tuneful}}\\

\midrule




\textsc{Brotli}&$-$\percent[n]{0.38}{1}{0.38$^\dagger$}{40}&\phantom{$-$}\percent[p]{0.17}{1}{0.17$^\star$}{40}&$-$\percent[n]{0.07}{1}{0.07$^{\,\,\,}$}{20}&$-$\percent[n]{0.26}{1}{0.26$^\dagger$}{40}&\phantom{$-$}\percent[p]{0.02}{1}{0.02$^{\,\,\,}$}{20}&\phantom{$-$}\percent[p]{0.27}{1}{0.27$^\dagger$}{40}&$-$\percent[n]{0.15}{1}{0.15$^\star$}{40}&\phantom{$-$}\percent[p]{0.00}{1}{0.00$^{\,\,\,}$}{20}&$-$\percent[n]{0.20}{1}{0.20$^\dagger$}{40}&$-$\percent[n]{0.01}{1}{0.01$^{\,\,\,}$}{20}&$-$\percent[n]{0.08}{1}{0.08$^{\,\,\,}$}{20}&$-$\percent[n]{0.05}{1}{0.05$^{\,\,\,}$}{20}&\phantom{$-$}\percent[p]{0.09}{1}{0.09$^{\,\,\,}$}{20}&\phantom{$-$}\percent[p]{0.36}{1}{0.36$^\dagger$}{40}&\phantom{$-$}\percent[p]{0.10}{1}{0.10$^{\,\,\,}$}{40}&\phantom{$-$}\percent[p]{0.07}{1}{0.07$^{\,\,\,}$}{20}\\
\bottomrule
\end{tabular}

\end{adjustbox}

\end{table*}

%% file: Tables/RQ3/rq3_bmbt_perf.tex
\begin{table*}[t!]
\caption{Spearman correlations between model accuracy and its resulting tuning performance for batch model-based tuners. The formats are the same as Table~\ref{tb:rq4_smbt_perf}.}
\label{tb:rq4_bmbt_eff}

\centering

\setlength{\tabcolsep}{0.7mm}
\centering
\footnotesize
\begin{adjustbox}{width=\textwidth,center} 

\begin{tabular}{lc|c|c|c|c|c|c|c|c||c|c|c|c|c|c|c|c|c}


\toprule

\multirow{2}{*}{\textbf{System}}&\multicolumn{9}{c||}{\textbf{MAPE and Performance}}&\multicolumn{9}{c}{\textbf{$\mu$RD and Performance}}\\

\cmidrule{2-19}

&\textbf{\texttt{BestConfig}}&\textbf{\texttt{Irace}}&\textbf{\texttt{GGA}}&\textbf{\texttt{ParamILS}}&\textbf{\texttt{Random}}&\textbf{\texttt{GA}}&\textbf{\texttt{SWAY}}&\textbf{\texttt{ConEx}}&\textbf{\texttt{Brute-force}}&\textbf{\texttt{BestConfig}}&\textbf{\texttt{Irace}}&\textbf{\texttt{GGA}}&\textbf{\texttt{ParamILS}}&\textbf{\texttt{Random}}&\textbf{\texttt{GA}}&\textbf{\texttt{SWAY}}&\textbf{\texttt{ConEx}}&\textbf{\texttt{Brute-force}}\\

\midrule




\textsc{Brotli}&-&-&-&-&-&-&-&-&\phantom{$-$}\percent[p]{0.19}{1}{0.19$^\star$}{40}&-&-&-&-&-&-&-&-&\phantom{$-$}\percent[p]{0.25}{1}{0.25$^\dagger$}{40}\\
\textsc{LLVM}&-&-&-&-&-&-&-&-&\phantom{$-$}\percent[p]{0.10}{1}{0.10$^{\,\,\,}$}{40}&-&-&-&-&-&-&-&-&\phantom{$-$}\percent[p]{0.14}{1}{0.14$^\star$}{40}\\
\textsc{Lrzip}&-&-&-&-&-&-&-&-&\phantom{$-$}\percent[p]{0.20}{1}{0.20$^\dagger$}{40}&-&-&-&-&-&-&-&-&\phantom{$-$}\percent[p]{0.17}{1}{0.17$^\star$}{40}\\
\textsc{xgboost}&-&-&-&-&-&-&-&-&\phantom{$-$}\percent[p]{0.24}{1}{0.24$^\dagger$}{40}&-&-&-&-&-&-&-&-&\phantom{$-$}\percent[p]{0.27}{1}{0.27$^\dagger$}{40}\\
\textsc{noc-CM-log}&-&-&-&-&-&-&-&-&\phantom{$-$}\percent[p]{0.51}{1}{0.51$^\dagger$}{60}&-&-&-&-&-&-&-&-&\phantom{$-$}\percent[p]{0.45}{1}{0.45$^\dagger$}{60}\\
\bottomrule
\end{tabular}

\end{adjustbox}

\end{table*}

%% file: Tables/RQ3/rq3_bmbt_eff.tex
\begin{table*}[t!]
\caption{Spearman correlations between model accuracy and its resulting tuning efficiency for batch model-based tuners. The formats are the same as Table~\ref{tb:rq4_smbt_perf}.}
\label{tb:rq4_bmbt_eff}

\centering

\setlength{\tabcolsep}{0.7mm}
\centering
\footnotesize
\begin{adjustbox}{width=\textwidth,center} 

\begin{tabular}{lc|c|c|c|c|c|c|c|c||c|c|c|c|c|c|c|c|c}


\toprule

\multirow{2}{*}{\textbf{System}}&\multicolumn{9}{c||}{\textbf{MAPE and Efficiency}}&\multicolumn{9}{c}{\textbf{$\mu$RD and Efficiency}}\\

\cmidrule{2-19}

&\textbf{\texttt{BestConfig}}&\textbf{\texttt{Irace}}&\textbf{\texttt{GGA}}&\textbf{\texttt{ParamILS}}&\textbf{\texttt{Random}}&\textbf{\texttt{GA}}&\textbf{\texttt{SWAY}}&\textbf{\texttt{ConEx}}&\textbf{\texttt{Brute-force}}&\textbf{\texttt{BestConfig}}&\textbf{\texttt{Irace}}&\textbf{\texttt{GGA}}&\textbf{\texttt{ParamILS}}&\textbf{\texttt{Random}}&\textbf{\texttt{GA}}&\textbf{\texttt{SWAY}}&\textbf{\texttt{ConEx}}&\textbf{\texttt{Brute-force}}\\

\midrule



\textsc{Brotli}&-&-&-&-&-&-&-&-&$-$\percent[n]{0.25}{1}{0.25$^\dagger$}{40}&-&-&-&-&-&-&-&-&$-$\percent[n]{0.16}{1}{0.16$^\star$}{40}\\
\textsc{LLVM}&-&-&-&-&-&-&-&-&\phantom{$-$}\percent[p]{0.01}{1}{0.01$^{\,\,\,}$}{20}&-&-&-&-&-&-&-&-&\phantom{$-$}\percent[p]{0.03}{1}{0.03$^{\,\,\,}$}{20}\\
\textsc{Lrzip}&-&-&-&-&-&-&-&-&\phantom{$-$}\percent[p]{0.01}{1}{0.01$^{\,\,\,}$}{20}&-&-&-&-&-&-&-&-&\phantom{$-$}\percent[p]{0.20}{1}{0.20$^\dagger$}{40}\\
\textsc{xgboost}&-&-&-&-&-&-&-&-&\phantom{$-$}\percent[p]{0.12}{1}{0.12$^\star$}{40}&-&-&-&-&-&-&-&-&\phantom{$-$}\percent[p]{0.07}{1}{0.07$^{\,\,\,}$}{20}\\
\textsc{noc-CM-log}&-&-&-&-&-&-&-&-&\phantom{$-$}\percent[p]{0.36}{1}{0.36$^\dagger$}{40}&-&-&-&-&-&-&-&-&\phantom{$-$}\percent[p]{0.32}{1}{0.32$^\dagger$}{40}\\
\bottomrule
\end{tabular}

\end{adjustbox}

\end{table*}

%% file: Figures/RQ5/smbt-mp.tex
\begin{tikzpicture}[scale=1] 

\begin{axis}[
    xbar, 
    bar width=.2cm,
     axis x line  = bottom,
    axis y line  = right,
    symbolic y coords={0,A,B,C,D,E,F,G,H,I,J,K}, 
    ytick=data, 
    width=3.5cm,
    height=5cm,
    legend image code/.code={%
        \draw[#1] (0cm,-0.1cm) rectangle (0.1cm,0.15cm);
    },
    legend style={at={(0.5,1.15)},
        anchor=north,
        legend columns=-1,
        font=\footnotesize,
        draw=none,
        fill=none,
        transpose legend, 
    },
    nodes near coords={\fontsize{7}{6.8}\selectfont \pgfmathprintnumber[fixed,precision=1,fixed zerofill]{\pgfplotspointmeta}$\pm$\pgfmathprintnumber[fixed,precision=1,fixed zerofill]{\label}},
    visualization depends on={\thisrow{y-min}-5 \as \offset},
    node near coords style={shift={(axis direction cs:\offset,0)},anchor=east},
    ytick align=inside,
    ylabel style={font=\small,yshift=-1em},
    grid style = {dashed},
    xmajorgrids=true, 
    xlabel={$+\Delta$},
    xmin=0, xmax=200, 
    xtick={50,100,150}, 
    yticklabels={0--10,10--20,20--30,30--40,40--50,50--60,60--70,70--80,80--90,90--100,\textgreater100}, 
    x dir=reverse,   
    every tick label/.append style={font=\footnotesize},
    xlabel style={font=\footnotesize}, 
    enlarge y limits={abs=0.2cm}, 
    ]

\addplot [fill=blue!50] plot [error bars/.cd,x dir=both,x explicit,error bar style={line join=mitre}] table[y=x,x=y,x error plus expr=\thisrow{y-max},x error minus expr=\thisrow{y-min},visualization depends on={\thisrow{y-max} \as \label}] {\mytable}; 


\end{axis}

\node at (2.48,3.6) {\footnotesize MAPE};

\begin{axis}[xshift=3cm,
    xbar, 
    bar width=.2cm,
     axis x line  = bottom,
    axis y line  = left,
    symbolic y coords={0,A,B,C,D,E,F,G,H,I,J,K}, 
       width=3.5cm,
    height=5cm,
    legend image code/.code={%
        \draw[#1] (0cm,-0.1cm) rectangle (0.1cm,0.15cm);
    },
    legend style={at={(0.5,1.15)},
        anchor=north,
        legend columns=-1,
        font=\footnotesize,
        draw=none,
        fill=none,
        transpose legend, 
    },
    nodes near coords={\fontsize{7}{6.8}\selectfont \pgfmathprintnumber[fixed,precision=1,fixed zerofill]{\pgfplotspointmeta}$\pm$\pgfmathprintnumber[fixed,precision=1,fixed zerofill]{\label}},
    visualization depends on={\thisrow{y-max}-5 \as \offset},
    node near coords style={shift={(axis direction cs:\offset,0)}},
    xtick=\empty,ytick=\empty,
    ytick align=inside,
    ylabel style={font=\small,yshift=-1em},
    grid style = {dashed},
    xmajorgrids=true, 
    xmin=0, xmax=200, 
    xtick={50,100,150}, 
    yticklabels = { },
     xlabel={$-\Delta$},
    every tick label/.append style={font=\footnotesize},
    xlabel style={font=\footnotesize}, 
    enlarge y limits={abs=0.2cm}, 
    ]

\addplot [fill=red!50]
    plot [error bars/.cd, x dir=both, x explicit,error bar style={},] 
    table[y=x,x=y,x error plus expr=\thisrow{y-max},x error minus expr=\thisrow{y-min},visualization depends on={\thisrow{y-max} \as \label}] {\mytablee}; 


\end{axis}

\end{tikzpicture}

%% file: Figures/RQ5/smbt-rp.tex
\begin{tikzpicture}[scale=1] 

\begin{axis}[
    xbar, 
    bar width=.2cm,
     axis x line  = bottom,
    axis y line  = right,
        symbolic y coords={0,A,B,C,D}, 
    ytick=data, 
    width=3.5cm,
    height=5cm,
    legend image code/.code={%
        \draw[#1] (0cm,-0.1cm) rectangle (0.1cm,0.15cm);
    },
    legend style={at={(0.5,1.15)},
        anchor=north,
        legend columns=-1,
        font=\footnotesize,
        draw=none,
        fill=none,
        transpose legend, 
    },
    nodes near coords={\fontsize{7}{6.8}\selectfont \pgfmathprintnumber[fixed,precision=1,fixed zerofill]{\pgfplotspointmeta}$\pm$\pgfmathprintnumber[fixed,precision=1,fixed zerofill]{\label}},
    visualization depends on={\thisrow{y-min}-1 \as \offset},
    node near coords style={shift={(axis direction cs:\offset,0)},anchor=east},
    ytick align=inside,
    ylabel style={font=\small,yshift=-1em},
    grid style = {dashed},
    xmajorgrids=true, 
    xlabel={$+\Delta$},
      xmin=0, xmax=50, 
    xtick={10,20,30,40}, 
    yticklabels={0--10,10--20,20--30,\textgreater30},  
    x dir=reverse,   
    every tick label/.append style={font=\footnotesize},
    xlabel style={font=\footnotesize}, 
    enlarge y limits={abs=0.2cm}, 
    ]

\addplot [fill=blue!50] plot [error bars/.cd,x dir=both,x explicit,error bar style={line join=mitre}] table[y=x,x=y,x error plus expr=\thisrow{y-max},x error minus expr=\thisrow{y-min},visualization depends on={\thisrow{y-max} \as \label}] {\rmytable}; 


\end{axis}

\node at (2.48,3.6) {\footnotesize $\mu$RD};
\node at (2.48,-0.25) {\footnotesize $\cdot 10^{-2}$};

\begin{axis}[xshift=3cm,
    xbar, 
    bar width=.2cm,
     axis x line  = bottom,
    axis y line  = left,
    symbolic y coords={0,A,B,C,D}, 
       width=3.5cm,
    height=5cm,
    legend image code/.code={%
        \draw[#1] (0cm,-0.1cm) rectangle (0.1cm,0.15cm);
    },
    legend style={at={(0.5,1.15)},
        anchor=north,
        legend columns=-1,
        font=\footnotesize,
        draw=none,
        fill=none,
        transpose legend, 
    },
    nodes near coords={\fontsize{7}{6.8}\selectfont \pgfmathprintnumber[fixed,precision=1,fixed zerofill]{\pgfplotspointmeta}$\pm$\pgfmathprintnumber[fixed,precision=1,fixed zerofill]{\label}},
    visualization depends on={\thisrow{y-max}-1 \as \offset},
    node near coords style={shift={(axis direction cs:\offset,0)}},
    xtick=\empty,ytick=\empty,
    ytick align=inside,
    ylabel style={font=\small,yshift=-1em},
    grid style = {dashed},
    xmajorgrids=true, 
    xmin=0, xmax=50, 
    xtick={10,20,30,40}, 
    yticklabels = { },
     xlabel={$-\Delta$},
    every tick label/.append style={font=\footnotesize},
    xlabel style={font=\footnotesize}, 
    enlarge y limits={abs=0.2cm}, 
    ]

\addplot [fill=red!50]
    plot [error bars/.cd, x dir=both, x explicit,error bar style={},] 
    table[y=x,x=y,x error plus expr=\thisrow{y-max},x error minus expr=\thisrow{y-min},visualization depends on={\thisrow{y-max} \as \label}] {\rmytablee}; 


\end{axis}

\end{tikzpicture}

%% file: Figures/RQ5/smbt-me.tex
\begin{tikzpicture}[scale=1] 

\begin{axis}[
    xbar, 
    bar width=.2cm,
     axis x line  = bottom,
    axis y line  = right,
    symbolic y coords={0,A,B,C,D,E,F,G,H,I,J,K}, 
    ytick=data, 
    width=3.5cm,
    height=5cm,
    legend image code/.code={%
        \draw[#1] (0cm,-0.1cm) rectangle (0.1cm,0.15cm);
    },
    legend style={at={(0.5,1.15)},
        anchor=north,
        legend columns=-1,
        font=\footnotesize,
        draw=none,
        fill=none,
        transpose legend, 
    },
    nodes near coords={\fontsize{7}{6.8}\selectfont \pgfmathprintnumber[fixed,precision=1,fixed zerofill]{\pgfplotspointmeta}$\pm$\pgfmathprintnumber[fixed,precision=1,fixed zerofill]{\label}},
    visualization depends on={\thisrow{y-min}-5 \as \offset},
    node near coords style={shift={(axis direction cs:\offset,0)},anchor=east},
    ytick align=inside,
    ylabel style={font=\small,yshift=-1em},
    grid style = {dashed},
    xmajorgrids=true, 
    xlabel={$+\Delta$},
    xmin=0, xmax=200, 
    xtick={50,100,150}, 
    yticklabels={0--10,10--20,20--30,30--40,40--50,50--60,60--70,70--80,80--90,90--100,\textgreater100}, 
    x dir=reverse,   
    every tick label/.append style={font=\footnotesize},
    xlabel style={font=\footnotesize}, 
    enlarge y limits={abs=0.2cm}, 
    ]

\addplot [fill=blue!50] plot [error bars/.cd,x dir=both,x explicit,error bar style={line join=mitre}] table[y=x,x=y,x error plus expr=\thisrow{y-max},x error minus expr=\thisrow{y-min},visualization depends on={\thisrow{y-max} \as \label}] {\mytablea}; 


\end{axis}

\node at (2.48,3.6) {\footnotesize MAPE};

\begin{axis}[xshift=3cm,
    xbar, 
    bar width=.2cm,
     axis x line  = bottom,
    axis y line  = left,
    symbolic y coords={0,A,B,C,D,E,F,G,H,I,J,K}, 
       width=3.5cm,
    height=5cm,
    legend image code/.code={%
        \draw[#1] (0cm,-0.1cm) rectangle (0.1cm,0.15cm);
    },
    legend style={at={(0.5,1.15)},
        anchor=north,
        legend columns=-1,
        font=\footnotesize,
        draw=none,
        fill=none,
        transpose legend, 
    },
    nodes near coords={\fontsize{7}{6.8}\selectfont \pgfmathprintnumber[fixed,precision=1,fixed zerofill]{\pgfplotspointmeta}$\pm$\pgfmathprintnumber[fixed,precision=1,fixed zerofill]{\label}},
    visualization depends on={\thisrow{y-max}-5 \as \offset},
    node near coords style={shift={(axis direction cs:\offset,0)}},
    xtick=\empty,ytick=\empty,
    ytick align=inside,
    ylabel style={font=\small,yshift=-1em},
    grid style = {dashed},
    xmajorgrids=true, 
    xmin=0, xmax=200, 
    xtick={50,100,150}, 
    yticklabels = { },
     xlabel={$-\Delta$},
    every tick label/.append style={font=\footnotesize},
    xlabel style={font=\footnotesize}, 
    enlarge y limits={abs=0.2cm}, 
    ]

\addplot [fill=red!50]
    plot [error bars/.cd, x dir=both, x explicit,error bar style={},] 
    table[y=x,x=y,x error plus expr=\thisrow{y-max},x error minus expr=\thisrow{y-min},visualization depends on={\thisrow{y-max} \as \label}] {\mytableaa}; 


\end{axis}

\end{tikzpicture}

%% file: Figures/RQ5/smbt-re.tex
\begin{tikzpicture}[scale=1] 

\begin{axis}[
    xbar, 
    bar width=.2cm,
     axis x line  = bottom,
    axis y line  = right,
        symbolic y coords={0,A,B,C,D}, 
    ytick=data, 
    width=3.5cm,
    height=5cm,
    legend image code/.code={%
        \draw[#1] (0cm,-0.1cm) rectangle (0.1cm,0.15cm);
    },
    legend style={at={(0.5,1.15)},
        anchor=north,
        legend columns=-1,
        font=\footnotesize,
        draw=none,
        fill=none,
        transpose legend, 
    },
    nodes near coords={\fontsize{7}{6.8}\selectfont \pgfmathprintnumber[fixed,precision=1,fixed zerofill]{\pgfplotspointmeta}$\pm$\pgfmathprintnumber[fixed,precision=1,fixed zerofill]{\label}},
    visualization depends on={\thisrow{y-min}-1 \as \offset},
    node near coords style={shift={(axis direction cs:\offset,0)},anchor=east},
    ytick align=inside,
    ylabel style={font=\small,yshift=-1em},
    grid style = {dashed},
    xmajorgrids=true, 
    xlabel={$+\Delta$},
      xmin=0, xmax=50, 
    xtick={10,20,30,40}, 
    yticklabels={0--10,10--20,20--30,\textgreater30},  
    x dir=reverse,   
    every tick label/.append style={font=\footnotesize},
    xlabel style={font=\footnotesize}, 
    enlarge y limits={abs=0.2cm}, 
    ]

\addplot [fill=blue!50] plot [error bars/.cd,x dir=both,x explicit,error bar style={line join=mitre}] table[y=x,x=y,x error plus expr=\thisrow{y-max},x error minus expr=\thisrow{y-min},visualization depends on={\thisrow{y-max} \as \label}] {\rmytablea}; 


\end{axis}

\node at (2.48,3.6) {\footnotesize $\mu$RD};
\node at (2.48,-0.25) {\footnotesize $\cdot 10^{-2}$};

\begin{axis}[xshift=3cm,
    xbar, 
    bar width=.2cm,
     axis x line  = bottom,
    axis y line  = left,
    symbolic y coords={0,A,B,C,D}, 
       width=3.5cm,
    height=5cm,
    legend image code/.code={%
        \draw[#1] (0cm,-0.1cm) rectangle (0.1cm,0.15cm);
    },
    legend style={at={(0.5,1.15)},
        anchor=north,
        legend columns=-1,
        font=\footnotesize,
        draw=none,
        fill=none,
        transpose legend, 
    },
    nodes near coords={\fontsize{7}{6.8}\selectfont \pgfmathprintnumber[fixed,precision=1,fixed zerofill]{\pgfplotspointmeta}$\pm$\pgfmathprintnumber[fixed,precision=1,fixed zerofill]{\label}},
    visualization depends on={\thisrow{y-max}-1 \as \offset},
    node near coords style={shift={(axis direction cs:\offset,0)}},
    xtick=\empty,ytick=\empty,
    ytick align=inside,
    ylabel style={font=\small,yshift=-1em},
    grid style = {dashed},
    xmajorgrids=true, 
    xmin=0, xmax=50, 
    xtick={10,20,30,40}, 
    yticklabels = { },
     xlabel={$-\Delta$},
    every tick label/.append style={font=\footnotesize},
    xlabel style={font=\footnotesize}, 
    enlarge y limits={abs=0.2cm}, 
    ]

\addplot [fill=red!50]
    plot [error bars/.cd, x dir=both, x explicit,error bar style={},] 
    table[y=x,x=y,x error plus expr=\thisrow{y-max},x error minus expr=\thisrow{y-min},visualization depends on={\thisrow{y-max} \as \label}] {\rmytableaa}; 


\end{axis}

\end{tikzpicture}

%% file: Figures/RQ5/bmbt-mp.tex
\begin{tikzpicture}[scale=1] 

\begin{axis}[
    xbar, 
    bar width=.2cm,
     axis x line  = bottom,
    axis y line  = right,
    symbolic y coords={0,A,B,C,D,E,F,G,H,I,J,K}, 
    ytick=data, 
    width=3.5cm,
    height=5cm,
    legend image code/.code={%
        \draw[#1] (0cm,-0.1cm) rectangle (0.1cm,0.15cm);
    },
    legend style={at={(0.5,1.15)},
        anchor=north,
        legend columns=-1,
        font=\footnotesize,
        draw=none,
        fill=none,
        transpose legend, 
    },
    nodes near coords={\fontsize{7}{6.8}\selectfont \pgfmathprintnumber[fixed,precision=1,fixed zerofill]{\pgfplotspointmeta}$\pm$\pgfmathprintnumber[fixed,precision=1,fixed zerofill]{\label}},
    visualization depends on={\thisrow{y-min}-5 \as \offset},
    node near coords style={shift={(axis direction cs:\offset,0)},anchor=east},
    ytick align=inside,
    ylabel style={font=\small,yshift=-1em},
    grid style = {dashed},
    xmajorgrids=true, 
    xlabel={$+\Delta$},
    xmin=0, xmax=200, 
    xtick={50,100,150}, 
    yticklabels={0--10,10--20,20--30,30--40,40--50,50--60,60--70,70--80,80--90,90--100,\textgreater100}, 
    x dir=reverse,   
    every tick label/.append style={font=\footnotesize},
    xlabel style={font=\footnotesize}, 
    enlarge y limits={abs=0.2cm}, 
    ]

\addplot [fill=blue!50] plot [error bars/.cd,x dir=both,x explicit,error bar style={line join=mitre}] table[y=x,x=y,x error plus expr=\thisrow{y-max},x error minus expr=\thisrow{y-min},visualization depends on={\thisrow{y-max} \as \label}] {\mytableb}; 


\end{axis}

\node at (2.48,3.6) {\footnotesize MAPE};

\begin{axis}[xshift=3cm,
    xbar, 
    bar width=.2cm,
     axis x line  = bottom,
    axis y line  = left,
    symbolic y coords={0,A,B,C,D,E,F,G,H,I,J,K}, 
       width=3.5cm,
    height=5cm,
    legend image code/.code={%
        \draw[#1] (0cm,-0.1cm) rectangle (0.1cm,0.15cm);
    },
    legend style={at={(0.5,1.15)},
        anchor=north,
        legend columns=-1,
        font=\footnotesize,
        draw=none,
        fill=none,
        transpose legend, 
    },
    nodes near coords={\fontsize{7}{6.8}\selectfont \pgfmathprintnumber[fixed,precision=1,fixed zerofill]{\pgfplotspointmeta}$\pm$\pgfmathprintnumber[fixed,precision=1,fixed zerofill]{\label}},
    visualization depends on={\thisrow{y-max}-5 \as \offset},
    node near coords style={shift={(axis direction cs:\offset,0)}},
    xtick=\empty,ytick=\empty,
    ytick align=inside,
    ylabel style={font=\small,yshift=-1em},
    grid style = {dashed},
    xmajorgrids=true, 
    xmin=0, xmax=200, 
    xtick={50,100,150}, 
    yticklabels = { },
     xlabel={$-\Delta$},
    every tick label/.append style={font=\footnotesize},
    xlabel style={font=\footnotesize}, 
    enlarge y limits={abs=0.2cm}, 
    ]

\addplot [fill=red!50]
    plot [error bars/.cd, x dir=both, x explicit,error bar style={},] 
    table[y=x,x=y,x error plus expr=\thisrow{y-max},x error minus expr=\thisrow{y-min},visualization depends on={\thisrow{y-max} \as \label}] {\mytablebb}; 


\end{axis}

\end{tikzpicture}

%% file: Figures/RQ5/bmbt-rp.tex
\begin{tikzpicture}[scale=1] 

\begin{axis}[
    xbar, 
    bar width=.2cm,
     axis x line  = bottom,
    axis y line  = right,
        symbolic y coords={0,A,B,C,D}, 
    ytick=data, 
    width=3.5cm,
    height=5cm,
    legend image code/.code={%
        \draw[#1] (0cm,-0.1cm) rectangle (0.1cm,0.15cm);
    },
    legend style={at={(0.5,1.15)},
        anchor=north,
        legend columns=-1,
        font=\footnotesize,
        draw=none,
        fill=none,
        transpose legend, 
    },
    nodes near coords={\fontsize{7}{6.8}\selectfont \pgfmathprintnumber[fixed,precision=1,fixed zerofill]{\pgfplotspointmeta}$\pm$\pgfmathprintnumber[fixed,precision=1,fixed zerofill]{\label}},
    visualization depends on={\thisrow{y-min}-1 \as \offset},
    node near coords style={shift={(axis direction cs:\offset,0)},anchor=east},
    ytick align=inside,
    ylabel style={font=\small,yshift=-1em},
    grid style = {dashed},
    xmajorgrids=true, 
    xlabel={$+\Delta$},
      xmin=0, xmax=50, 
    xtick={10,20,30,40}, 
    yticklabels={0--10,10--20,20--30,\textgreater30},  
    x dir=reverse,   
    every tick label/.append style={font=\footnotesize},
    xlabel style={font=\footnotesize}, 
    enlarge y limits={abs=0.2cm}, 
    ]

\addplot [fill=blue!50] plot [error bars/.cd,x dir=both,x explicit,error bar style={line join=mitre}] table[y=x,x=y,x error plus expr=\thisrow{y-max},x error minus expr=\thisrow{y-min},visualization depends on={\thisrow{y-max} \as \label}] {\rmytableb}; 


\end{axis}

\node at (2.48,3.6) {\footnotesize $\mu$RD};
\node at (2.48,-0.25) {\footnotesize $\cdot 10^{-2}$};

\begin{axis}[xshift=3cm,
    xbar, 
    bar width=.2cm,
     axis x line  = bottom,
    axis y line  = left,
    symbolic y coords={0,A,B,C,D}, 
       width=3.5cm,
    height=5cm,
    legend image code/.code={%
        \draw[#1] (0cm,-0.1cm) rectangle (0.1cm,0.15cm);
    },
    legend style={at={(0.5,1.15)},
        anchor=north,
        legend columns=-1,
        font=\footnotesize,
        draw=none,
        fill=none,
        transpose legend, 
    },
    nodes near coords={\fontsize{7}{6.8}\selectfont \pgfmathprintnumber[fixed,precision=1,fixed zerofill]{\pgfplotspointmeta}$\pm$\pgfmathprintnumber[fixed,precision=1,fixed zerofill]{\label}},
    visualization depends on={\thisrow{y-max}-1 \as \offset},
    node near coords style={shift={(axis direction cs:\offset,0)}},
    xtick=\empty,ytick=\empty,
    ytick align=inside,
    ylabel style={font=\small,yshift=-1em},
    grid style = {dashed},
    xmajorgrids=true, 
    xmin=0, xmax=50, 
    xtick={10,20,30,40}, 
    yticklabels = { },
     xlabel={$-\Delta$},
    every tick label/.append style={font=\footnotesize},
    xlabel style={font=\footnotesize}, 
    enlarge y limits={abs=0.2cm}, 
    ]

\addplot [fill=red!50]
    plot [error bars/.cd, x dir=both, x explicit,error bar style={},] 
    table[y=x,x=y,x error plus expr=\thisrow{y-max},x error minus expr=\thisrow{y-min},visualization depends on={\thisrow{y-max} \as \label}] {\rmytablebb}; 


\end{axis}

\end{tikzpicture}

%% file: Figures/RQ5/bmbt-me.tex
\begin{tikzpicture}[scale=1] 

\begin{axis}[
    xbar, 
    bar width=.2cm,
     axis x line  = bottom,
    axis y line  = right,
    symbolic y coords={0,A,B,C,D,E,F,G,H,I,J,K}, 
    ytick=data, 
    width=3.5cm,
    height=5cm,
    legend image code/.code={%
        \draw[#1] (0cm,-0.1cm) rectangle (0.1cm,0.15cm);
    },
    legend style={at={(0.5,1.15)},
        anchor=north,
        legend columns=-1,
        font=\footnotesize,
        draw=none,
        fill=none,
        transpose legend, 
    },
    nodes near coords={\fontsize{7}{6.8}\selectfont \pgfmathprintnumber[fixed,precision=1,fixed zerofill]{\pgfplotspointmeta}$\pm$\pgfmathprintnumber[fixed,precision=1,fixed zerofill]{\label}},
    visualization depends on={\thisrow{y-min}-5 \as \offset},
    node near coords style={shift={(axis direction cs:\offset,0)},anchor=east},
    ytick align=inside,
    ylabel style={font=\small,yshift=-1em},
    grid style = {dashed},
    xmajorgrids=true, 
    xlabel={$+\Delta$},
    xmin=0, xmax=200, 
    xtick={50,100,150}, 
    yticklabels={0--10,10--20,20--30,30--40,40--50,50--60,60--70,70--80,80--90,90--100,\textgreater100}, 
    x dir=reverse,   
    every tick label/.append style={font=\footnotesize},
    xlabel style={font=\footnotesize}, 
    enlarge y limits={abs=0.2cm}, 
    ]

\addplot [fill=blue!50] plot [error bars/.cd,x dir=both,x explicit,error bar style={line join=mitre}] table[y=x,x=y,x error plus expr=\thisrow{y-max},x error minus expr=\thisrow{y-min},visualization depends on={\thisrow{y-max} \as \label}] {\mytablec}; 


\end{axis}

\node at (2.48,3.6) {\footnotesize MAPE};

\begin{axis}[xshift=3cm,
    xbar, 
    bar width=.2cm,
     axis x line  = bottom,
    axis y line  = left,
    symbolic y coords={0,A,B,C,D,E,F,G,H,I,J,K}, 
       width=3.5cm,
    height=5cm,
    legend image code/.code={%
        \draw[#1] (0cm,-0.1cm) rectangle (0.1cm,0.15cm);
    },
    legend style={at={(0.5,1.15)},
        anchor=north,
        legend columns=-1,
        font=\footnotesize,
        draw=none,
        fill=none,
        transpose legend, 
    },
    nodes near coords={\fontsize{7}{6.8}\selectfont \pgfmathprintnumber[fixed,precision=1,fixed zerofill]{\pgfplotspointmeta}$\pm$\pgfmathprintnumber[fixed,precision=1,fixed zerofill]{\label}},
    visualization depends on={\thisrow{y-max}-5 \as \offset},
    node near coords style={shift={(axis direction cs:\offset,0)}},
    xtick=\empty,ytick=\empty,
    ytick align=inside,
    ylabel style={font=\small,yshift=-1em},
    grid style = {dashed},
    xmajorgrids=true, 
    xmin=0, xmax=200, 
    xtick={50,100,150}, 
    yticklabels = { },
     xlabel={$-\Delta$},
    every tick label/.append style={font=\footnotesize},
    xlabel style={font=\footnotesize}, 
    enlarge y limits={abs=0.2cm}, 
    ]

\addplot [fill=red!50]
    plot [error bars/.cd, x dir=both, x explicit,error bar style={},] 
    table[y=x,x=y,x error plus expr=\thisrow{y-max},x error minus expr=\thisrow{y-min},visualization depends on={\thisrow{y-max} \as \label}] {\mytablecc}; 


\end{axis}

\end{tikzpicture}

%% file: Figures/RQ5/bmbt-re.tex
\begin{tikzpicture}[scale=1] 

\begin{axis}[
    xbar, 
    bar width=.2cm,
     axis x line  = bottom,
    axis y line  = right,
        symbolic y coords={0,A,B,C,D}, 
    ytick=data, 
    width=3.5cm,
    height=5cm,
    legend image code/.code={%
        \draw[#1] (0cm,-0.1cm) rectangle (0.1cm,0.15cm);
    },
    legend style={at={(0.5,1.15)},
        anchor=north,
        legend columns=-1,
        font=\footnotesize,
        draw=none,
        fill=none,
        transpose legend, 
    },
    nodes near coords={\fontsize{7}{6.8}\selectfont \pgfmathprintnumber[fixed,precision=1,fixed zerofill]{\pgfplotspointmeta}$\pm$\pgfmathprintnumber[fixed,precision=1,fixed zerofill]{\label}},
    visualization depends on={\thisrow{y-min}-1 \as \offset},
    node near coords style={shift={(axis direction cs:\offset,0)},anchor=east},
    ytick align=inside,
    ylabel style={font=\small,yshift=-1em},
    grid style = {dashed},
    xmajorgrids=true, 
    xlabel={$+\Delta$},
      xmin=0, xmax=50, 
    xtick={10,20,30,40}, 
    yticklabels={0--10,10--20,20--30,\textgreater30},  
    x dir=reverse,   
    every tick label/.append style={font=\footnotesize},
    xlabel style={font=\footnotesize}, 
    enlarge y limits={abs=0.2cm}, 
    ]

\addplot [fill=blue!50] plot [error bars/.cd,x dir=both,x explicit,error bar style={line join=mitre}] table[y=x,x=y,x error plus expr=\thisrow{y-max},x error minus expr=\thisrow{y-min},visualization depends on={\thisrow{y-max} \as \label}] {\rmytablec}; 


\end{axis}

\node at (2.48,3.6) {\footnotesize $\mu$RD};
\node at (2.48,-0.25) {\footnotesize $\cdot 10^{-2}$};

\begin{axis}[xshift=3cm,
    xbar, 
    bar width=.2cm,
     axis x line  = bottom,
    axis y line  = left,
    symbolic y coords={0,A,B,C,D}, 
       width=3.5cm,
    height=5cm,
    legend image code/.code={%
        \draw[#1] (0cm,-0.1cm) rectangle (0.1cm,0.15cm);
    },
    legend style={at={(0.5,1.15)},
        anchor=north,
        legend columns=-1,
        font=\footnotesize,
        draw=none,
        fill=none,
        transpose legend, 
    },
    nodes near coords={\fontsize{7}{6.8}\selectfont \pgfmathprintnumber[fixed,precision=1,fixed zerofill]{\pgfplotspointmeta}$\pm$\pgfmathprintnumber[fixed,precision=1,fixed zerofill]{\label}},
    visualization depends on={\thisrow{y-max}-1 \as \offset},
    node near coords style={shift={(axis direction cs:\offset,0)}},
    xtick=\empty,ytick=\empty,
    ytick align=inside,
    ylabel style={font=\small,yshift=-1em},
    grid style = {dashed},
    xmajorgrids=true, 
    xmin=0, xmax=50, 
    xtick={10,20,30,40}, 
    yticklabels = { },
     xlabel={$-\Delta$},
    every tick label/.append style={font=\footnotesize},
    xlabel style={font=\footnotesize}, 
    enlarge y limits={abs=0.2cm}, 
    ]

\addplot [fill=red!50]
    plot [error bars/.cd, x dir=both, x explicit,error bar style={},] 
    table[y=x,x=y,x error plus expr=\thisrow{y-max},x error minus expr=\thisrow{y-min},visualization depends on={\thisrow{y-max} \as \label}] {\rmytablecc}; 


\end{axis}

\end{tikzpicture}

%% file: discussion.tex
\section{Discussion: A New Perspective of Configuration Landscape Analysis}
\label{sec:discussion}

To understand the reasons behind some of the most surprising results obtained from the study, we analyze examples from the perspective of fitness landscape analysis~\cite{DBLP:series/sci/PitzerA12,DBLP:conf/seams/Chen22} over the configuration space that provides insights into the spatial information and ``difficulty'' of the tuning problems.

Since the model guides the tuner, essentially we transit from tuning in the real configuration landscape of the system to a landscape emulated by the model, hence comparing how well the model-built landscape resembles the real one can provide much richer information beyond the accuracy metrics. To that end, we use both intuitive landscape visualization and metrics in the analysis.

In addition to the above, we have also provide a theoretical hypothesis as to what is the key barrier that make the model less effective for configuration tuning.

\subsection{Landscape Visualization}

In this work, we leverage Heatmap, which is one of the most straightforward methods, to visualize the configuration landscape~\cite{DBLP:journals/ec/KerschkeWPGDTE19}. Since the landscape is naturally multi-dimensional, we use multi-dimensional scaling (MDS) to convert all the dimensions of configuration options by creating two new dimensions. The benefit of MDS is that it preserves the global distance between each pair of data samples, which is important for analyzing the overall structure of the landscape. As such, in a configuration landscape Heatmap, the two axes represent the reduced dimensions while the color denotes the performance values. We also normalize the performance value in the Heatmap for each of the real and model-built landscapes to make the similarity/difference more intuitive. This is because a model would still emulate an excellent landscape compared with the real one as long as their relative structure and distribution are similar, even though the absolute configurations' performance is rather different. 


\subsection{Landscape Metrics}

\subsubsection{Fitness Distance Correlation (FDC)} 

In general, FDC assesses how close the relation between performance value and distance to the nearest optimum in the configuration space~\cite{DBLP:conf/icga/JonesF95}, which quantifies the overall guidance that the configuration landscape can offer for a tuner~\cite{DBLP:journals/tsmc/TavaresPC08}. Formally, FDC (denoted as $\varrho$) is computed as:
\begin{align}
	 \varrho(f,d) = {1 \over {\sigma_f \sigma_d p}} \sum^p_{i=1} (f_i - \overline{f}) (d_i - \overline{d})
	 \label{Eq:fdc}
\end{align}
where $p$ is the number of configurations measured. $f_i = f(x_i)$ is the performance value for the $i$th configuration and $d_i = d_{opt}(x_i)$ is the shortest Hamming distance of such a configuration to its nearest global optimum. $\overline{f}$ ($\overline{d}$)
and $\sigma_f$ ($\sigma_d$) are the mean and standard deviation, respectively.

Intuitively, FDC is the Pearson correlation between $f$ and $d$, ranging within $[-1,1]$ where $1$ and $-1$ imply the strongest monotonically positive and negative correlation, respectively; $0$ indicates no correlation can be detected. For a performance that needs minimization, when $0 < \varrho \leq 1$, the configuration turns better (smaller performance value) as the shortest distance to a global optimum reduces. This means that, when FDC becomes closer to $1$, the guidance provided to a tuner is stronger and it is more likely to exist a path towards a global optimum via configurations with decreasing performance values, making the tuning potentially easier. In contrast, $-1 \leq \varrho < 0$ indicates the opposite.

\subsubsection{Deviation between Predicted and True Global Optimum} 

Since a modeled landscape might result in a different global optimum(a) compared with the real one, we need to understand to what extent the guidance (by FDC) in the landscape of a model makes sense with respect to the real landscape. To this end, we assess the deviation between the global optimum estimated by the model and the true global optimum in the real landscape by means of their Hamming distance and performance gap, denoted as $\Delta d$ and $\Delta p$ respectively. If the real performance of the predicted global optimum configuration, as identified in the modeled landscape, is closer to that of the true global optimum in the real landscape on both configuration and performance, then we know that guiding the tuner toward the predicted global optimum in the modeled landscape is likely to be more effective. 


\subsubsection{Correlation Length} 

Comparing FDC and the real performance of the global optimum still cannot account for the local paths between local optima, i.e., the ruggedness of the landscape. As a result, we additionally measure the Correlation Length ($\ell$)~\cite{stadler1996landscapes}, calculated as below (the notations are the same as that for Equation~\ref{Eq:fdc}):
\begin{align}
		\ell(p,s) = - (\ln|{1 \over {\sigma_f^2 (p-s)}} \sum^{p-s}_{i=1} (f_i - \overline{f}) (f_{i+s} - \overline{f})|)^{-1} 
		 \label{Eq:rugg}
\end{align}
$\ell(p,s)$ is essentially a normalized autocorrelation function of neighboring configurations' performance values explored, in which $s$ denotes the Hamming distance of a neighbor and $p$ is the total number of measured configurations. We set $s$ according to the nearest possible neighboring configurations measured in our experiments. The higher the value of $\ell$, the smoother the landscape, as the performance of adjacently sampled configurations is more correlated. Otherwise, it indicates a more rugged surface~\cite{stadler1996landscapes}, which means it the easier to trap a tuner.

\subsubsection{What Constitutes a Good Model-emulated Landscape?}

Given the above metrics, an excellent emulation of the configuration landscape created by a model should have close FDC and deviation of the predicted global optimum to the true one with respect to the real landscape, hence it provides similar guidance to the tuners. At the same time, the correlation length should be higher, because this leads to a smoother landscape which can be easier for a tuner to explore. It is important to note that a higher correlation length in an emulated landscape is only useful when the guidance it provides is also similar to that of the real landscape, or otherwise the tuning could be directed to the wrong optimum region, even though the entire landscape might be easier for a tuner to tackle.

\subsection{Observational Analysis}

\subsubsection{Why is the Model Useful (Useless) to the Tuning Quality?}



It is worth noting that, while models are generally not effective in tuning the quality of batch model-based tuners under the maximum budget we consider in this work, they can still reduce the overhead significantly, which could be helpful for certain scenarios. For example, when the available resource is rather limited and only very little configuration can be measured, then, since in \textbf{RQ4}, we show that the model accuracy does not strongly correlate with the tuning results, a much less accurate model (with less data) might still lead to similar tuning quality to the case when the training data is more sufficient. This, in contrast, would much more significantly degrade the quality of the model-free counterpart if they were to run under the same budget.

\begin{figure}[t!]
\centering
\subfloat[Initial \texttt{DT} landscape (seq.)\\FDC: $\varrho=-0.786$ \\Hamming distance: $\Delta d=12$\\Performance gap: $\Delta p=240.3$\\Correlation length: $\ell=0.357$]{\includegraphics[width=0.49\columnwidth]{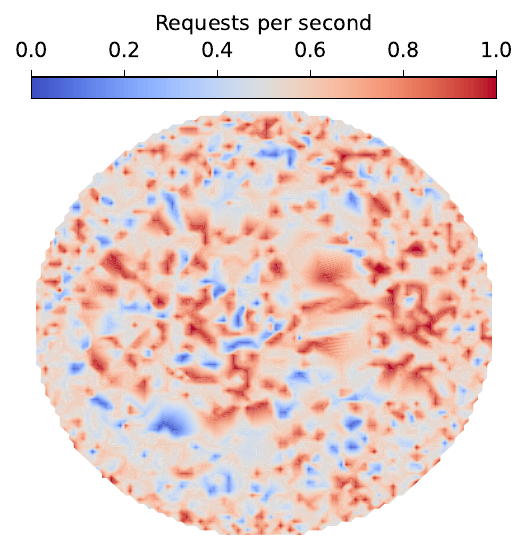}}
~\hfill
\subfloat[Final \texttt{DT} landscape (seq.)\\FDC: $\varrho=-0.046$\\Hamming distance: $\Delta d=11$\\Performance gap: $\Delta p=55.4$\\Correlation length: $\ell=0.312$]{\includegraphics[width=0.49\columnwidth]{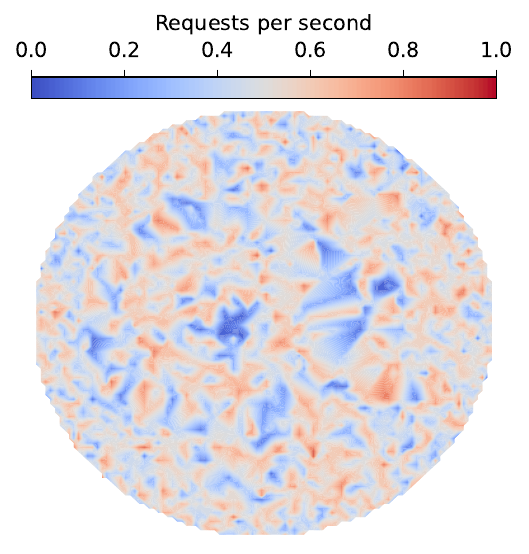}}

\subfloat[\texttt{DT} landscape (batch)\\ FDC: $\varrho=-0.107$\\Hamming distance: $\Delta d=12$\\Performance gap: $\Delta p=168.2$\\Correlation length: $\ell=0.213$]{\includegraphics[width=0.49\columnwidth]{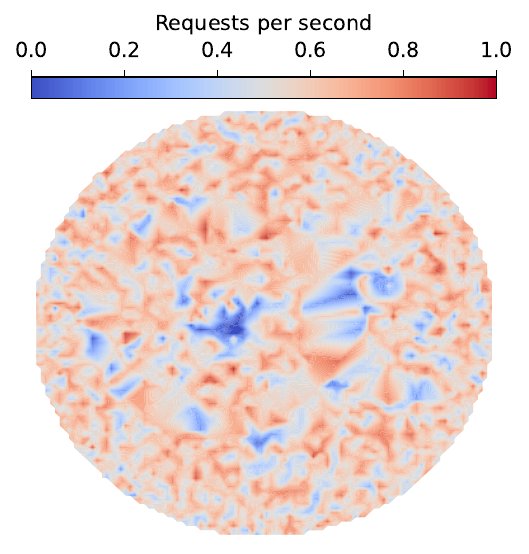}}
~\hfill
\subfloat[Real landscape\\FDC: $\varrho=-0.002$\\Hamming distance: $\Delta d=0$\\Performance gap: $\Delta p=0$\\Correlation length: $\ell=0.270$]{\includegraphics[width=0.49\columnwidth]{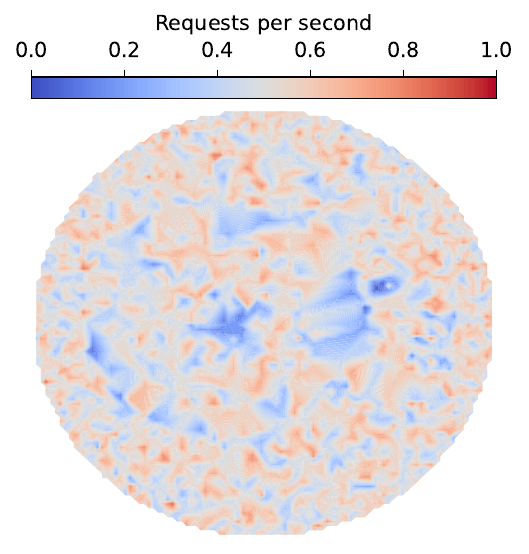}}

\caption{Configuration landscapes derived from the model for the sequential model-based tuner (a and b), the model for the batch model-based tuner (c), and the real configuration space (d) for \textsc{Tomcat}.} 
\label{fig:dis1}
\end{figure}

To understand why the models are more helpful for the sequential model-based tuners over the batch counterparts from \textbf{RQ1}, Figure~\ref{fig:dis1} illustrates a typical example using Heatmaps based on the two MDS reduced dimensions for \textsc{Tomcat} under \texttt{DT}. Since \textbf{RQ3} and \textbf{RQ4} reveal that the model accuracy can be misleading, we compare how well the model can emulate the real configuration landscape. We see that visually, the initial model under a sequential model-based tuner (i.e., only be trained by the hot-start samples) is less similar to the real landscape compared with the model trained by a batch (Figure~\ref{fig:dis1}a and Figure~\ref{fig:dis1}d vs. Figure~\ref{fig:dis1}c and Figure~\ref{fig:dis1}d). This can also be reflected in the metrics: compared with the batch model, the initial model is more distant from the real landscape on nearly all metrics. 

However, when the model under sequential model-based tuners is progressively updated, the landscape it emulated would eventually become much closer to the real one: If we compare Figure~\ref{fig:dis1}b and Figure~\ref{fig:dis1}d, we observe that now, the Heatmap of the final model is clearly more similar to the real landscape than the batch counterpart, especially on the peaks and troughs, which is also the case over all metrics (despite that they are trained under similar training size, i.e., 282 vs. 288). In particular, the closer FDC and deviation of global optimum imply that the landscape of the final model provides tuning guidance that is closer to the real one compared with its batch counterpart. In the meantime, we see that the final model has a higher correlation length than the real landscape while the batch model emulates lower. This suggests that the progressively updated final model exhibits less severe local optimum issues and hence is easier for the tuners (especially those that leverage local search), at the same time, the guidance it provides is also more similar to the real landscape. Therefore, the progressively updated model under sequential model-based tuners can emulate the tuning landscape better (and easier) than its batch counterpart, explaining why the models tend to be more useful for the sequential model-based tuners but tend to be useless for the batch ones.

\begin{figure}[t!]
\centering

\subfloat[\texttt{HINNPerf} landscape\\$\mu$RD: $0.142$\\FDC: $\varrho=0.57$\\Hamming distance: $\Delta d=3$\\Performance gap: $\Delta p=1.84$\\Correlation length: $\ell=1.760$]{\includegraphics[width=0.49\columnwidth]{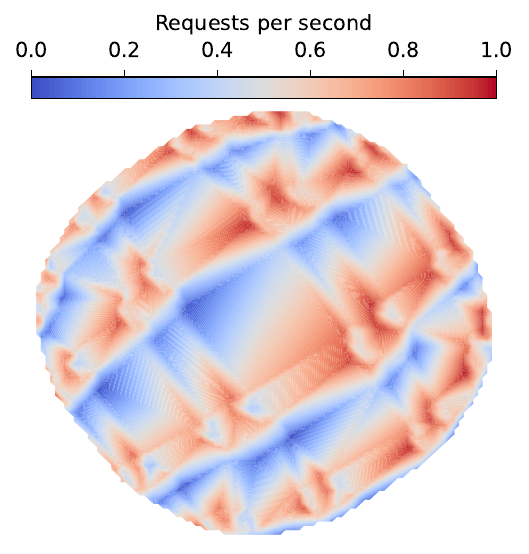}}
~\hfill
\subfloat[\texttt{DT} landscape\\$\mu$RD: $0.068$\\FDC: $\varrho=0.56$\\Hamming distance: $\Delta d=2$\\Performance gap: $\Delta p=1.86$\\Correlation length: $\ell=1.741$]{\includegraphics[width=0.49\columnwidth]{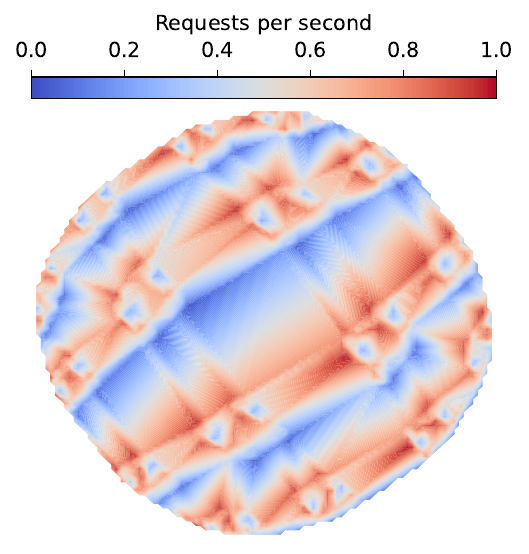}}

\caption{Configuration landscapes derived from the models for a batch model-based tuner for \textsc{LLVM}, showing a large $\mu$RD improvement (from a to b) might not lead to significant deviation on the modeled configuration landscape.} 
\label{fig:dis2}
\end{figure}

\subsubsection{Why Better Model Accuracy Does Not Always Lead to Superior Tuning Quality?}

While it is counter-intuitive to observe the results for \textbf{RQ3} and \textbf{RQ4}, they are indeed possible from the perspective of landscape analysis. Figure~\ref{fig:dis2} shows a case of \textsc{LLVM}, in which we see that the $\mu$RD can be improved from 0.142 to 0.068 by simply changing \texttt{HINNPerf} to \texttt{DT} on a batch model-based tuner (from Figure~\ref{fig:dis2}a to Figure~\ref{fig:dis2}b)---a $2.09\times$ improvement. However, the tuning quality when pairing those two models with different tuners is generally indistinguishable. This is because, visually, we see that their landscape appears to be quite similar. This is also reflected by the landscape metrics, e.g., there is nearly no change in the FDC and correlation length. The above is a typical example of the key reason that the changes in model accuracy often do not strongly correlate with the changes in tuning quality.

\begin{figure}[t!]
\centering

\subfloat[Final \texttt{GP}\\MAPE: $25.7\%$\\FDC: $\varrho=0.0792$\\Hamming dis.: $\Delta d=5$\\Perf. gap: $\Delta p=37.763$\\Corr. length: $\ell=1.206$]{\includegraphics[width=0.33\columnwidth]{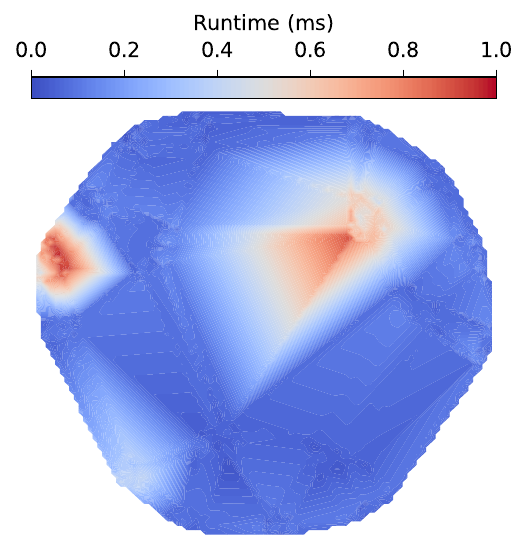}}
~\hfill  
\subfloat[Final \texttt{HINNPerf}\\MAPE: $47.4\%$\\FDC: $\varrho=0.158$\\Hamming dis.: $\Delta d=7$\\Perf. gap: $\Delta p=23.892$\\Corr. length: $\ell=1.484$]{\includegraphics[width=0.33\columnwidth]{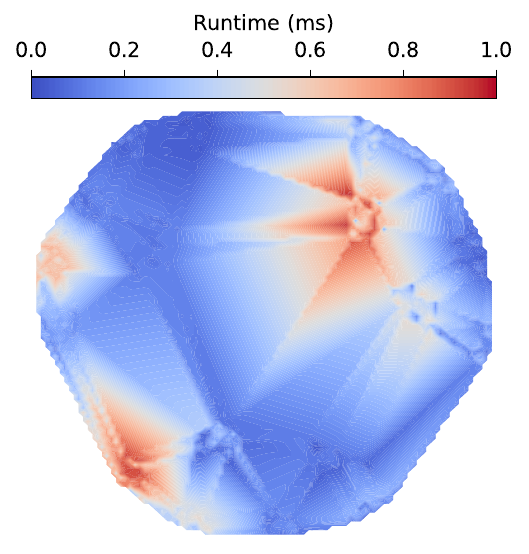}}
~\hfill
\subfloat[Real landscape\\MAPE: $0$\\FDC: $\varrho=0.174$\\Hamming dis.: $\Delta d=0$\\Perf. gap: $\Delta p=0$\\Corr. length: $\ell=1.400$]{\includegraphics[width=0.33\columnwidth]{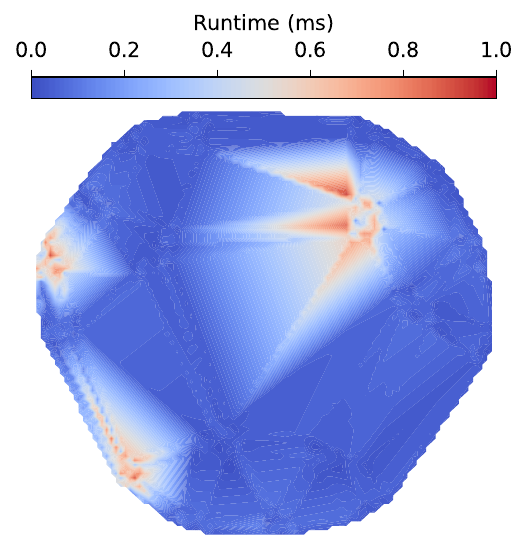}}

\caption{Configuration landscapes derived from the final models for a sequential model-based tuner for \textsc{hsmgp}, showing a large MAPE improvement (from b to a) might also lead to a significantly more distant configuration landscape modeled compared with the real one (c).} 
\label{fig:dis3}
\end{figure}

One very surprising observation is that better model accuracy might negatively correlate with the tuning results. Figure~\ref{fig:dis3} shows one of such common examples. Here, for a sequential model-based tuner that tunes \textsc{HSMGP}, the MAPE can be improved from 47.4\% with $1.84\times$ to 25.7\% by using \texttt{GP} (Figure~\ref{fig:dis3}a) instead of \texttt{HINNPerf} (Figure~\ref{fig:dis3}b), but the tuning quality in the former is often significantly worsened over different tuners. This is because both the visual interpretation, patterns of local optima regions, and nearly all the metrics of landscape indicate that \texttt{HINNPerf} emulates the real landscape (Figure~\ref{fig:dis3}c) better than the \texttt{GP}. In particular, \texttt{HINNPerf}'s FDC and deviation between predicted and true global optimum are closer to those of \texttt{GP}, meaning that the \texttt{HINNPerf} landscape offers more effective guidance in the tuning. The higher correlation length in \texttt{HINNPerf} landscape than the real landscape and that of the \texttt{GP} also suggests that the former can also relieve some tuning difficulties caused by the local optima on the real landscape (as it is smoother) while providing closed guidance towards the true global optimum. \texttt{GP}, in contrast, makes the landscape harder by, e.g., creating unnecessary local optima. 


All the above examples imply one rule: the model accuracy is not a reliable indication of the emulated landscape within which the tuner would explore, hence cannot properly reflect the better or worse when comparing the tuners.

\subsubsection{Why A Smaller Accuracy Change is Needed to Significantly Influence the Tuning Under Higher Accuracy?}

While \textbf{RQ5} reveals how much accuracy changes are needed to significantly improve the tuning quality, it additionally demonstrates an interesting pattern: worse accuracy needs a larger change to do so compared with better accuracy. This can also be explained from the perspective of configuration landscape analysis.  

\begin{figure}[t!]
\centering

\subfloat[\texttt{DT} landscape\\MAPE: $0.8\%$\\FDC: $\varrho=0.32$\\Hamming distance: $\Delta d=3.7$\\Performance gap: $\Delta p=13.80$\\Correlation length: $\ell=1.54$]{\includegraphics[width=0.49\columnwidth]{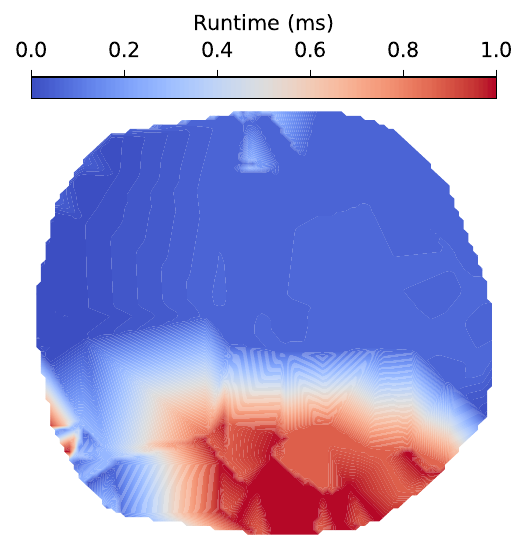}}
~\hfill
\subfloat[\texttt{DECART} landscape\\MAPE: $0.7\%$\\FDC: $\varrho=0.58$\\Hamming distance: $\Delta d=3.5$\\Performance gap: $\Delta p=1.09$\\Correlation length: $\ell=1.77$]{\includegraphics[width=0.49\columnwidth]{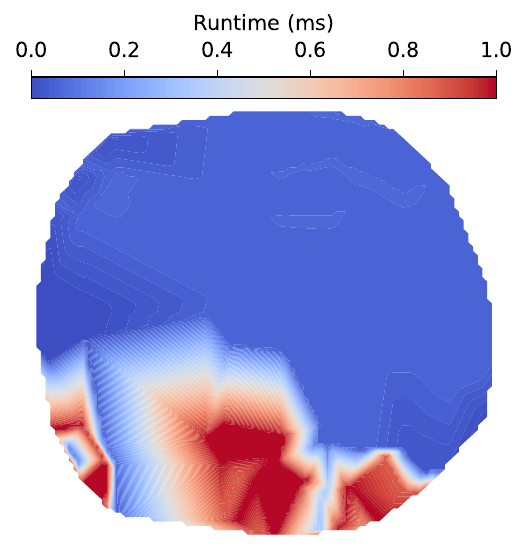}}

\caption{Configuration landscapes derived from the models for a batch model-based tuner on \textsc{HSQLDB}, showing a small MAPE reduction in an already excellent accuracy (from a to b) might lead to significant deviation on the modeled configuration landscape, which results in dramatic improvement on the tuning quality.} 
\label{fig:dis4}
\end{figure}

Figure~\ref{fig:dis4} is a common example from \textbf{RQ5} where the MAPE of the two models is excellent, i.e., both are less than 1\%. Here, when switching from \texttt{DT} (Figure~\ref{fig:dis4}a) to \texttt{DECART} (Figure~\ref{fig:dis4}b), the MAPE only decreases 0.1\%, but there is a statistically significant improvement in the tuning quality thereof. In fact, although the change exhibited in accuracy is trivial, we see that the deviation in the landscape visualization, as well as the metrics, are clear and substantial, especially on the FDC and performance gap. This explains why a rather small change on a model with some relatively good accuracy might still lead to considerable tuning quality enrichment. 

\begin{figure}[t!]
\centering

\subfloat[\texttt{LR} landscape\\MAPE: $111.8\%$\\FDC: $\varrho=0.73$\\Hamming distance: $\Delta d=6.9$\\Performance gap: $\Delta p=0.06$\\Correlation length: $\ell=1.73$]{\includegraphics[width=0.49\columnwidth]{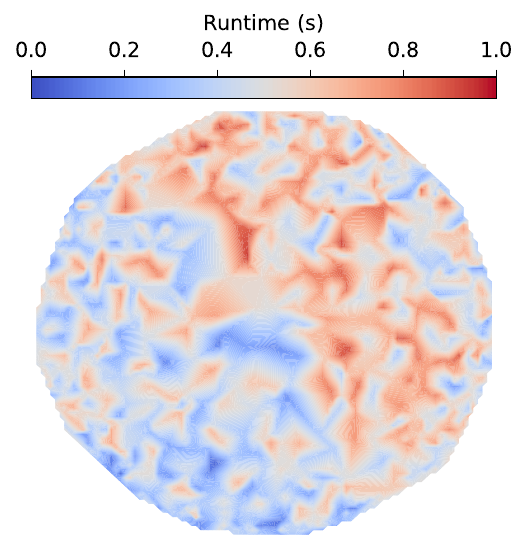}}
~\hfill
\subfloat[\texttt{SPLConqueror} landscape\\MAPE: $48.2\%$\\FDC: $\varrho=0.72$\\Hamming distance: $\Delta d=6.9$\\Performance gap: $\Delta p=0.08$\\Correlation length: $\ell=1.68$]{\includegraphics[width=0.49\columnwidth]{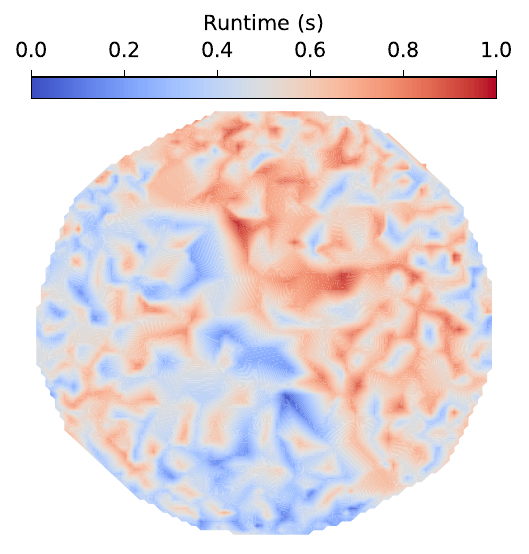}}

\caption{Configuration landscapes derived from the models for a batch model-based tuner for \textsc{SQLite}, showing a large MAPE reduction in a very bad accuracy (from a to b) might lead to trivial deviation on the modeled configuration landscape, resulting non-significant tuning quality change.} 
\label{fig:dis5}
\end{figure}

On the other hand, a drastic improvement in a model with badly performed accuracy might still lead to a trivial change in the tuning quality. Figure~\ref{fig:dis5} shows a typical example, in which we see that changing form \texttt{LR} to \texttt{SPLConqueror} on \textsc{SQLite} can improve the accuracy from 111.8\% to 48.2\%---a 63.6\% deviation (from Figure~\ref{fig:dis5}a to Figure~\ref{fig:dis5}b). However, with such a large accuracy change, we note that the landscapes are almost identical, i.e., both the metrics and visualization show no obvious deviations. 

\subsection{Theoretical Hypothesis}
\label{sec:theory-dis}

Landscape analysis helps us to better explain our observations based on empirical data, here, we additionally provide a theoretical hypothesis as to the fundamental causes of the results in this study. In particular, we conjecture the reason that a model cannot emulate the landscape well while being less useful to the tuning is due to the presence of high sparsity and ruggedness in the configuration landscape (i.e., similar configurations can cause drastically different impacts on the performance), which can be discussed from both the system view and modeling view below.

\subsubsection{System View}

A key cause of the sparsity and ruggedness in the configuration landscape is the design intention of many configurable systems, for which we identify three categories of options/interactions:

\begin{itemize}
    \item \textbf{Bottlenecks-related options:} These options, when changed to certain value ranges, can lead to drastic performance shifts. For example, in \textsc{Tomcat}, the option \texttt{maxKeepAliveRequests} controls the maximum number of requests that an HTTP connection can handle before it is closed. An overly high value of \texttt{maxKeepAliveRequests} may lead to excessive resource consumption and response delays. On the other hand, too small values cause a large number of concurrent connections that last for a long time and can exhaust the thread pool, thus making the system struggle.
    
    \item \textbf{Architecture-level options:} There are options that control the architectural selection of underlying algorithms, which can cause large and sudden changes in the performance. For example,
    e.g., in \textsc{Exastencils}, the option \texttt{explorationId} is used to store a series of categorical symbols that are associated with certain ``exploration" algorithms to be exploited, leading to largely deviated performance among the optional values. 
    
    
    \item \textbf{Dependent options:} There might be dependencies between two options, which can constrain their setting and leave ``holes'' in the configuration landscape. For instance, in \textsc{Hadoop}, the options \texttt{mapreduceJobReduces} and \texttt{mapreduceFrameworkName} have strong dependencies. Specifically, when \texttt{mapreduceFrameworkName} is set to \texttt{local}, \texttt{mapreduceJobReduces} will be ignored because the job runs locally and cannot be parallelized or scheduled in a distributed manner, thereby causing sparsity and ruggedness.
\end{itemize}

Since building the model is often a purely data-driven practice, domain information related to configurable systems is rarely considered. This prevents the model from properly emulating the configuration landscape, despite performing well on the accuracy metrics. 

\subsubsection{Modeling View}

From the modeling perspective, another reason that causes the model to be less helpful to the tuning is due to their local overfitting issue. Specifically, since we now know that the accuracy metrics can be misleading, highly ``accurate'' models can be detrimental to configuration tuning due to their tendency to overfit too narrowly on specific regions of the configuration landscape, prioritizing those regions of the configuration space with local optima or placing excessive emphasis on a small number of influential options. As a result, the model might be measured as ``excellent'' using misleading accuracy metrics, but it does not effectively generalize to the broader space. This failure to generalize can lead to a misrepresentation of the true configuration landscape, causing the tuner to miss potentially superior configurations. The model might also unnecessarily trap the tuner in local optima and prevent it from exploring potentially promising regions.


%% file: lessons.tex
\section{Lessons Learned and Opportunities}
\label{sec:lessons}

Our study provides evidence that challenges the general belief presented in Section~\ref{sec:background}, demonstrating its inaccuracy and misleading nature. Specifically, the results also allow us to learn several lessons on different aspects that help to provide insights for future research opportunities, which we discuss below.

\subsection{On How to Use Model in Tuning}

From \textbf{\textit{Finding 1}} and \textbf{\textit{Finding 2}}, we learn that

\keybox{
\textit{\textbf{Lesson 1:} Depending on how the model is applied, in general, it might or might not be useful for tuning quality. The community should, therefore, emphasize how the models can be progressively updated throughout the tuning (in sequential model-based tuners) rather than batch-train them in prior (batch model-based tuners).}
}

It is interesting to reveal that, compared with the model-free counterpart in terms of tuning quality, models tend to be more useful for sequential model-based tuners under sufficient and fair budget (\textbf{\textit{Finding 1}}) while having negligible or even harmful impacts on the batch model-based tuners (\textbf{\textit{Finding 2}})---which is still not fully comply with the general belief. This suggests that, while using the models would certainly help to relieve the cost and is desirable for some scenarios (e.g., when the possible budget is small), the way how they are used within the tuning is important for tuning quality. In particular, this also influences the designs of certain tuners: it is inappropriate to assume that a model-free tuner can be seamlessly and arbitrarily paired with a surrogate model (i.e., the batch model-based tuners) while retaining the same tuning effectiveness. Instead, a more sophisticated interplay between model updates and tuner guidance, such as those in the sequential model-based tuner, is required. To that end, a potential future opportunity lies in:

\keybox{
\textit{\textbf{Opportunity 1:} Efficient online model updating/learning for configuration tuning.}
}

Models can be updated within the existing sequential model-based tuners, but it is mostly achieved by retraining the entire model. This might be possible for simple models with little data, e.g., \texttt{LR} and \texttt{DT}, but soon would become infeasible for large models with increasing data size, e.g., \texttt{DeepPerf} and \texttt{DaL}, since every retraining can pose considerable overhead~\cite{DBLP:conf/icse/Chen19b}. Indeed, a complete model training from scratch can take up to an hour~\cite{DaL} for a complex model. As such, we envisage that there will be an increasing need for online model updating/learning strategies, tailored to the characteristics of configuration data and the tuner, that learns the newly measured configurations without discarding the current model and retaining it from scratch.


\subsection{On the Optimality of Model Choices}

A straightforward conclusion obtained from \textbf{\textit{Finding 3}} is

\keybox{
\textit{\textbf{Lesson 2:} The originally chosen models in the sequential model-based tuners are still far from being optimal for the tuning, hence researchers should follow a more thorough investigation of the model choice.}
}

It is surprising to observe that those models, albeit claimed to be the best choice in their corresponding sequential model-based tuners, are mostly sub-optimal for tuning quality. There are still considerable chances where they can be seamlessly substituted with another model to form a new tuner-model pair that achieves significant tuning improvement, thereby researchers should follow a more systematic justification when choosing the model. This suggests an interesting future research opportunity:

\keybox{
\textit{\textbf{Opportunity 2:} Automated model and tuner construction throughout configuration tuning.}
}

Since our results show that the best model (for tuning quality) differs depending on the system and tuner, we envisage that the optimal tuner-model pair for sequential model-based tuners would only become clear during tuning, and future tuners can benefit from a more loosely designed architecture. This, in turn, suggests a new bi-level optimization formulation of the configuration tuning problem wherein the first level, the goal is to find the best tuner-model pair while the second level simultaneously focuses on tuning the configuration under different tuner-model pairs.

\subsection{On the Relation between Model Accuracy and Tuning Quality}

A key insight we obtained from \textbf{\textit{Finding 4}} and \textbf{\textit{Finding 5}} is

\keybox{
\textit{\textbf{Lesson 3:} A more accurate model does not generally imply better tuning quality, hence the community should shift away from pure accuracy-driven research of surrogate models for configuration tuning.}
}

It is remarkable to note that, unlike what is implied in the general belief, the model accuracy rarely exhibits a strong positive correlation to its effectiveness on tuning quality, meaning that seeking a more accurate model does not often lead to better quality for tuning. In fact, in most cases, the model accuracy does not influence the tuning or can even retain a negative correlation. As explained in Section~\ref{sec:discussion}, this is due to the nature of the configuration landscape: the failure of the model to learn certain landscape properties that can significantly impact a tuner, e.g., fitness guidance, ruggedness, and the severity of local optima issues in the tuning landscape. Yet, these properties cannot be captured/measured by the accuracy metrics. The key message we obtained is that the current accuracy-driven research on modeling configuration is potentially misleading, and we should not rely on accuracy alone to judge the model's effectiveness for configuration turning. This, therefore, raises a promising research thread:

\keybox{
\textit{\textbf{Opportunity 3:} Additional proxies, alongside accuracy, for measuring the usefulness of the model in model-based configuration tuning.}
}

One such technique, as we have demonstrated in Section~\ref{sec:discussion}, is fitness landscape analysis~\cite{DBLP:series/sci/PitzerA12}, which contains many proxies to assess the behavior of a tuner. For example, as we have shown, the severity of local optima can be measured by correlation length~\cite{stadler1996landscapes}, directly quantifying the shape and difficulty of the surrogate landscape (as well as to what extent it differs from the real landscape). This, if used together with the accuracy, can better reflect how the model might help when used for the actual configuration tuning.

From our theoretical discussion in Section~\ref{sec:theory-dis}, we urge the community to revisit our strategy in building configuration performance models:

\keybox{
\textit{\textbf{Opportunity 4:} Configuration performance modeling should additionally consider code patterns that cause sparsity and ruggedness.}
}

Indeed, there have been recent efforts working towards a similar initiative~\cite{DaL,HINNPerf,DBLP:conf/icse/VelezJSAK21,DBLP:journals/ase/VelezJSSAK20,DeepPerf}. However, studies~\cite{DaL,HINNPerf,DeepPerf} that aim to tackle sparsity still originate from a data distribution point of view; while works~\cite{DBLP:journals/ase/VelezJSSAK20,DBLP:conf/icse/VelezJSAK21} that leverage information of code in building the model have not consider those patterns that cause sparsity and ruggedness. As such, a better synergy between those two categories of efforts is still necessary for future research.

\subsection{On the Significance of Model Accuracy Change}

\textbf{\textit{Finding 6}} provides us with an intuitive understanding: 

\keybox{
\textit{\textbf{Lesson 4:} The range of accuracy value influences how much change in the accuracy of the model can be non-trivially beneficial to the tuning quality. Therefore, for newly proposed models where MAPE or $\mu$RD are used, one should at least consolidate their claims on model effectiveness by comparing their accuracy changes with the average $\Delta$ discovered in this work.}
}

Indeed, we often witness from existing work that researchers claim ``the model increase by $x\%$ MAPE over the others'' as an indication of its usefulness for tasks like configuration tuning. \textbf{\textit{Finding 4}} and \textbf{\textit{Finding 5}} have already suggested this can be misleading, hence we should avoid making such a claim as the indication of the model's effectiveness for configuration tuning. \textbf{\textit{Finding 6}} further offers evidence that approximates to what extent those claims tend to be inappropriate. For example, suppose that, under sequential model-based tuning, there is a state-of-the-art model $B$ that has 30\% MAPE while a newly proposed model $A$ performs 25\%---a 5\% improvement. For that case, Figure~\ref{fig:rq5-1} shows that, since $B$'s MAPE is within $[30,40]$, the improvements (from $B$ to $A$) in accuracy need to be at least 13\% on average in order to create significantly better tuning results, and hence the superiority of the proposed model $A$ might not be meaningful from the model's usefulness standpoint. This leads to an interesting research opportunity:

\keybox{
\textit{\textbf{Opportunity 5:} A more thorough procedure for quantifying the meaningfulness of model accuracy's change, answering the question of ``how much accuracy change is practically meaningful?''.}
}

Our finding calls for a more systematic evaluation procedure for model assessment to draw conclusions on how much difference tends to be useful for practical configuration tuning. This goes beyond the introduction of metrics, paving the way for a more complete engineering methodology in configuration performance modeling.










%% file: threats.tex
\section{Threats to Validity}
\label{sec:threats}

This section elaborates on the potential threats to the validity of our study.

\subsection{Internal Threats} 

Internal threats are the potential biases introduced by the bias of parameter settings that may impact the accuracy or reliability of the findings, including:

\begin{itemize}
\item \textbf{Tuning Budget:} The tuning budget influences the tuning quality and can be system-specific. In this study, we follow a common strategy used in previous work~\cite{DBLP:journals/tosem/ChenL23a,DBLP:journals/tosem/ChenL23,DBLP:journals/ase/GerasimouCT18}: run all the tuners for each system and record the number of measurements that lead to reasonable convergence. We set the biggest budget found across the tuner to ensure fairness. Although this still does not guarantee true convergence, it serves as a sensibly good approximation considering the expensive measurements in configuration tuning.

\item \textbf{Hot-start size:} To ensure reliability, we use the most common and largest hot-start size to train the model for sequential model-based tuners. This number serves as a trade-off between exploration and exploitation: a too-large one would cause the tuning budget to become too limited while a too-small one might cause the model to severely mislead the tuning. Indeed, we cannot say for certain that this setting is optimal, but it tends to be the most pragmatic setting for our study.

\item \textbf{Training sample size:} We also follow what has been used in existing work~\cite{DeepPerf,DaL,HINNPerf} to choose the size for batch-training the model under batch model-based tuners. Again, we set the largest size found in the literature and it has been proven to be one of the most appropriate sample sizes in configuration performance modeling research.

\item \textbf{Other model/tuner settings:} For all models and tuners studied, we set the same parameter settings as used in their corresponding work, or follow the same hyperparameter tuning strategy, e.g., in \texttt{DeepPerf}.

\end{itemize}

\subsection{Construct Threats} Threats to construct validity may lie in the following aspects:

\begin{itemize}
\item \textbf{Metrics:} To ensure coverage, our studies consider two types of metrics (accuracy and tuning quality), each involving more than one metric that is of diverse nature. For example, MAPE and $\mu$RD are representative of their residual and ranked accuracy; while the performance and efficiency are the two most important factors in configuration tuning. All of those are commonly used in prior work~\cite{DeepPerf,DaL,HINNPerf,bestconfig,flash,DBLP:conf/sigsoft/NairMSA17,DBLP:conf/sigsoft/0001L21,DBLP:conf/icse/GaoZ0LY21}. Nevertheless, studying additional metrics might lead to additional insights.

\item \textbf{Stochastic bias:} Most of the models and tuners are stochastic in nature, therefore it is important to assess the statistical significance. To mitigate such, we repeat each experiment 30 runs and use Mann-Whitney U-Test and Scott-Knott ESD Test for pairwise and multiple comparisons, respectively. We also use Spearman correlation to analyze the result with interpretation widely used for software engineering problems~\cite{DBLP:conf/icse/Chen19b,DBLP:journals/tse/WattanakriengkraiWKTTIM23}.

\end{itemize}

\subsection{External Threats} External validity refers to limitations in the generalizability of the findings to other populations, such as the selection of subject systems and training samples:

\begin{itemize}
    \item \textbf{Software systems:} In this study, we consider 29 configurable systems that are of diverse domains, performance attributes, and scales, leading to one of the largest empirical studies to date in this field. Nevertheless, we acknowledge that this collection of systems is not exhaustive and the inclusion of more systems might prove to be fruitful.
    
    \item \textbf{Tuners:} To ensure comprehensive coverage, we consider a total of 17 tuners, including 8 sequential model-based ones and 9 model-free/batch model-based ones. These include sample methods, e.g., \texttt{Random}, and complex ones that involve different dimension/space reduction mechanisms, e.g., \texttt{BOCA}. Indeed, there are always newly proposed tuners that can further consolidate our findings.

    \item \textbf{Models:} We consider a list of 10 models that are commonly studied in the community. Again, while this list is not exhaustive, the results have already provided strong evidence that additional models are unlikely to invalidate the conclusions drawn. 
    
\end{itemize}

%% file: related.tex
\section{Related Work}
\label{sec:related}
In this section, we discuss the studies that are relevant to this work.

\subsection{Surrogate Models for Configuration Performance}


Researchers have proposed different configuration surrogate models to guide configuration tuning, ranging from statistical models~\cite{SPL, DECART, flash} to more recent deep learning-based ones~\cite{DeepPerf, HINNPerf, DaL, DBLP:journals/corr/sempl,DBLP:journals/tse/ChenB17}.

Among others, \texttt{SPLConqueror}~\cite{SPL} combines linear regression with different sampling methods to capture configuration option interactions. \texttt{DECART}~\cite{DECART} improves upon CART with hyperparameter tuning and an efficient resampling method. However, these models face challenges when dealing with sparse datasets---a common property for configurable software systems. To overcome these limitations, deep learning-based approaches have developed rapidly for learning configuration performance in the last decade. For example, \texttt{DeepPerf}~\cite{DeepPerf}, a deep neural network model, effectively addresses feature sparsity in all configurable systems by utilizing $L_{1}$ regularization to reduce the prediction error. \texttt{DaL}~\cite{DaL} divides the sparse configuration data into distinct and more focused subsets and trains a local deep neural network for each subset, which has been shown to achieve better accuracy. 


There has also been some discussion on the suitability of models for different systems. For example, Zhao et al.~\cite{zhao2023automatic} points out that \texttt{GP} can only model continuous options well but \texttt{RF} is capable of handling both continuous and categorical options. However, \texttt{RF} often does not guarantee high accuracy.

All the above discussion and work attempts to improve model accuracy based on the belief that the higher the accuracy, the better it will help with configuration tuning.


\subsection{Configuration Tuners}
Over the past decade, researchers have proposed and used various search algorithms to find better-performing configurations with a limited budget \cite{atconf,bestconfig,BOCA,conex,flash,Ottertune,llamatune,restune,lifeadaptive2025}. 


\subsubsection{Model-Free Tuners}

Among the model-free tuners where no surrogate model is used, \texttt{BestConfig}~\cite{bestconfig} utilizes the divide-and-diverge sampling method and the recursive bound-and-search algorithm to explore the sample space. The key is to leverage the sparse nature of the configuration space. \texttt{ConEx}~\cite{conex} is yet another example that leverages the Markov Chain Monte Carlo sampling strategy, paired with a simulated annealing-liked meta-heuristics, to tune the configuration. The designs are derived from the fact that the configuration space exhibits many local optima that need to be escaped.

Some other tuners see configuration tuning as a black box, hence leveraging a search algorithm that works on any optimization problem. For example, \texttt{GGA}~\cite{gga} tunes configuration by using a gender separation-based meta-heuristic genetic algorithm, which specifically emphasizes different selection pressures to males and females to expedite the search. \texttt{SWAY}~\cite{Sway} is an effort from the software engineering community that seeks to focus on the ``exploration'' of the space first: it selects a large set of configurations and then samples within that set to narrow down to the optimal one. The key motivation for those tuners that ignore models is primarily due to the assumption that ``a model cannot be trained accurately enough to help with the tuning''.

\subsubsection{Model-Based Tuners}

As mentioned, the model-free tuners can be seamlessly paired with any surrogate models that are pre-trained in advance to form batch model-based tuners, as what has been commonly followed in the field~\cite{DBLP:conf/sigsoft/0001L21,DBLP:journals/corr/abs-2112-07303,DBLP:journals/tosem/ChenLBY18,shi2024efficient}. For example, Chen et al.~\cite{DBLP:journals/corr/abs-2112-07303} state that their tuner can be paired with any models to facilitate cheaper tuning. Shi et al.~\cite{shi2024efficient} also leverage linear regression as a surrogate to boost the ability to explore in the tuner.

Sequential model-based tuners, in contrast, require more sophisticated search methods as the model would also need to be updated throughout the tuning. A vast majority of the examples exist but most of them primarily differ on the chosen model and how it is updated~\cite{Ottertune,restune,flash,BOCA}. For example, \texttt{OtterTune}~\cite{Ottertune} and \texttt{ResTune}~\cite{restune} have selected \texttt{GP} as their surrogate models while \texttt{FLASH}~\cite{flash} uses \texttt{DT} as a surrogate model to accelerate the search. Some other tuners consolidate the operators during the tuning, e.g., \texttt{BOCA}~\cite{BOCA} leverages Random Forest to identify the most important configuration options to serve as the key in the tuning and equip its sampling with a decay function, which gradually reduces the exploration of those non-important options. 

Yet, despite the popularity of model-based tuners, they mostly assume that a more accurate model is the dominant factor in designing a better tuner.



\subsection{White-box Model/Tuning}

Apart from the above black-box models/tuners, indeed, existing white-box models or tuning that relies on those models~\cite{DBLP:journals/ase/VelezJSSAK20,DBLP:conf/icse/VelezJSAK21} are useful in explaining certain internal knowledge about the system to humans or can guide the tuning to some extent (e.g., answering questions such as which options or the relevant code is more influential?). However, they still cannot provide additional means to measure the usefulness of the model during the tuning. For example, white box models still cannot quantify aspects such as fitness guidance, ruggedness, and the severity of local optima issues in the tuning landscape, all of which are important factors in the effectiveness of a configuration tuner. Therefore, as long as the model still acts as an active guide in those approaches and one solely measures the usefulness of the model via accuracy, then the misleading knowledge would still exist. Other white box tuning approaches that do not leverage the model often only work for specific systems (e.g., \texttt{MysqlTuner}~\cite{mysqltuner}), and hence not generalizable to the general configuration tuning problem we focus on.

The above is the key motivation behind the introduction of fitness landscape analysis~\cite{DBLP:series/sci/PitzerA12} in this work---a whole paradigm that provides insights about why/how the model is useful to the tuner or not. As such, the landscape analysis works for both black and white-box models/tuners. It is true that, at the current stage, the landscape metrics have not linked to the internal structure of the systems, e.g., the nature of their options; this is certainly one of our future works for proposing a more thorough analysis framework for configuration tuning deriving from fitness landscape analysis.

\subsection{Empirical Studies on Configuration}

Various empirical studies exist on software configuration from several different perspectives.



\subsubsection{General Configuration Studies}

In general, most empirical studies on configuration focus on understanding its characteristics. Among others, Xu et al.~\cite{DBLP:conf/sigsoft/XuJFZPT15} demonstrate that the number of configuration options has increased significantly over the years, and as such, most developers have not yet exploited the full benefits offered by those configurations. Zhang et al.~\cite{DBLP:conf/icse/ZhangHLL0X21} investigate how configurations evolve among system versions, from which some patterns were discovered. Other studies focus on understanding the consequence of inappropriate configuration, e.g., those that lead to performance bugs~\cite{DBLP:conf/esem/HanY16}.

While those works do not target model and tuning, they provide insights about the characteristics of configurable software systems and hence are orthogonal to this study.

\subsubsection{Studies for Configuration Learning}

Since leveraging the surrogate models is often considered a promising way to relive the expansiveness of configuration tuning, many studies have been conducted on different aspects of building such a model~\cite{DBLP:conf/splc/0003APJ21,DBLP:conf/msr/GongC22,DBLP:conf/kbse/JamshidiSVKPA17,DBLP:conf/icse/Chen19b,DBLP:journals/ase/CaoBZWWSLZ24}. For example, Gong and Chen~\cite{DBLP:conf/msr/GongC22} study the impact of encoding on the model accuracy, respectively. Jamshidi et al.~\cite{DBLP:conf/kbse/JamshidiSVKPA17} investigate how a model learned in one environment, e.g., hardware or version, can be transferred into the other while maintaining a good level of accuracy. Chen~\cite{DBLP:conf/icse/Chen19b} seeks to understand what is the best way to update a model when a new configuration is measured, considering either complete retraining or incremental learning. Indeed, those works emphasize the model accuracy as the key metric, but they are not concerned about how the model can be used for configuration tuning.

\subsubsection{Studies for Tuning without Models}

Empirical studies exist for examining the model-free tuners and have not investigated the usefulness of models for tuning. For example, Liao et al.~\cite{DBLP:journals/tosem/LiaoLSM22} present a study comparing the tuners for deep learning systems. Their findings confirm the necessity of configuration/hyperparameter tuning. Chen and Li~\cite{DBLP:journals/tosem/ChenL23a} also perform an empirical study on configuration tuning without a surrogate model. In particular, the goal is to understand whether some requirements should be considered as part of the tuning objectives, and under what circumstances they might be beneficial or harmful.


\subsubsection{Prior Understandings of Models for Tuning}

Some studies have focused on comparing model-based tuners explicitly, covering general configurable systems~\cite{sayyad2013parameter} or on specific domains such as database systems~\cite{zhang2022facilitating,van2021inquiry} and defect predictors~\cite{DBLP:conf/icse/LiX0WT20}. For example, Zhang et al.~\cite{zhang2022facilitating} study and compare different tuners in the database system domain. However, their emphasis is on examining the best tuner along with their models without systematically exploring the influence of the model on the tuning process. Aken et al. \cite{van2021inquiry} replace the model chosen by \texttt{OtterTune} with alternative models and compare their tuning performance, aiming to find the most effective model to be used with \texttt{OtterTune}. Nevertheless, again, their study neither summarizes how the accuracy of models can impact the tuning nor the importance of using a model. 




Overall, existing research has acknowledged that the differently chosen model can influence the tuning results, but there is still a lack of clear understanding regarding the true benefit of using models; the correlation between models (and their accuracy) to the tuning; and how accurate the model needs to be in order to create significant improvements to tuning---all of which are what we seek to study in this work.

%% file: conclusion.tex
\section{Conclusion}
\label{sec:conclusion}

This paper challenges a long-believing yet unconfirmed stereotype for model-based configuration tuning: the higher the model accuracy, the better the tuning results, and vice versa. We do that via conducting one of the largest scale empirical studies to date, consisting of 10 models, 17 tuners, and 29 systems, leading to 13,612 cases. The findings suggest that the models might not be useful depending on how they are used with the tuners and their accuracy can lie: it is not uncommon that more accurate models cannot lead to better tuning quality. We also reveal that the chosen model for most previously proposed tuners is far from optimal and document to what extent the model accuracy needs to change in order to significantly improve the tuning, according to different accuracy ranges. Among others, our key message is:


\begin{tcbitemize}[%
    raster columns=1, 
    raster rows=1
    ]
  \tcbitem[myhbox={}{Key takeaway}] \textit{We should take one step back from the natural “accuracy is all” belief for model-based configuration tuning.}
\end{tcbitemize}

We provide discussions on the rationale behind the observations from a new perspective of configuration landscape analysis. Those, together with the lessons learned, allow us to outline several promising research opportunities:

\begin{itemize}
    \item Efficient online model updating/learning for
configuration tuning.
\item Automated model and tuner construction
throughout configuration tuning.
\item Metrics beyond accuracy to measure model effectiveness to the tuning.
\item Sparsity/ruggedness relevant code patterns-driven configuration performance learning and modeling.
\item Procedure for quantifying the meaningfulness of model accuracy’s change.
\end{itemize}

This paper is merely a starting point of a series of fruitful future directions in this field. We hope that our findings will spark a dialogue on the reasoning behind the models' usefulness for the tuning process and hence further promote more fruitful research on software configuration tuning.

